\documentclass[prl,aps,superscriptaddress,twocolumn]{revtex4}
\usepackage{hyperref}
\usepackage{color} 
\usepackage{bm}
\usepackage{dcolumn}
\usepackage{times}
\usepackage{amssymb} 
\usepackage{amsmath} 
\usepackage{ragged2e}
\usepackage{graphicx}
\usepackage{epstopdf}
\usepackage{cancel}
\usepackage[english]{babel}
\usepackage{mathrsfs}
\usepackage{textcomp}
\pdfoutput=1
\usepackage{tikz}
\usepackage{pgfplots}
\usepgfplotslibrary{groupplots}
\usetikzlibrary{external}
\usepackage{amsthm}
\usepackage{amsfonts}
\usepackage{braket} 

\usepackage{cleveref}
\begin{document}

\title{
Frustrated extended Bose-Hubbard model and deconfined quantum critical points with optical lattices at the anti-magic wavelength 
}

\author{Niccol\`o Baldelli}
\email{niccolo.baldelli@icfo.eu}
\thanks{These two authors contributed equally}
\affiliation{ICFO - Institut de Ci\`encies Fot\`oniques, The Barcelona Institute of Science and Technology, 08860 Castelldefels (Barcelona), Spain}

\author{Cesar R. Cabrera}
\thanks{These two authors contributed equally}
\affiliation{Institut f\"ur Laserphysik, Universit\"at Hamburg,
Luruper Chaussee 149, 22761 Hamburg, Germany}

\author{Sergi Juli\`a-Farr\'e}
\affiliation{ICFO - Institut de Ci\`encies Fot\`oniques, The Barcelona Institute of Science and Technology, 08860 Castelldefels (Barcelona), Spain}

\author{Monika Aidelsburger}
\affiliation{Max-Planck-Institut f\"ur Quantenoptik, 85748 Garching, Germany}
\affiliation{Ludwig-Maximilians-Universit\"at M\"unchen, Schellingstr. 4, D-80799 Munich, Germany}
\affiliation{Munich Center for Quantum Science and Technology (MCQST), Schellingstr. 4, D-80799 Munich, Germany}

\author{Luca Barbiero}
\email{luca.barbiero@polito.it}
\affiliation{Institute for Condensed Matter Physics and Complex Systems,
DISAT, Politecnico di Torino, I-10129 Torino, Italy}

\date{\today}

\begin{abstract}
The study of geometrically frustrated many-body quantum systems is of central importance to uncover novel quantum mechanical effects. We design a scheme where ultracold bosons trapped in a one-dimensional state-dependent optical lattice are modeled by a frustrated Bose-Hubbard Hamiltonian. A derivation of the Hamiltonian parameters based on Cesium atoms, further show large tunability of contact and nearest-neighbour interactions. For pure contact repulsion, we discover the presence of two phases peculiar to frustrated quantum magnets: the bond-order-wave insulator with broken inversion symmetry and a chiral superfluid. When the nearest-neighbour repulsion becomes sizeable, a further density-wave insulator with broken translational symmetry can appear. We show that the phase transition between the two spontaneously-symmetry-broken phases is continuous, thus representing a one-dimensional deconfined quantum critical point not captured by the Landau–Ginzburg-Wilson symmetry-breaking paradigm. Our results provide a solid ground to unveil the novel quantum physics induced by the interplay of non-local interactions, geometrical frustration, and quantum fluctuations.

\end{abstract}

\maketitle

\paragraph{Introduction.}
Geometrically-frustrated many-body quantum systems~\cite{Lhuillier2001,Lacroix2011} represent a fruitful research field where a plethora of novel phases of matter has been unveiled. Paradigmatic examples are different topological insulators~\cite{Kane2005,Fujimoto2009,Wang2011,Wang2012} and superconductors~\cite{Qi2011,Sato_2017}, spin liquids~\cite{Balents2010,Zhou2017,Szasz2020}, and valence bond solids~\cite{ANDERSON1973153,Majumdar1969,Haldane1982}. Nevertheless, because of the deep complexity generated by competing interactions, frustration, and quantum fluctuations, various scenarios are still poorly understood. In this respect, numerical studies are highly demanding and often affected by finite size effects~\cite{Lauchli2011}. At the same time, possible sample imperfections and limited detection probes narrow the efficiency of solid-state experimental platforms~\cite{Ramirez1994,Starykh_2015}. \\
\indent The tunability and control offered by quantum simulation experiments based on ultracold atoms in optical lattices~\cite{gross2017} provide a promising alternative for understanding the behavior of a large variety of physical systems~\cite{Lewenstein07}. However, theoretical proposals to engineer ultracold frustrated synthetic materials are mainly based on direct implementations of specific geometries~\cite{Damski2005,Eckardt_2010,Glaetzle2015,Zhang2015,Yamamoto2020} and alternative approaches are scarce~\cite{Anisimovas2016,Cabedo2020,barbiero2022}. Moreover, experimental realizations of frustrated quantum systems relying on optical lattice engineering successfully achieved weakly interacting~\cite{An2018,Leung2020,Wang2021,Brown2022}, classical~\cite{Struck2011,Struck2013,saugmann22} or kinetically frustrated~\cite{Yang2021,mongkolkiattichai2022,xu2022doping,prichard2023directly,lebrat2023observation} regimes. Noticeably, these experiments investigated configurations with purely local couplings, while realizations of geometrically frustrated systems with beyond-contact interactions have not yet been achieved.\\
\begin{figure}
    \centering
    \includegraphics[width=\linewidth]{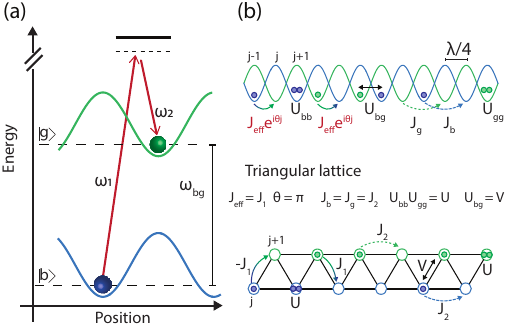}
    \caption{Experimental scheme to realize Eq.~\eqref{ham1}. (a) State-dependent optical lattice at the anti-magic-wavelength $\lambda$. Due to the opposite polarizability, both states experience a trapping potential with opposite strength. Raman-assisted tunneling between neighbouring sites is induced using a two-photon Raman transition between states $\ket{b}$ and $\ket{g}$ for $\omega_{bg}=\omega_1-\omega_2$ the energy difference between the two states on neighboring lattice sites.
    (b) The state-dependent lattice can be seen as two shifted sub-lattices with an effective lattice spacing of $\lambda/4$ (for a retro-reflected configuration), intra/interspecies tunneling $J_b,J_g/J_{\text{eff}}e^{\imath\theta j}$ and interaction $U_{bb},U_{gg}/U_{bg}$ (upper panel). (c) Choosing $J_b=J_g=J_2$, $J_{\text{eff}}=J_1$, $\theta=\pi$, $U_{bb}=U_{gg}=U$ and $U_{bg}=V$ this scheme is modeled by the frustrated extended Bose-Hubbard model defined in Eq.~\eqref{ham2}. }
    \label{fig1}
\end{figure}
Interactions with an extended range are a key ingredient to create spontaneous-symmetry-broken (SSB) states of matter with broken translational symmetry. The occurrence of these phases in various physical systems~\cite{gruner1988,Chester1970,GLENDENNING2001393} has sparked significant experimental interest leading to their observation in atomic frustration-free atom-cavity~\cite{landig2016,helson2022} setups, continuum-~\cite{tanzi2019,guo2019,chomaz2019} and lattice-dipolar~\cite{su2023} systems, as well as in out-of-equilibrium configurations~\cite{Guardado2021,Zahn2022}. While Rydberg atoms in optical tweezer arrays allow engineering long-range couplings also in frustrated geometries~\cite{Scholl2021,semeghini2021} and polar molecules in optical lattices represent a promising platform in this direction \cite{Christakis2023}, their effectiveness remains limited to the study of spin-$1/2$ systems. In this respect, proposals to investigate Hamiltonians with an enlarged Hilbert space, where beyond-contact repulsion and geometric frustration strongly compete are absent.\\
As illustrated schematically in Fig. 1, we design a realistic experimental setup, where a frustrated extended BoseHubbard Hamiltonian is realized using an atomic mixture in a one-dimensional (1D) lattice at the anti-magic wavelength~\cite{Anisimovas2016,QuantumWalk,PhysRevA.94.013620}. At this wavelength, the two different components experience the same polarizability with opposite signs, generating an opposite potential for each of the components. While this technique can be employed with several atomic species \cite{mandel_2003,forster_microwave_2009,belmechri_microwave_2013,riegger_2018,oppong_2022,heinz_2020,hohn_state-dependent_2023} with minimal heating, we provide specific values for an implementation with Cesium atoms. Here, convenient inter- and intra-species Feshbach resonances~\cite{CesiumResonances2019} enable the engineering of frustrated quantum systems with tunable contact and nearest-neighbour (NN) interactions without requiring large electric or magnetic dipole moments. Remarkably, in the presence of only local repulsion we recover the phases predicted to occur in frustrated triangular quantum magnets~\cite{Drechsler2007,SATO2011,Furukawa2012,Wolter2012,Orlova2017,Grams2022}. The latter include chiral superfluidity (CSF) and a site-inversion SSB bond-ordered-wave (BOW) insulator. When the NN repulsion is turned on, the BOW phase is destabilized in favor of a translational SSB insulator, namely a density wave (DW). By performing variational-uniform-matrix-product-states (VUMPS) calculations~\cite{Haegeman2013,Zauner2018}, we find the transition between the two SSB insulators (DW-BOW) to be continuous and thus going beyond the Landau–Ginzburg-Wilson symmetry-breaking paradigm~\cite{Landau_ssb,wilson_ssb}, which would instead predict a discontinuous first-order phase transition. As pioneering works demonstrated~\cite{Senthil2004,Senthil2004v2,senthil2023deconfined}, quantum fluctuations can indeed give rise to second-order continuous phase transitions between different ordered SSB phases where the gap vanishes in one specific point: the deconfined quantum critical point (DQCP). Because of their deep quantum nature combined with possible exotic properties like fractional excitations and emergent gauge fields, an exceptional theoretical effort has unveiled the presence of DQCPs in a large variety of 2D spin~\cite{Sandvik2007,Jiang_2008,motrunich2008,Lou2009,Banerjee2010,Sandvik2010,Harada2013,Chen2013,Nahum2015,Shao2016,Lee2019,song2023deconfined} and fermionic~\cite{Zi2019,Assaad2016,liu2022,Yuan2023} models as well as in 3D~\cite{Charrier2008,Sreejith2015}, 1D~\cite{Jiang2019,Roberts2019,Huang2019,Mudry2019,Roberts2021,lee2022}, and 0D~\cite{prembabu2022} two-level systems. Moreover, unique evidence of their possible existence has been provided in recent solid-state experiments~\cite{Zayed2017,Guo2020,cui2022,Tao2022}. Here, we prove that DQCPs can be accurately investigated with ultracold atoms in an optical lattice.  
\paragraph{Frustrated extended Bose-Hubbard (FEBH).}
As shown in Fig.~\ref{fig1} (a), we consider a two-component Bose gas trapped in a 1D state-dependent optical lattice with $L$ sites. The two atomic species, hereafter defined as $\ket{b}$ and $\ket{g}$, experience opposite polarizability and, because of the anti-magic wavelength condition, remain localized in two sub-lattices formed respectively by the intensity maxima and minima of the periodic potential. This configuration thus mimics an effective discrete geometry with $\tilde{L}=2L$ sites and lattice spacing $\lambda/4$ (for retro-reflected lattices), see Fig.~\ref{fig1} (b). Since the two sub-lattices have by definition the same depth, the $\ket{b}$- and $\ket{g}$-bosons experience the same direct hopping amplitude $J_b=J_g=J$. On the other hand, intra- $U_{bb}$, $U_{gg}$ and inter-species $U_{bg}$ interactions are potentially different and tunable. Furthermore, tunable Raman-assisted tunneling processes $J_{\text{eff}} e^{\imath\theta j}$ connect consecutive sites of different sub-lattices; as a consequence, one tunneling event is accompanied by converting one internal state into the other [Fig.~\ref{fig1} (a)] and $j$ is the $\tilde{L}$-lattice site index. Here $J_{\text{eff}}$ and $\theta$ are given by the intensity and wavevector of the Raman coupling \cite{Jaksch_2003,Gerbier_2010}. This setup is accurately modeled by the Hamiltonian
\begin{eqnarray}
\begin{split}
H=&-\sum_{j}\Big[J(a_{j}^\dagger a_{j+2}+h.c.)\\&+J_{\text{eff}}\left(e^{i\theta (2j-1)}a_{2j-1}^\dagger a_{2j}+e^{-i\theta (2j)}a_{2j}^\dagger a_{2j+1}+h.c.\right)\Big]\\
&+\sum_{j}\Big[\frac{U_{bb}}{2}n_{2j-1}(n_{2j-1}-1)+\frac{U_{gg}}{2}n_{2j}(n_{2j}-1)\Big]\\
&+U_{bg}\sum_in_in_{i+1},
\end{split}
\label{ham1}
\end{eqnarray}
where $a_{j}^\dagger/a_{j}$ is a bosonic creation/annihilation operator. To provide a more concrete implementation, we focus on two internal states of Cesium $\ket{b} \equiv \ket{F=3, m_F =3}$ and $\ket{g} \equiv \ket{F=3, m_F =2}$ where the interactions are tunable~\cite{SM}. A further essential aspect of the proposed configuration is its reliance on adiabatic state preparation. For instance, it is possible to prepare an initial state with $N$ bosons in the $\ket{b}$ state forming a Mott insulator with $N=N_b=L$. By subsequently introducing $J_{\text{eff}}e^{\imath\theta j}$, it becomes possible to populate the $\ket{g}$ state and therefore reach the regime of half-filling $\bar{n}=N/\tilde{L}=0.5$, with $N=N_b+N_g$, which is particularly suitable to explore SSB phases~\cite{su2023}. Finally, an adiabatic lowering of the lattice depth gives rise to a finite direct tunneling $J$.\\ As a specific example, we fix $U_{gg}=U_{bb}=U$ and $\theta=\pi$ so that, after renaming $J=J_2$, $J_{\text{eff}}=J_1$ and $U_{bg}=V$, Eq.~\eqref{ham1} becomes 
\begin{eqnarray}
\begin{split}
    H_{\text{FEBH}}=&-\sum_{j}\Big[J_2(a^{\dagger}_j a_{j+2}+h.c.)+J_1(-1)^j(a^{\dagger}_j a_{j+1}+h.c.)\Big]\\
    &+\frac{U}{2}\sum_{j}n_j(n_j-1)+V\sum_jn_jn_{j+1},
\end{split}
\label{ham2}
\end{eqnarray}
with $J_1,J_2>0$. As shown in Fig.~\ref{fig1} (b), Eq.~(\ref{ham2}) describes a frustrated extended Bose-Hubbard model in a triangular ladder at half filling, where the staggered nature of $J_1$ gives rise to the effective geometrical frustration and $U$ ($V$) accounts for the contact (NN) interaction. Although various versions and regimes of similar models have been theoretically studied~\cite{dallatorre2006,Dhar2012,Greschner2013,Zaletel2014,Mishra2015,Romen2018,Fraxanet2022,Roy2022,Halati2023}, configurations with staggered NN tunneling both with and without NN interaction have not been investigated. 
\paragraph{Effective frustrated quantum magnet.}
\begin{figure}
    \centering
    \includegraphics{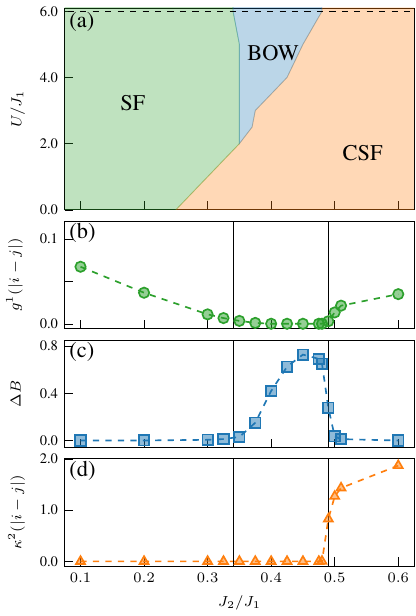}
    \caption{(a) Phase diagram of $H_{\text{FEBH}}$ Eq.~\eqref{ham2} as a function of $J_2/J_1$ and $U/J_1$, showing the superfluid (SF), bond-order-wave (BOW) and chiral superfluid (CSF) phases; (b) superfluid correlator $g^1(|i-j|)$ for $|i-j|=100$; (c) bond-order-wave order parameter $\Delta B$; (d) chiral superfluid correlator $\kappa^2(|i-j|)$ for $|i-j|=100$; The vertical continuous lines in (b)-(d) are the estimated transition points for $U/J_1=6$ [dashed line in (a)]. All the figures refer to the configuration where the total density $\bar{n}=0.5$ and the NN repulsion $V=0$. The VUMPS simulation have been performed by using a bond dimension $\chi$=400}
    \label{fig2}
\end{figure}
We begin our analysis by considering the $V=0$ case. As shown in Fig.~\ref{fig2}(a), in this regime $H_{\text{FEBH}}$ hosts three different phases. For weak frustration, namely for small $J_2/J_1$, we detect a gapless superfluid (SF), captured by the quasi long-range-order (LRO) of the correlator defined as
\begin{equation}
g^1(|i-j|)=\langle b_i^\dagger b_j\rangle,
\label{sf}
\end{equation} 
see Fig.~\ref{fig2}(b). On the other hand, $g^1(|i-j|)$ vanishes exponentially for strong enough $U$ and larger $J_2/J_1$. This behavior signals the appearance of a gapped phase~\cite{Giamarchi2004} which, as shown in Fig.~\ref{fig2}(c), is characterized by a finite value of the local order parameter~\footnote{Notice that the $+$ between the two operators is required because of the specific gauge constraint in which we are working, namely by the staggered $J_1$.}
\begin{equation}
\Delta B=\frac{1}{L}\sum_j\langle B_j+B_{j+1}\rangle,
\label{bow}
\end{equation}
where $B_j=(b_j^\dagger b_{j+1}+b_{j+1}^\dagger b_{j})$. Specifically, $\Delta B\neq 0$ demonstrates the presence of a dimerized BOW phase with broken site-inversion symmetry. A notable observation is that this lattice dimerization bears a striking resemblance to the Peierls instability~\cite{peierls_96}. In particular, while in real materials the effective dimerization is generated by the electron-phonon coupling, here it is induced by the combination of finite interaction, quantum fluctuation and geometrical frustration. It is further relevant to mention that the symmetry-protected topological nature of 1D~\cite{Julia2022} and 2D~\cite{Fraxanet2023} BOW phases have been recently discovered. In the strongly frustrated regime of large $J_2/J_1$, the BOW gives way to a new gapless state where $g^1(|i-j|)$ exhibits quasi LRO. The findings in Fig.~\ref{fig2}(d) illustrate that this regime can be classified as a CSF captured by the LRO of the correlator defined as
\begin{eqnarray}
\kappa^2(|i-j|)=\langle \kappa_i \kappa_j\rangle,
\label{csf}
\end{eqnarray}
where $\kappa_j=-\frac{\imath}{2}(b^\dagger_j b_{j+1} - b^\dagger_{j+1}b_j)$ is the vector chiral order parameter~\cite{Greschner2013,Zaletel2014}. This point shows this CSF to be characterized by alternated finite currents between NN sites, thus resembling an effective vortex-antivortex antiferromagnetic crystal with staggered loop currents around each effective triangular plaquette. As shown in Refs~\cite{SATO2011,Furukawa2012}, the three phases discussed above appear in the phase diagram of triangular Heisenberg Hamiltonians, which are believed to accurately model specific frustrated quantum magnets~\cite{Drechsler2007,Wolter2012,Orlova2017,Grams2022}. We confirm this strong analogy by considering the limit of infinite contact repulsion $U=\infty$, where Eq.~\eqref{ham2} maps exactly onto a magnetic frustrated Heisenberg model~\cite{SM}. Although Eqs.~(\ref{bow}) and (\ref{csf}) may suggest that detecting CSF and BOW phases requires demanding protocols to extract hopping amplitudes, we propose in~\cite{SM} an alternative scheme. Here, the CSF can be revealed through measurements of the momentum distribution, while the BOW can be detected through a string correlator that only requires local density measurements. Our scheme thus presents itself as a valuable and alternative method for studying and comprehending frustrated quantum magnets.
\paragraph{Deconfined quantum critical points.}
\begin{figure*}
    \centering
    \includegraphics{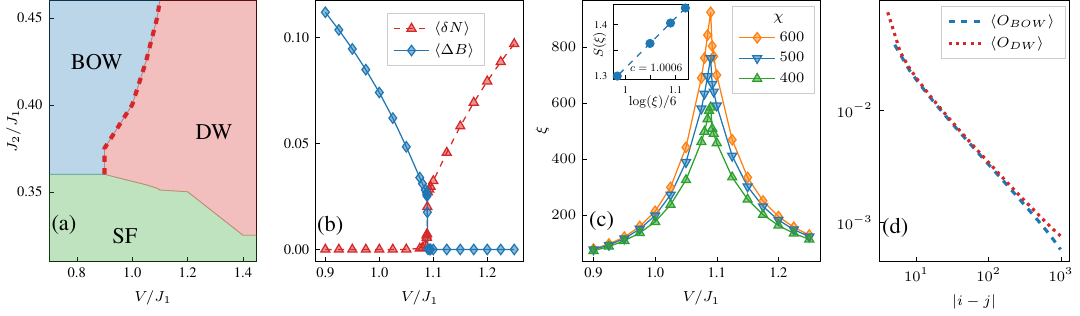}
    \caption{Effect of the NN repulsion $V$ in the Hamiltonian~\eqref{ham2}. For all the panels, we fix $U/J_1=6$ and $\bar{n}=0.5$. (a) Phase diagram of $H_{\text{EFBH}}$ in the $V/J_1-J_2/J_1$ plane, using $\chi=400$. (b) $\Delta B$ and $\delta N$ as a function of $V/J_1$ for $J_2/J_1=0.45$ and $\chi = 600$. (c) The correlation length $\xi$ as a function of $V/J_1$ for different bond-dimension $\chi$ and fixed $J_2/J_1=0.45$. Inset: scaling of the entanglement entropy $S(\xi)$ as a function of $\xi$ at the critical point for bond dimensions $\chi = 400,500,600$ showing the extrapolated central charge $c=1$. (d) Decay of $O_{\text{DW}}$ and $O_{\text{BOW}}$ at the critical point for fixed $J_2/J_1=0.45$.}
    \label{fig3}
\end{figure*}
The scattering properties of Cs atoms also make it possible to engineer sizable non-local repulsion in the range $V/U<0.3$, see~\cite{SM}. In order to be able to reach relatively large values of $V$, we fix  $U/J_1=6$ [see the dashed line in Fig.~\ref{fig2}(a)] and we concentrate on the regime of weak and intermediate frustration ~\footnote{We checked that the CSF is not affected by the presence of $V$.}. Our VUMPS analysis in Fig.~\ref{fig3}(a) finds the SF stable with respect to the addition of moderate $V$. On the contrary, for a gradual increase of the NN repulsion the system undergoes a phase transition from the BOW to a different SSB insulator identified by the local order parameter
\begin{equation}
\delta N=\frac{1}{L}\sum_j(-1)^j(n_j-\bar{n}).
\label{dw}
\end{equation}
 A finite $\delta N$, which can be accurately probed through quantum gas microscopy~\cite{Gross2021}, reflects the spatial alternation between empty and singly occupied sites, revealing the appearance of a DW characterized by broken translational symmetry.  Phase transitions between two gapped SSB phases are usually described by the Landau–Ginzburg-Wilson theory~\cite{Landau_ssb, wilson_ssb}. The latter predicts the existence of a first-order transition where the gap never vanishes and it exhibits a discontinuity between two finite values. Interestingly, configurations deep in quantum regimes can deviate from this paradigm. Quantum fluctuations can indeed give rise to continuous phase transitions between different SSB states, where the gap vanishes only at a single point: the deconfined quantum critical point~\cite{Senthil2004,Senthil2004v2}. The numerical detection of DQCPs is highly challenging. On the one hand, their complete characterization requires algorithms directly mimicking the thermodynamic limit and, on the other hand, the diverging correlation length $\xi$ occurring at criticality requires a specific entanglement scaling~\cite{Pollmann2009}. In this respect, VUMPS are particularly suitable. This advantage stems from the fact that the variational optimization is performed on a unit cell directly in the thermodynamic limit. In this way it is possible to avoid the slower and non-monotonous convergence to the variational optimum~\cite{Zauner2018} peculiar to algorithms involving a gradual growth of the system size. Thanks to this technique, our results in Figs.~\ref{fig3}(b)-(d) clearly demonstrate the BOW-DW transition to be continuous and therefore the presence of 1D DQCPs. Specifically, we find that $\Delta B$ and $\delta N$ vanish continuously exactly at the same $V/J_1$, see Fig.~\ref{fig3}(b). In order to rule out the presence of a weakly first-order phase transition, we extract the correlation length $\xi$~\footnote{The correlation length is extracted through the relation $\xi = -N/\log(\lambda_2)$ where $N$ is the number of sites of a unit cell and $\lambda_2$ is second highest eigenvalue of the transfer matrix.}. Within a matrix-product-state (MPS) formalism, $\xi$ must not depend on the bond dimension $\chi$ in the presence of a finite gap. On the other hand, a $\chi$-dependence in the form of a cusp should be observed only at the critical point~\cite{Roberts2019,Huang2019,lee2022}. Fig.~\ref{fig3}(c) accurately confirms this behavior. In the inset of Fig.~\ref{fig3}(c) we further extract the central charge $c$ through the relation $S=c\log(\xi(\chi))/6$, where for MPS around a critical point it is proven that $ \xi(\chi) \sim \chi^{\kappa}$ with $\kappa/6=(c (\sqrt{12/c}+1))^{-1}$~\cite{Pollmann2009}.
Conformal field theories rigorously demonstrate that the extracted $c=1$ implies indeed the presence of a critical regime which, in this case, is further characterized by an emergent $U(1)$ symmetry. In order to enforce our results, we calculate the decay of the correlations functions
\begin{equation}
O_{\text{BOW}}(|i-j|)=\langle(B_i+B_{i+1})(B_j+B_{j+1})\rangle,
\label{bowcorr}
\end{equation} 
\begin{equation}
O_{\text{DW}}(|i-j|)=\langle(n_i-\bar{n})(n_j-\bar{n})\rangle.
\label{cdwcorr}
\end{equation}
Here, we expect LRO of $O_{\text{BOW}}/O_{\text{DW}}$ uniquely in the BOW/DW phase while both should vanish algebraically at criticality. Fig.~\ref{fig3}(d) precisely shows the expected power-law decay. This point, that we have been able to demonstrate thanks to the fact that VUMPS mimic the thermodynamic limit, unambiguously proves the critical nature of this transition point. Finally, we point out that, as imposed in DQCPs~\cite{Roberts2019}, the two correlation functions clearly vanish in the thermodynamic limit with the same exponent. 
\paragraph{Conclusions.}
We designed an experimental scheme based on two-component bosonic atoms in an optical lattice at the anti-magic wavelength. We modeled this setup through a single-band frustrated extended Bose-Hubbard model where contact and nearest-neighbour interactions, geometrical frustration and quantum fluctuation strongly compete. For contact interaction only, we demonstrated that states of matter peculiar to frustrated quantum magnets, namely chiral superfluids and spontaneously-symmetry-broken bond-order-wave insulators, can be created and accurately probed. For strong nearest-neighbour repulsion, a new spontaneously-symmetry-broken density wave insulator occurs. We proved that the phase transition between the two spontaneously-symmetry-broken phases is continuous, thus representing an elusive quantum mechanical effect: one-dimensional deconfined quantum critical points not captured by the Landau–Ginzburg-Wilson symmetry-breaking paradigm.\\ \indent From an experimental perspective, our findings offer an alternative approach to engineer beyond-contact interactions with two crucial advantages. Firstly, it allows to naturally incorporate the effect of  geometrical frustration by means of an adiabatic state preparation and therefore to realistically explore deep quantum regimes. Secondly, it does not require the use of particles with strong magnetic or electric dipole moment. From the theory perspective, we performed an advanced numerical analysis of an unexplored version of the paradigmatic Bose-Hubbard Hamiltonian. This has revealed its richness of intriguing quantum mechanical effects. Specifically, on one side this model is able to mimic triangular frustrated quantum magnets and, on the other, it represents an example of a soft-core bosonic system exhibiting deconfined-quantum-critical-points. Our work provides valuable insights and opens up new avenues for studying and comprehending strongly interacting frustrated quantum systems. 
\begin{acknowledgments}
We thank D. Burba, M. Capone, I. Carusotto, F. Cinti, E. Demler, F. Ferlaino, M. Lewenstein, G. Juzeliunas, G. Modugno, I. B. Spielman, L. Tanzi, C. Weitenberg, O. Zilberberg and the Cs team at LMU for discussions. M.A. received funding from the Deutsche Forschungsgemeinschaft (DFG, German Research Foundation) via Research Unit FOR 2414 under project number 277974659, via Research Unit FOR 5522 under project number 499180199, under Germany’s Excellence Strategy – EXC-2111 – 390814868 (MCQST) and under Horizon Europe programme HORIZON-CL4-2022-QUANTUM-02-SGA via the project 101113690 (PASQuanS2.1). L. B. acknowledges Politecnico di Torino for the starting package grant number 54$\_$RSG21BL01. C.R.C has received funding from the European Union's Horizon 2020 research and innovation programme under the Marie Skłodowska-Curie grant agreement No 897142. N.B. acknowledges support from a ``la Caixa" Foundation fellowship (ID 100010434, code LCF/BQ/DI20/11780033). ICFO group acknowledges support from: ERC AdG NOQIA; MICIN/AEI (PGC2018-0910.13039/501100011033, CEX2019-000910-S/10.13039/501100011033, Plan National FIDEUA PID2019-106901GB-I00, FPI; MICIIN with funding from European Union NextGenerationEU (PRTR-C17.I1): QUANTERA MAQS PCI2019-111828-2); MCIN/AEI/ 10.13039/501100011033 and by the ``European Union NextGeneration EU/PRTR'' QUANTERA DYNAMITE PCI2022-132919 within the QuantERA II Programme that has received funding from the European Union’s Horizon 2020 research and innovation programme under Grant Agreement No 101017733 Proyectos de I+D+I ``Retos Colaboración'' QUSPIN RTC2019-007196-7); Fundaci\'o Cellex; Fundaci\'o Mir-Puig; Generalitat de Catalunya (European Social Fund FEDER and CERCA program, AGAUR Grant No. 2021 SGR 01452, QuantumCAT \ U16-011424, co-funded by ERDF Operational Program of Catalonia 2014-2020); Barcelona Supercomputing Center MareNostrum (FI-2023-1-0013); EU (PASQuanS2.1, 101113690); EU Horizon 2020 FET-OPEN OPTOlogic (Grant No 899794); EU Horizon Europe Program (Grant Agreement 101080086 — NeQST), National Science Centre, Poland (Symfonia Grant No. 2016/20/W/ST4/00314); ICFO Internal ``QuantumGaudi'' project; European Union's Horizon 2020 research and innovation program under the Marie-Skłodowska-Curie grant agreement No 101029393 (STREDCH) and No 847648 (``La Caixa” Junior Leaders fellowships ID100010434: LCF/BQ/PI19/11690013, LCF/BQ/PI20/11760031, LCF/BQ/PR20/11770012, LCF/BQ/PR21/11840013). Views and opinions expressed are, however, those of the author(s) only and do not necessarily reflect those of the European Union, European Commission, European Climate, Infrastructure and Environment Executive Agency (CINEA), or any other granting authority. Neither the European Union nor any granting authority can be held responsible for them.
\end{acknowledgments}

\bibliographystyle{apsrev4-2}
\bibliography{biblio.bib}

\begin{thebibliography}{136}%
\makeatletter
\providecommand \@ifxundefined [1]{%
 \@ifx{#1\undefined}
}%
\providecommand \@ifnum [1]{%
 \ifnum #1\expandafter \@firstoftwo
 \else \expandafter \@secondoftwo
 \fi
}%
\providecommand \@ifx [1]{%
 \ifx #1\expandafter \@firstoftwo
 \else \expandafter \@secondoftwo
 \fi
}%
\providecommand \natexlab [1]{#1}%
\providecommand \enquote  [1]{``#1''}%
\providecommand \bibnamefont  [1]{#1}%
\providecommand \bibfnamefont [1]{#1}%
\providecommand \citenamefont [1]{#1}%
\providecommand \href@noop [0]{\@secondoftwo}%
\providecommand \href [0]{\begingroup \@sanitize@url \@href}%
\providecommand \@href[1]{\@@startlink{#1}\@@href}%
\providecommand \@@href[1]{\endgroup#1\@@endlink}%
\providecommand \@sanitize@url [0]{\catcode `\\12\catcode `\$12\catcode `\&12\catcode `\#12\catcode `\^12\catcode `\_12\catcode `\%12\relax}%
\providecommand \@@startlink[1]{}%
\providecommand \@@endlink[0]{}%
\providecommand \url  [0]{\begingroup\@sanitize@url \@url }%
\providecommand \@url [1]{\endgroup\@href {#1}{\urlprefix }}%
\providecommand \urlprefix  [0]{URL }%
\providecommand \Eprint [0]{\href }%
\providecommand \doibase [0]{https://doi.org/}%
\providecommand \selectlanguage [0]{\@gobble}%
\providecommand \bibinfo  [0]{\@secondoftwo}%
\providecommand \bibfield  [0]{\@secondoftwo}%
\providecommand \translation [1]{[#1]}%
\providecommand \BibitemOpen [0]{}%
\providecommand \bibitemStop [0]{}%
\providecommand \bibitemNoStop [0]{.\EOS\space}%
\providecommand \EOS [0]{\spacefactor3000\relax}%
\providecommand \BibitemShut  [1]{\csname bibitem#1\endcsname}%
\let\auto@bib@innerbib\@empty
\bibitem [{\citenamefont {Lhuillier}\ and\ \citenamefont {Misguich}(2001)}]{Lhuillier2001}%
  \BibitemOpen
  \bibfield  {author} {\bibinfo {author} {\bibfnamefont {C.}~\bibnamefont {Lhuillier}}\ and\ \bibinfo {author} {\bibfnamefont {G.}~\bibnamefont {Misguich}},\ }\bibinfo {title} {Frustrated quantum magnets},\ in\ \href {https://doi.org/10.1007/3-540-45649-X_6} {\emph {\bibinfo {booktitle} {High Magnetic Fields: Applications in Condensed Matter Physics and Spectroscopy}}},\ \bibinfo {editor} {edited by\ \bibinfo {editor} {\bibfnamefont {C.}~\bibnamefont {Berthier}}, \bibinfo {editor} {\bibfnamefont {L.~P.}\ \bibnamefont {L{\'e}vy}},\ and\ \bibinfo {editor} {\bibfnamefont {G.}~\bibnamefont {Martinez}}}\ (\bibinfo  {publisher} {Springer Berlin Heidelberg},\ \bibinfo {address} {Berlin, Heidelberg},\ \bibinfo {year} {2001})\ pp.\ \bibinfo {pages} {161--190}\BibitemShut {NoStop}%
\bibitem [{\citenamefont {Lacroix}\ \emph {et~al.}(2011)\citenamefont {Lacroix}, \citenamefont {Mendels},\ and\ \citenamefont {Mila}}]{Lacroix2011}%
  \BibitemOpen
  \bibfield  {author} {\bibinfo {author} {\bibfnamefont {C.}~\bibnamefont {Lacroix}}, \bibinfo {author} {\bibfnamefont {P.}~\bibnamefont {Mendels}},\ and\ \bibinfo {author} {\bibfnamefont {F.}~\bibnamefont {Mila}},\ }\href@noop {} {\emph {\bibinfo {title} {Introduction to frustrated magnetism}}}\ (\bibinfo  {publisher} {Springer Ser. Solid-State Sci},\ \bibinfo {year} {2011})\BibitemShut {NoStop}%
\bibitem [{\citenamefont {Kane}\ and\ \citenamefont {Mele}(2005)}]{Kane2005}%
  \BibitemOpen
  \bibfield  {author} {\bibinfo {author} {\bibfnamefont {C.~L.}\ \bibnamefont {Kane}}\ and\ \bibinfo {author} {\bibfnamefont {E.~J.}\ \bibnamefont {Mele}},\ }\href {https://doi.org/10.1103/PhysRevLett.95.226801} {\bibfield  {journal} {\bibinfo  {journal} {Phys. Rev. Lett.}\ }\textbf {\bibinfo {volume} {95}},\ \bibinfo {pages} {226801} (\bibinfo {year} {2005})}\BibitemShut {NoStop}%
\bibitem [{\citenamefont {Fujimoto}(2009)}]{Fujimoto2009}%
  \BibitemOpen
  \bibfield  {author} {\bibinfo {author} {\bibfnamefont {S.}~\bibnamefont {Fujimoto}},\ }\href {https://doi.org/10.1103/PhysRevLett.103.047203} {\bibfield  {journal} {\bibinfo  {journal} {Phys. Rev. Lett.}\ }\textbf {\bibinfo {volume} {103}},\ \bibinfo {pages} {047203} (\bibinfo {year} {2009})}\BibitemShut {NoStop}%
\bibitem [{\citenamefont {Wang}\ \emph {et~al.}(2011{\natexlab{a}})\citenamefont {Wang}, \citenamefont {Gu}, \citenamefont {Gong},\ and\ \citenamefont {Sheng}}]{Wang2011}%
  \BibitemOpen
  \bibfield  {author} {\bibinfo {author} {\bibfnamefont {Y.-F.}\ \bibnamefont {Wang}}, \bibinfo {author} {\bibfnamefont {Z.-C.}\ \bibnamefont {Gu}}, \bibinfo {author} {\bibfnamefont {C.-D.}\ \bibnamefont {Gong}},\ and\ \bibinfo {author} {\bibfnamefont {D.~N.}\ \bibnamefont {Sheng}},\ }\href {https://doi.org/10.1103/PhysRevLett.107.146803} {\bibfield  {journal} {\bibinfo  {journal} {Phys. Rev. Lett.}\ }\textbf {\bibinfo {volume} {107}},\ \bibinfo {pages} {146803} (\bibinfo {year} {2011}{\natexlab{a}})}\BibitemShut {NoStop}%
\bibitem [{\citenamefont {Wang}\ \emph {et~al.}(2012)\citenamefont {Wang}, \citenamefont {Yao}, \citenamefont {Gu}, \citenamefont {Gong},\ and\ \citenamefont {Sheng}}]{Wang2012}%
  \BibitemOpen
  \bibfield  {author} {\bibinfo {author} {\bibfnamefont {Y.-F.}\ \bibnamefont {Wang}}, \bibinfo {author} {\bibfnamefont {H.}~\bibnamefont {Yao}}, \bibinfo {author} {\bibfnamefont {Z.-C.}\ \bibnamefont {Gu}}, \bibinfo {author} {\bibfnamefont {C.-D.}\ \bibnamefont {Gong}},\ and\ \bibinfo {author} {\bibfnamefont {D.~N.}\ \bibnamefont {Sheng}},\ }\href {https://doi.org/10.1103/PhysRevLett.108.126805} {\bibfield  {journal} {\bibinfo  {journal} {Phys. Rev. Lett.}\ }\textbf {\bibinfo {volume} {108}},\ \bibinfo {pages} {126805} (\bibinfo {year} {2012})}\BibitemShut {NoStop}%
\bibitem [{\citenamefont {Qi}\ and\ \citenamefont {Zhang}(2011)}]{Qi2011}%
  \BibitemOpen
  \bibfield  {author} {\bibinfo {author} {\bibfnamefont {X.-L.}\ \bibnamefont {Qi}}\ and\ \bibinfo {author} {\bibfnamefont {S.-C.}\ \bibnamefont {Zhang}},\ }\href {https://doi.org/10.1103/RevModPhys.83.1057} {\bibfield  {journal} {\bibinfo  {journal} {Rev. Mod. Phys.}\ }\textbf {\bibinfo {volume} {83}},\ \bibinfo {pages} {1057} (\bibinfo {year} {2011})}\BibitemShut {NoStop}%
\bibitem [{\citenamefont {Sato}\ and\ \citenamefont {Ando}(2017)}]{Sato_2017}%
  \BibitemOpen
  \bibfield  {author} {\bibinfo {author} {\bibfnamefont {M.}~\bibnamefont {Sato}}\ and\ \bibinfo {author} {\bibfnamefont {Y.}~\bibnamefont {Ando}},\ }\href {https://doi.org/10.1088/1361-6633/aa6ac7} {\bibfield  {journal} {\bibinfo  {journal} {Rep. Prog. Phys.}\ }\textbf {\bibinfo {volume} {80}},\ \bibinfo {pages} {076501} (\bibinfo {year} {2017})}\BibitemShut {NoStop}%
\bibitem [{\citenamefont {Balents}()}]{Balents2010}%
  \BibitemOpen
  \bibfield  {author} {\bibinfo {author} {\bibfnamefont {L.}~\bibnamefont {Balents}},\ }\href {https://doi.org/10.1038/nature08917} {\bibfield  {journal} {\bibinfo  {journal} {Nature}\ }\textbf {\bibinfo {volume} {464}},\ \bibinfo {pages} {199}}\BibitemShut {NoStop}%
\bibitem [{\citenamefont {Zhou}\ \emph {et~al.}(2017)\citenamefont {Zhou}, \citenamefont {Kanoda},\ and\ \citenamefont {Ng}}]{Zhou2017}%
  \BibitemOpen
  \bibfield  {author} {\bibinfo {author} {\bibfnamefont {Y.}~\bibnamefont {Zhou}}, \bibinfo {author} {\bibfnamefont {K.}~\bibnamefont {Kanoda}},\ and\ \bibinfo {author} {\bibfnamefont {T.-K.}\ \bibnamefont {Ng}},\ }\href {https://doi.org/10.1103/RevModPhys.89.025003} {\bibfield  {journal} {\bibinfo  {journal} {Rev. Mod. Phys.}\ }\textbf {\bibinfo {volume} {89}},\ \bibinfo {pages} {025003} (\bibinfo {year} {2017})}\BibitemShut {NoStop}%
\bibitem [{\citenamefont {Szasz}\ \emph {et~al.}(2020)\citenamefont {Szasz}, \citenamefont {Motruk}, \citenamefont {Zaletel},\ and\ \citenamefont {Moore}}]{Szasz2020}%
  \BibitemOpen
  \bibfield  {author} {\bibinfo {author} {\bibfnamefont {A.}~\bibnamefont {Szasz}}, \bibinfo {author} {\bibfnamefont {J.}~\bibnamefont {Motruk}}, \bibinfo {author} {\bibfnamefont {M.~P.}\ \bibnamefont {Zaletel}},\ and\ \bibinfo {author} {\bibfnamefont {J.~E.}\ \bibnamefont {Moore}},\ }\href {https://doi.org/10.1103/PhysRevX.10.021042} {\bibfield  {journal} {\bibinfo  {journal} {Phys. Rev. X}\ }\textbf {\bibinfo {volume} {10}},\ \bibinfo {pages} {021042} (\bibinfo {year} {2020})}\BibitemShut {NoStop}%
\bibitem [{\citenamefont {Anderson}(1973)}]{ANDERSON1973153}%
  \BibitemOpen
  \bibfield  {author} {\bibinfo {author} {\bibfnamefont {P.}~\bibnamefont {Anderson}},\ }\href {https://doi.org/https://doi.org/10.1016/0025-5408(73)90167-0} {\bibfield  {journal} {\bibinfo  {journal} {Mater. Res. Bull.}\ }\textbf {\bibinfo {volume} {8}},\ \bibinfo {pages} {153} (\bibinfo {year} {1973})}\BibitemShut {NoStop}%
\bibitem [{\citenamefont {Majumdar}\ and\ \citenamefont {Ghosh}(1969)}]{Majumdar1969}%
  \BibitemOpen
  \bibfield  {author} {\bibinfo {author} {\bibfnamefont {C.~K.}\ \bibnamefont {Majumdar}}\ and\ \bibinfo {author} {\bibfnamefont {D.~K.}\ \bibnamefont {Ghosh}},\ }\href {https://doi.org/10.1063/1.1664978} {\bibfield  {journal} {\bibinfo  {journal} {J. Math. Phys}\ }\textbf {\bibinfo {volume} {10}},\ \bibinfo {pages} {1388} (\bibinfo {year} {1969})}\BibitemShut {NoStop}%
\bibitem [{\citenamefont {Haldane}(1982)}]{Haldane1982}%
  \BibitemOpen
  \bibfield  {author} {\bibinfo {author} {\bibfnamefont {F.~D.~M.}\ \bibnamefont {Haldane}},\ }\href {https://doi.org/10.1103/PhysRevB.25.4925} {\bibfield  {journal} {\bibinfo  {journal} {Phys. Rev. B}\ }\textbf {\bibinfo {volume} {25}},\ \bibinfo {pages} {4925} (\bibinfo {year} {1982})}\BibitemShut {NoStop}%
\bibitem [{\citenamefont {L{\"a}uchli}(2011)}]{Lauchli2011}%
  \BibitemOpen
  \bibfield  {author} {\bibinfo {author} {\bibfnamefont {A.~M.}\ \bibnamefont {L{\"a}uchli}},\ }\bibinfo {title} {Numerical simulations of frustrated systems},\ in\ \href {https://doi.org/10.1007/978-3-642-10589-0_18} {\emph {\bibinfo {booktitle} {Introduction to Frustrated Magnetism: Materials, Experiments, Theory}}},\ \bibinfo {editor} {edited by\ \bibinfo {editor} {\bibfnamefont {C.}~\bibnamefont {Lacroix}}, \bibinfo {editor} {\bibfnamefont {P.}~\bibnamefont {Mendels}},\ and\ \bibinfo {editor} {\bibfnamefont {F.}~\bibnamefont {Mila}}}\ (\bibinfo  {publisher} {Springer Berlin Heidelberg},\ \bibinfo {address} {Berlin, Heidelberg},\ \bibinfo {year} {2011})\ pp.\ \bibinfo {pages} {481--511}\BibitemShut {NoStop}%
\bibitem [{\citenamefont {Ramirez}(1994)}]{Ramirez1994}%
  \BibitemOpen
  \bibfield  {author} {\bibinfo {author} {\bibfnamefont {A.~P.}\ \bibnamefont {Ramirez}},\ }\href {https://doi.org/10.1146/annurev.ms.24.080194.002321} {\bibfield  {journal} {\bibinfo  {journal} {Annu. rev. mater. sci.}\ }\textbf {\bibinfo {volume} {24}},\ \bibinfo {pages} {453} (\bibinfo {year} {1994})}\BibitemShut {NoStop}%
\bibitem [{\citenamefont {Starykh}(2015)}]{Starykh_2015}%
  \BibitemOpen
  \bibfield  {author} {\bibinfo {author} {\bibfnamefont {O.~A.}\ \bibnamefont {Starykh}},\ }\href {https://doi.org/10.1088/0034-4885/78/5/052502} {\bibfield  {journal} {\bibinfo  {journal} {Rep. Prog. Phys.}\ }\textbf {\bibinfo {volume} {78}},\ \bibinfo {pages} {052502} (\bibinfo {year} {2015})}\BibitemShut {NoStop}%
\bibitem [{\citenamefont {Gross}\ and\ \citenamefont {Bloch}(2017)}]{gross2017}%
  \BibitemOpen
  \bibfield  {author} {\bibinfo {author} {\bibfnamefont {C.}~\bibnamefont {Gross}}\ and\ \bibinfo {author} {\bibfnamefont {I.}~\bibnamefont {Bloch}},\ }\href {https://doi.org/10.1126/science.aal3837} {\bibfield  {journal} {\bibinfo  {journal} {Science}\ }\textbf {\bibinfo {volume} {357}},\ \bibinfo {pages} {995} (\bibinfo {year} {2017})}\BibitemShut {NoStop}%
\bibitem [{\citenamefont {Lewenstein}\ \emph {et~al.}(2007)\citenamefont {Lewenstein}, \citenamefont {Sanpera}, \citenamefont {Ahufinger}, \citenamefont {Damski}, \citenamefont {Sen(De)},\ and\ \citenamefont {Sen}}]{Lewenstein07}%
  \BibitemOpen
  \bibfield  {author} {\bibinfo {author} {\bibfnamefont {M.}~\bibnamefont {Lewenstein}}, \bibinfo {author} {\bibfnamefont {A.}~\bibnamefont {Sanpera}}, \bibinfo {author} {\bibfnamefont {V.}~\bibnamefont {Ahufinger}}, \bibinfo {author} {\bibfnamefont {B.}~\bibnamefont {Damski}}, \bibinfo {author} {\bibfnamefont {A.}~\bibnamefont {Sen(De)}},\ and\ \bibinfo {author} {\bibfnamefont {U.}~\bibnamefont {Sen}},\ }\href {https://doi.org/10.1080/00018730701223200} {\bibfield  {journal} {\bibinfo  {journal} {Adv. Phys.}\ }\textbf {\bibinfo {volume} {56}},\ \bibinfo {pages} {243} (\bibinfo {year} {2007})}\BibitemShut {NoStop}%
\bibitem [{\citenamefont {Damski}\ \emph {et~al.}(2005)\citenamefont {Damski}, \citenamefont {Fehrmann}, \citenamefont {Everts}, \citenamefont {Baranov}, \citenamefont {Santos},\ and\ \citenamefont {Lewenstein}}]{Damski2005}%
  \BibitemOpen
  \bibfield  {author} {\bibinfo {author} {\bibfnamefont {B.}~\bibnamefont {Damski}}, \bibinfo {author} {\bibfnamefont {H.}~\bibnamefont {Fehrmann}}, \bibinfo {author} {\bibfnamefont {H.-U.}\ \bibnamefont {Everts}}, \bibinfo {author} {\bibfnamefont {M.}~\bibnamefont {Baranov}}, \bibinfo {author} {\bibfnamefont {L.}~\bibnamefont {Santos}},\ and\ \bibinfo {author} {\bibfnamefont {M.}~\bibnamefont {Lewenstein}},\ }\href {https://doi.org/10.1103/PhysRevA.72.053612} {\bibfield  {journal} {\bibinfo  {journal} {Phys. Rev. A}\ }\textbf {\bibinfo {volume} {72}},\ \bibinfo {pages} {053612} (\bibinfo {year} {2005})}\BibitemShut {NoStop}%
\bibitem [{\citenamefont {Eckardt}\ \emph {et~al.}(2010)\citenamefont {Eckardt}, \citenamefont {Hauke}, \citenamefont {Soltan-Panahi}, \citenamefont {Becker}, \citenamefont {Sengstock},\ and\ \citenamefont {Lewenstein}}]{Eckardt_2010}%
  \BibitemOpen
  \bibfield  {author} {\bibinfo {author} {\bibfnamefont {A.}~\bibnamefont {Eckardt}}, \bibinfo {author} {\bibfnamefont {P.}~\bibnamefont {Hauke}}, \bibinfo {author} {\bibfnamefont {P.}~\bibnamefont {Soltan-Panahi}}, \bibinfo {author} {\bibfnamefont {C.}~\bibnamefont {Becker}}, \bibinfo {author} {\bibfnamefont {K.}~\bibnamefont {Sengstock}},\ and\ \bibinfo {author} {\bibfnamefont {M.}~\bibnamefont {Lewenstein}},\ }\href {https://doi.org/10.1209/0295-5075/89/10010} {\bibfield  {journal} {\bibinfo  {journal} {EPL}\ }\textbf {\bibinfo {volume} {89}},\ \bibinfo {pages} {10010} (\bibinfo {year} {2010})}\BibitemShut {NoStop}%
\bibitem [{\citenamefont {Glaetzle}\ \emph {et~al.}(2015)\citenamefont {Glaetzle}, \citenamefont {Dalmonte}, \citenamefont {Nath}, \citenamefont {Gross}, \citenamefont {Bloch},\ and\ \citenamefont {Zoller}}]{Glaetzle2015}%
  \BibitemOpen
  \bibfield  {author} {\bibinfo {author} {\bibfnamefont {A.~W.}\ \bibnamefont {Glaetzle}}, \bibinfo {author} {\bibfnamefont {M.}~\bibnamefont {Dalmonte}}, \bibinfo {author} {\bibfnamefont {R.}~\bibnamefont {Nath}}, \bibinfo {author} {\bibfnamefont {C.}~\bibnamefont {Gross}}, \bibinfo {author} {\bibfnamefont {I.}~\bibnamefont {Bloch}},\ and\ \bibinfo {author} {\bibfnamefont {P.}~\bibnamefont {Zoller}},\ }\href {https://doi.org/10.1103/PhysRevLett.114.173002} {\bibfield  {journal} {\bibinfo  {journal} {Phys. Rev. Lett.}\ }\textbf {\bibinfo {volume} {114}},\ \bibinfo {pages} {173002} (\bibinfo {year} {2015})}\BibitemShut {NoStop}%
\bibitem [{\citenamefont {Zhang}\ and\ \citenamefont {Jo}(2015)}]{Zhang2015}%
  \BibitemOpen
  \bibfield  {author} {\bibinfo {author} {\bibfnamefont {T.}~\bibnamefont {Zhang}}\ and\ \bibinfo {author} {\bibfnamefont {G.-B.}\ \bibnamefont {Jo}},\ }\href {https://doi.org/10.1038/srep16044} {\bibfield  {journal} {\bibinfo  {journal} {Sci. Rep.}\ }\textbf {\bibinfo {volume} {5}},\ \bibinfo {pages} {16044} (\bibinfo {year} {2015})}\BibitemShut {NoStop}%
\bibitem [{\citenamefont {Yamamoto}\ \emph {et~al.}(2020)\citenamefont {Yamamoto}, \citenamefont {Fukuhara},\ and\ \citenamefont {Danshita}}]{Yamamoto2020}%
  \BibitemOpen
  \bibfield  {author} {\bibinfo {author} {\bibfnamefont {D.}~\bibnamefont {Yamamoto}}, \bibinfo {author} {\bibfnamefont {T.}~\bibnamefont {Fukuhara}},\ and\ \bibinfo {author} {\bibfnamefont {I.}~\bibnamefont {Danshita}},\ }\href {https://doi.org/10.1038/s42005-020-0323-5} {\bibfield  {journal} {\bibinfo  {journal} {Commun. Phys.}\ }\textbf {\bibinfo {volume} {3}} (\bibinfo {year} {2020})}\BibitemShut {NoStop}%
\bibitem [{\citenamefont {Anisimovas}\ \emph {et~al.}(2016{\natexlab{a}})\citenamefont {Anisimovas}, \citenamefont {Ra\ifmmode \check{c}\else \v{c}\fi{}i\ifmmode~\bar{u}\else \={u}\fi{}nas}, \citenamefont {Str\"ater}, \citenamefont {Eckardt}, \citenamefont {Spielman},\ and\ \citenamefont {Juzeli\ifmmode~\bar{u}\else \={u}\fi{}nas}}]{Anisimovas2016}%
  \BibitemOpen
  \bibfield  {author} {\bibinfo {author} {\bibfnamefont {E.}~\bibnamefont {Anisimovas}}, \bibinfo {author} {\bibfnamefont {M.}~\bibnamefont {Ra\ifmmode \check{c}\else \v{c}\fi{}i\ifmmode~\bar{u}\else \={u}\fi{}nas}}, \bibinfo {author} {\bibfnamefont {C.}~\bibnamefont {Str\"ater}}, \bibinfo {author} {\bibfnamefont {A.}~\bibnamefont {Eckardt}}, \bibinfo {author} {\bibfnamefont {I.~B.}\ \bibnamefont {Spielman}},\ and\ \bibinfo {author} {\bibfnamefont {G.}~\bibnamefont {Juzeli\ifmmode~\bar{u}\else \={u}\fi{}nas}},\ }\href {https://doi.org/10.1103/PhysRevA.94.063632} {\bibfield  {journal} {\bibinfo  {journal} {Phys. Rev. A}\ }\textbf {\bibinfo {volume} {94}},\ \bibinfo {pages} {063632} (\bibinfo {year} {2016}{\natexlab{a}})}\BibitemShut {NoStop}%
\bibitem [{\citenamefont {Cabedo}\ \emph {et~al.}(2020)\citenamefont {Cabedo}, \citenamefont {Claramunt}, \citenamefont {Mompart}, \citenamefont {Ahufinger},\ and\ \citenamefont {Celi}}]{Cabedo2020}%
  \BibitemOpen
  \bibfield  {author} {\bibinfo {author} {\bibfnamefont {J.}~\bibnamefont {Cabedo}}, \bibinfo {author} {\bibfnamefont {J.}~\bibnamefont {Claramunt}}, \bibinfo {author} {\bibfnamefont {J.}~\bibnamefont {Mompart}}, \bibinfo {author} {\bibfnamefont {V.}~\bibnamefont {Ahufinger}},\ and\ \bibinfo {author} {\bibfnamefont {A.}~\bibnamefont {Celi}},\ }\href {https://doi.org/10.1140/epjd/e2020-10129-1} {\bibfield  {journal} {\bibinfo  {journal} {Eur. Phys. J. D}\ }\textbf {\bibinfo {volume} {74}},\ \bibinfo {pages} {123} (\bibinfo {year} {2020})}\BibitemShut {NoStop}%
\bibitem [{\citenamefont {Barbiero}\ \emph {et~al.}(2022)\citenamefont {Barbiero}, \citenamefont {Cabedo}, \citenamefont {Lewenstein}, \citenamefont {Tarruell},\ and\ \citenamefont {Celi}}]{barbiero2022}%
  \BibitemOpen
  \bibfield  {author} {\bibinfo {author} {\bibfnamefont {L.}~\bibnamefont {Barbiero}}, \bibinfo {author} {\bibfnamefont {J.}~\bibnamefont {Cabedo}}, \bibinfo {author} {\bibfnamefont {M.}~\bibnamefont {Lewenstein}}, \bibinfo {author} {\bibfnamefont {L.}~\bibnamefont {Tarruell}},\ and\ \bibinfo {author} {\bibfnamefont {A.}~\bibnamefont {Celi}},\ }\href@noop {} {} (\bibinfo {year} {2022}),\ \Eprint {https://arxiv.org/abs/2212.06112} {arXiv:2212.06112 [cond-mat.quant-gas]} \BibitemShut {NoStop}%
\bibitem [{\citenamefont {An}\ \emph {et~al.}(2018)\citenamefont {An}, \citenamefont {Meier},\ and\ \citenamefont {Gadway}}]{An2018}%
  \BibitemOpen
  \bibfield  {author} {\bibinfo {author} {\bibfnamefont {F.~A.}\ \bibnamefont {An}}, \bibinfo {author} {\bibfnamefont {E.~J.}\ \bibnamefont {Meier}},\ and\ \bibinfo {author} {\bibfnamefont {B.}~\bibnamefont {Gadway}},\ }\href {https://doi.org/10.1103/PhysRevX.8.031045} {\bibfield  {journal} {\bibinfo  {journal} {Phys. Rev. X}\ }\textbf {\bibinfo {volume} {8}},\ \bibinfo {pages} {031045} (\bibinfo {year} {2018})}\BibitemShut {NoStop}%
\bibitem [{\citenamefont {Leung}\ \emph {et~al.}(2020)\citenamefont {Leung}, \citenamefont {Schwarz}, \citenamefont {Chang}, \citenamefont {Brown}, \citenamefont {Unnikrishnan},\ and\ \citenamefont {Stamper-Kurn}}]{Leung2020}%
  \BibitemOpen
  \bibfield  {author} {\bibinfo {author} {\bibfnamefont {T.-H.}\ \bibnamefont {Leung}}, \bibinfo {author} {\bibfnamefont {M.~N.}\ \bibnamefont {Schwarz}}, \bibinfo {author} {\bibfnamefont {S.-W.}\ \bibnamefont {Chang}}, \bibinfo {author} {\bibfnamefont {C.~D.}\ \bibnamefont {Brown}}, \bibinfo {author} {\bibfnamefont {G.}~\bibnamefont {Unnikrishnan}},\ and\ \bibinfo {author} {\bibfnamefont {D.}~\bibnamefont {Stamper-Kurn}},\ }\href {https://doi.org/10.1103/PhysRevLett.125.133001} {\bibfield  {journal} {\bibinfo  {journal} {Phys. Rev. Lett.}\ }\textbf {\bibinfo {volume} {125}},\ \bibinfo {pages} {133001} (\bibinfo {year} {2020})}\BibitemShut {NoStop}%
\bibitem [{\citenamefont {Wang}\ \emph {et~al.}(2011{\natexlab{b}})\citenamefont {Wang}, \citenamefont {Luo}, \citenamefont {Liu}, \citenamefont {Liu}, \citenamefont {Hemmerich},\ and\ \citenamefont {Xu}}]{Wang2021}%
  \BibitemOpen
  \bibfield  {author} {\bibinfo {author} {\bibfnamefont {X.-Q.}\ \bibnamefont {Wang}}, \bibinfo {author} {\bibfnamefont {G.-Q.}\ \bibnamefont {Luo}}, \bibinfo {author} {\bibfnamefont {J.-Y.}\ \bibnamefont {Liu}}, \bibinfo {author} {\bibfnamefont {W.~V.}\ \bibnamefont {Liu}}, \bibinfo {author} {\bibfnamefont {A.}~\bibnamefont {Hemmerich}},\ and\ \bibinfo {author} {\bibfnamefont {Z.-F.}\ \bibnamefont {Xu}},\ }\href {https://doi.org/10.1038/s41586-021-03702-0} {\bibfield  {journal} {\bibinfo  {journal} {Nature}\ }\textbf {\bibinfo {volume} {596}},\ \bibinfo {pages} {227} (\bibinfo {year} {2011}{\natexlab{b}})}\BibitemShut {NoStop}%
\bibitem [{\citenamefont {Brown}\ \emph {et~al.}(2022)\citenamefont {Brown}, \citenamefont {Chang}, \citenamefont {Schwarz}, \citenamefont {Leung}, \citenamefont {Kozii}, \citenamefont {Avdoshkin}, \citenamefont {Moore},\ and\ \citenamefont {Stamper-Kurn}}]{Brown2022}%
  \BibitemOpen
  \bibfield  {author} {\bibinfo {author} {\bibfnamefont {C.~D.}\ \bibnamefont {Brown}}, \bibinfo {author} {\bibfnamefont {S.-W.}\ \bibnamefont {Chang}}, \bibinfo {author} {\bibfnamefont {M.~N.}\ \bibnamefont {Schwarz}}, \bibinfo {author} {\bibfnamefont {T.-H.}\ \bibnamefont {Leung}}, \bibinfo {author} {\bibfnamefont {V.}~\bibnamefont {Kozii}}, \bibinfo {author} {\bibfnamefont {A.}~\bibnamefont {Avdoshkin}}, \bibinfo {author} {\bibfnamefont {J.~E.}\ \bibnamefont {Moore}},\ and\ \bibinfo {author} {\bibfnamefont {D.}~\bibnamefont {Stamper-Kurn}},\ }\href {https://doi.org/10.1126/science.abm6442} {\bibfield  {journal} {\bibinfo  {journal} {Science}\ }\textbf {\bibinfo {volume} {377}},\ \bibinfo {pages} {1319} (\bibinfo {year} {2022})}\BibitemShut {NoStop}%
\bibitem [{\citenamefont {Struck}\ \emph {et~al.}(2011)\citenamefont {Struck}, \citenamefont {\"Olschl\"ager}, \citenamefont {Targat}, \citenamefont {Soltan-Panahi}, \citenamefont {Eckardt}, \citenamefont {Lewenstein}, \citenamefont {Windpassinger},\ and\ \citenamefont {Sengstock}}]{Struck2011}%
  \BibitemOpen
  \bibfield  {author} {\bibinfo {author} {\bibfnamefont {J.}~\bibnamefont {Struck}}, \bibinfo {author} {\bibfnamefont {C.}~\bibnamefont {\"Olschl\"ager}}, \bibinfo {author} {\bibfnamefont {R.~L.}\ \bibnamefont {Targat}}, \bibinfo {author} {\bibfnamefont {P.}~\bibnamefont {Soltan-Panahi}}, \bibinfo {author} {\bibfnamefont {A.}~\bibnamefont {Eckardt}}, \bibinfo {author} {\bibfnamefont {M.}~\bibnamefont {Lewenstein}}, \bibinfo {author} {\bibfnamefont {P.}~\bibnamefont {Windpassinger}},\ and\ \bibinfo {author} {\bibfnamefont {K.}~\bibnamefont {Sengstock}},\ }\href {https://doi.org/10.1126/science.1207239} {\bibfield  {journal} {\bibinfo  {journal} {Science}\ }\textbf {\bibinfo {volume} {333}},\ \bibinfo {pages} {996} (\bibinfo {year} {2011})}\BibitemShut {NoStop}%
\bibitem [{\citenamefont {Struck}\ \emph {et~al.}(2013)\citenamefont {Struck}, \citenamefont {Weinberg}, \citenamefont {Ölschläger}, \citenamefont {Windpassinger}, \citenamefont {Simonet}, \citenamefont {Sengstock}, \citenamefont {Höppner}, \citenamefont {Hauke}, \citenamefont {Eckardt}, \citenamefont {Lewenstein},\ and\ \citenamefont {Mathey}}]{Struck2013}%
  \BibitemOpen
  \bibfield  {author} {\bibinfo {author} {\bibfnamefont {J.}~\bibnamefont {Struck}}, \bibinfo {author} {\bibfnamefont {M.}~\bibnamefont {Weinberg}}, \bibinfo {author} {\bibfnamefont {C.}~\bibnamefont {Ölschläger}}, \bibinfo {author} {\bibfnamefont {P.}~\bibnamefont {Windpassinger}}, \bibinfo {author} {\bibfnamefont {J.}~\bibnamefont {Simonet}}, \bibinfo {author} {\bibfnamefont {K.}~\bibnamefont {Sengstock}}, \bibinfo {author} {\bibfnamefont {R.}~\bibnamefont {Höppner}}, \bibinfo {author} {\bibfnamefont {P.}~\bibnamefont {Hauke}}, \bibinfo {author} {\bibfnamefont {A.}~\bibnamefont {Eckardt}}, \bibinfo {author} {\bibfnamefont {M.}~\bibnamefont {Lewenstein}},\ and\ \bibinfo {author} {\bibfnamefont {L.}~\bibnamefont {Mathey}},\ }\href {https://doi.org/10.1038/nphys2750} {\bibfield  {journal} {\bibinfo  {journal} {Nat. Phys.}\ }\textbf {\bibinfo {volume} {9}},\ \bibinfo {pages} {738} (\bibinfo {year} {2013})}\BibitemShut {NoStop}%
\bibitem [{\citenamefont {Saugmann}\ \emph {et~al.}(2022)\citenamefont {Saugmann}, \citenamefont {Vargas}, \citenamefont {Kiefer}, \citenamefont {Hachman}, \citenamefont {Eichberger}, \citenamefont {Hemmerich},\ and\ \citenamefont {Larson}}]{saugmann22}%
  \BibitemOpen
  \bibfield  {author} {\bibinfo {author} {\bibfnamefont {P.}~\bibnamefont {Saugmann}}, \bibinfo {author} {\bibfnamefont {J.}~\bibnamefont {Vargas}}, \bibinfo {author} {\bibfnamefont {Y.}~\bibnamefont {Kiefer}}, \bibinfo {author} {\bibfnamefont {M.}~\bibnamefont {Hachman}}, \bibinfo {author} {\bibfnamefont {R.}~\bibnamefont {Eichberger}}, \bibinfo {author} {\bibfnamefont {A.}~\bibnamefont {Hemmerich}},\ and\ \bibinfo {author} {\bibfnamefont {J.}~\bibnamefont {Larson}},\ }\href {https://doi.org/10.1103/PhysRevA.106.L041302} {\bibfield  {journal} {\bibinfo  {journal} {Phys. Rev. A}\ }\textbf {\bibinfo {volume} {106}},\ \bibinfo {pages} {L041302} (\bibinfo {year} {2022})}\BibitemShut {NoStop}%
\bibitem [{\citenamefont {Yang}\ \emph {et~al.}(2021)\citenamefont {Yang}, \citenamefont {Liu}, \citenamefont {Mongkolkiattichai},\ and\ \citenamefont {Schauss}}]{Yang2021}%
  \BibitemOpen
  \bibfield  {author} {\bibinfo {author} {\bibfnamefont {J.}~\bibnamefont {Yang}}, \bibinfo {author} {\bibfnamefont {L.}~\bibnamefont {Liu}}, \bibinfo {author} {\bibfnamefont {J.}~\bibnamefont {Mongkolkiattichai}},\ and\ \bibinfo {author} {\bibfnamefont {P.}~\bibnamefont {Schauss}},\ }\href {https://doi.org/10.1103/PRXQuantum.2.020344} {\bibfield  {journal} {\bibinfo  {journal} {PRX Quantum}\ }\textbf {\bibinfo {volume} {2}},\ \bibinfo {pages} {020344} (\bibinfo {year} {2021})}\BibitemShut {NoStop}%
\bibitem [{\citenamefont {Mongkolkiattichai}\ \emph {et~al.}(2022)\citenamefont {Mongkolkiattichai}, \citenamefont {Liu}, \citenamefont {Garwood}, \citenamefont {Yang},\ and\ \citenamefont {Schauss}}]{mongkolkiattichai2022}%
  \BibitemOpen
  \bibfield  {author} {\bibinfo {author} {\bibfnamefont {J.}~\bibnamefont {Mongkolkiattichai}}, \bibinfo {author} {\bibfnamefont {L.}~\bibnamefont {Liu}}, \bibinfo {author} {\bibfnamefont {D.}~\bibnamefont {Garwood}}, \bibinfo {author} {\bibfnamefont {J.}~\bibnamefont {Yang}},\ and\ \bibinfo {author} {\bibfnamefont {P.}~\bibnamefont {Schauss}},\ }\href@noop {} {} (\bibinfo {year} {2022}),\ \Eprint {https://arxiv.org/abs/2210.14895} {arXiv:2210.14895 [cond-mat.quant-gas]} \BibitemShut {NoStop}%
\bibitem [{\citenamefont {Xu}\ \emph {et~al.}(2022)\citenamefont {Xu}, \citenamefont {Kendrick}, \citenamefont {Kale}, \citenamefont {Gang}, \citenamefont {Ji}, \citenamefont {Scalettar}, \citenamefont {Lebrat},\ and\ \citenamefont {Greiner}}]{xu2022doping}%
  \BibitemOpen
  \bibfield  {author} {\bibinfo {author} {\bibfnamefont {M.}~\bibnamefont {Xu}}, \bibinfo {author} {\bibfnamefont {L.~H.}\ \bibnamefont {Kendrick}}, \bibinfo {author} {\bibfnamefont {A.}~\bibnamefont {Kale}}, \bibinfo {author} {\bibfnamefont {Y.}~\bibnamefont {Gang}}, \bibinfo {author} {\bibfnamefont {G.}~\bibnamefont {Ji}}, \bibinfo {author} {\bibfnamefont {R.~T.}\ \bibnamefont {Scalettar}}, \bibinfo {author} {\bibfnamefont {M.}~\bibnamefont {Lebrat}},\ and\ \bibinfo {author} {\bibfnamefont {M.}~\bibnamefont {Greiner}},\ }\href@noop {} {} (\bibinfo {year} {2022}),\ \Eprint {https://arxiv.org/abs/2212.13983} {arXiv:2212.13983 [cond-mat.quant-gas]} \BibitemShut {NoStop}%
\bibitem [{\citenamefont {Prichard}\ \emph {et~al.}(2023)\citenamefont {Prichard}, \citenamefont {Spar}, \citenamefont {Morera}, \citenamefont {Demler}, \citenamefont {Yan},\ and\ \citenamefont {Bakr}}]{prichard2023directly}%
  \BibitemOpen
  \bibfield  {author} {\bibinfo {author} {\bibfnamefont {M.~L.}\ \bibnamefont {Prichard}}, \bibinfo {author} {\bibfnamefont {B.~M.}\ \bibnamefont {Spar}}, \bibinfo {author} {\bibfnamefont {I.}~\bibnamefont {Morera}}, \bibinfo {author} {\bibfnamefont {E.}~\bibnamefont {Demler}}, \bibinfo {author} {\bibfnamefont {Z.~Z.}\ \bibnamefont {Yan}},\ and\ \bibinfo {author} {\bibfnamefont {W.~S.}\ \bibnamefont {Bakr}},\ }\href@noop {} {} (\bibinfo {year} {2023}),\ \Eprint {https://arxiv.org/abs/2308.12951} {arXiv:2308.12951 [cond-mat.quant-gas]} \BibitemShut {NoStop}%
\bibitem [{\citenamefont {Lebrat}\ \emph {et~al.}(2023)\citenamefont {Lebrat}, \citenamefont {Xu}, \citenamefont {Kendrick}, \citenamefont {Kale}, \citenamefont {Gang}, \citenamefont {Seetharaman}, \citenamefont {Morera}, \citenamefont {Khatami}, \citenamefont {Demler},\ and\ \citenamefont {Greiner}}]{lebrat2023observation}%
  \BibitemOpen
  \bibfield  {author} {\bibinfo {author} {\bibfnamefont {M.}~\bibnamefont {Lebrat}}, \bibinfo {author} {\bibfnamefont {M.}~\bibnamefont {Xu}}, \bibinfo {author} {\bibfnamefont {L.~H.}\ \bibnamefont {Kendrick}}, \bibinfo {author} {\bibfnamefont {A.}~\bibnamefont {Kale}}, \bibinfo {author} {\bibfnamefont {Y.}~\bibnamefont {Gang}}, \bibinfo {author} {\bibfnamefont {P.}~\bibnamefont {Seetharaman}}, \bibinfo {author} {\bibfnamefont {I.}~\bibnamefont {Morera}}, \bibinfo {author} {\bibfnamefont {E.}~\bibnamefont {Khatami}}, \bibinfo {author} {\bibfnamefont {E.}~\bibnamefont {Demler}},\ and\ \bibinfo {author} {\bibfnamefont {M.}~\bibnamefont {Greiner}},\ }\href@noop {} {} (\bibinfo {year} {2023}),\ \Eprint {https://arxiv.org/abs/2308.12269} {arXiv:2308.12269 [cond-mat.quant-gas]} \BibitemShut {NoStop}%
\bibitem [{\citenamefont {Gr\"uner}(1988)}]{gruner1988}%
  \BibitemOpen
  \bibfield  {author} {\bibinfo {author} {\bibfnamefont {G.}~\bibnamefont {Gr\"uner}},\ }\href {https://doi.org/10.1103/RevModPhys.60.1129} {\bibfield  {journal} {\bibinfo  {journal} {Rev. Mod. Phys.}\ }\textbf {\bibinfo {volume} {60}},\ \bibinfo {pages} {1129} (\bibinfo {year} {1988})}\BibitemShut {NoStop}%
\bibitem [{\citenamefont {Chester}(1970)}]{Chester1970}%
  \BibitemOpen
  \bibfield  {author} {\bibinfo {author} {\bibfnamefont {G.~V.}\ \bibnamefont {Chester}},\ }\href {https://doi.org/10.1103/PhysRevA.2.256} {\bibfield  {journal} {\bibinfo  {journal} {Phys. Rev. A}\ }\textbf {\bibinfo {volume} {2}},\ \bibinfo {pages} {256} (\bibinfo {year} {1970})}\BibitemShut {NoStop}%
\bibitem [{\citenamefont {Glendenning}(2001)}]{GLENDENNING2001393}%
  \BibitemOpen
  \bibfield  {author} {\bibinfo {author} {\bibfnamefont {N.~K.}\ \bibnamefont {Glendenning}},\ }\href {https://doi.org/https://doi.org/10.1016/S0370-1573(00)00080-6} {\bibfield  {journal} {\bibinfo  {journal} {Phys. Rep.}\ }\textbf {\bibinfo {volume} {342}},\ \bibinfo {pages} {393} (\bibinfo {year} {2001})}\BibitemShut {NoStop}%
\bibitem [{\citenamefont {Landig}\ \emph {et~al.}(2016)\citenamefont {Landig}, \citenamefont {Hruby}, \citenamefont {Dogra}, \citenamefont {Landini}, \citenamefont {Mottl}, \citenamefont {Donner},\ and\ \citenamefont {Esslinger}}]{landig2016}%
  \BibitemOpen
  \bibfield  {author} {\bibinfo {author} {\bibfnamefont {R.}~\bibnamefont {Landig}}, \bibinfo {author} {\bibfnamefont {L.}~\bibnamefont {Hruby}}, \bibinfo {author} {\bibfnamefont {N.}~\bibnamefont {Dogra}}, \bibinfo {author} {\bibfnamefont {M.}~\bibnamefont {Landini}}, \bibinfo {author} {\bibfnamefont {R.}~\bibnamefont {Mottl}}, \bibinfo {author} {\bibfnamefont {T.}~\bibnamefont {Donner}},\ and\ \bibinfo {author} {\bibfnamefont {T.}~\bibnamefont {Esslinger}},\ }\href {https://doi.org/10.1038/nature17409} {\bibfield  {journal} {\bibinfo  {journal} {Nature}\ }\textbf {\bibinfo {volume} {532}},\ \bibinfo {pages} {476–479} (\bibinfo {year} {2016})}\BibitemShut {NoStop}%
\bibitem [{\citenamefont {Helson}\ \emph {et~al.}(2022)\citenamefont {Helson}, \citenamefont {Zwettler}, \citenamefont {Mivehvar}, \citenamefont {Colella}, \citenamefont {Roux}, \citenamefont {Konishi}, \citenamefont {Ritsch},\ and\ \citenamefont {Brantut}}]{helson2022}%
  \BibitemOpen
  \bibfield  {author} {\bibinfo {author} {\bibfnamefont {V.}~\bibnamefont {Helson}}, \bibinfo {author} {\bibfnamefont {T.}~\bibnamefont {Zwettler}}, \bibinfo {author} {\bibfnamefont {F.}~\bibnamefont {Mivehvar}}, \bibinfo {author} {\bibfnamefont {E.}~\bibnamefont {Colella}}, \bibinfo {author} {\bibfnamefont {K.}~\bibnamefont {Roux}}, \bibinfo {author} {\bibfnamefont {H.}~\bibnamefont {Konishi}}, \bibinfo {author} {\bibfnamefont {H.}~\bibnamefont {Ritsch}},\ and\ \bibinfo {author} {\bibfnamefont {J.-P.}\ \bibnamefont {Brantut}},\ }\href@noop {} {} (\bibinfo {year} {2022}),\ \Eprint {https://arxiv.org/abs/2212.04402} {arXiv:2212.04402 [cond-mat.quant-gas]} \BibitemShut {NoStop}%
\bibitem [{\citenamefont {Tanzi}\ \emph {et~al.}(2019)\citenamefont {Tanzi}, \citenamefont {Roccuzzo}, \citenamefont {Lucioni}, \citenamefont {Famà}, \citenamefont {Fioretti}, \citenamefont {Gabbanini}, \citenamefont {Modugno}, \citenamefont {Recati},\ and\ \citenamefont {Stringari}}]{tanzi2019}%
  \BibitemOpen
  \bibfield  {author} {\bibinfo {author} {\bibfnamefont {L.}~\bibnamefont {Tanzi}}, \bibinfo {author} {\bibfnamefont {S.~M.}\ \bibnamefont {Roccuzzo}}, \bibinfo {author} {\bibfnamefont {E.}~\bibnamefont {Lucioni}}, \bibinfo {author} {\bibfnamefont {F.}~\bibnamefont {Famà}}, \bibinfo {author} {\bibfnamefont {A.}~\bibnamefont {Fioretti}}, \bibinfo {author} {\bibfnamefont {C.}~\bibnamefont {Gabbanini}}, \bibinfo {author} {\bibfnamefont {G.}~\bibnamefont {Modugno}}, \bibinfo {author} {\bibfnamefont {A.}~\bibnamefont {Recati}},\ and\ \bibinfo {author} {\bibfnamefont {S.}~\bibnamefont {Stringari}},\ }\href {https://doi.org/10.1038/s41586-019-1568-6} {\bibfield  {journal} {\bibinfo  {journal} {Nature}\ }\textbf {\bibinfo {volume} {574}},\ \bibinfo {pages} {382–385} (\bibinfo {year} {2019})}\BibitemShut {NoStop}%
\bibitem [{\citenamefont {Guo}\ \emph {et~al.}(2019)\citenamefont {Guo}, \citenamefont {Böttcher}, \citenamefont {Hertkorn}, \citenamefont {Schmidt}, \citenamefont {Wenzel}, \citenamefont {Büchler}, \citenamefont {Langen},\ and\ \citenamefont {Pfau}}]{guo2019}%
  \BibitemOpen
  \bibfield  {author} {\bibinfo {author} {\bibfnamefont {M.}~\bibnamefont {Guo}}, \bibinfo {author} {\bibfnamefont {F.}~\bibnamefont {Böttcher}}, \bibinfo {author} {\bibfnamefont {J.}~\bibnamefont {Hertkorn}}, \bibinfo {author} {\bibfnamefont {J.-N.}\ \bibnamefont {Schmidt}}, \bibinfo {author} {\bibfnamefont {M.}~\bibnamefont {Wenzel}}, \bibinfo {author} {\bibfnamefont {H.~P.}\ \bibnamefont {Büchler}}, \bibinfo {author} {\bibfnamefont {T.}~\bibnamefont {Langen}},\ and\ \bibinfo {author} {\bibfnamefont {T.}~\bibnamefont {Pfau}},\ }\href {https://doi.org/10.1038/s41586-019-1569-5} {\bibfield  {journal} {\bibinfo  {journal} {Nature}\ }\textbf {\bibinfo {volume} {574}},\ \bibinfo {pages} {386–389} (\bibinfo {year} {2019})}\BibitemShut {NoStop}%
\bibitem [{\citenamefont {Chomaz}\ \emph {et~al.}(2019)\citenamefont {Chomaz}, \citenamefont {Petter}, \citenamefont {Ilzh\"ofer}, \citenamefont {Natale}, \citenamefont {Trautmann}, \citenamefont {Politi}, \citenamefont {Durastante}, \citenamefont {van Bijnen}, \citenamefont {Patscheider}, \citenamefont {Sohmen}, \citenamefont {Mark},\ and\ \citenamefont {Ferlaino}}]{chomaz2019}%
  \BibitemOpen
  \bibfield  {author} {\bibinfo {author} {\bibfnamefont {L.}~\bibnamefont {Chomaz}}, \bibinfo {author} {\bibfnamefont {D.}~\bibnamefont {Petter}}, \bibinfo {author} {\bibfnamefont {P.}~\bibnamefont {Ilzh\"ofer}}, \bibinfo {author} {\bibfnamefont {G.}~\bibnamefont {Natale}}, \bibinfo {author} {\bibfnamefont {A.}~\bibnamefont {Trautmann}}, \bibinfo {author} {\bibfnamefont {C.}~\bibnamefont {Politi}}, \bibinfo {author} {\bibfnamefont {G.}~\bibnamefont {Durastante}}, \bibinfo {author} {\bibfnamefont {R.~M.~W.}\ \bibnamefont {van Bijnen}}, \bibinfo {author} {\bibfnamefont {A.}~\bibnamefont {Patscheider}}, \bibinfo {author} {\bibfnamefont {M.}~\bibnamefont {Sohmen}}, \bibinfo {author} {\bibfnamefont {M.~J.}\ \bibnamefont {Mark}},\ and\ \bibinfo {author} {\bibfnamefont {F.}~\bibnamefont {Ferlaino}},\ }\href {https://doi.org/10.1103/PhysRevX.9.021012} {\bibfield  {journal} {\bibinfo  {journal} {Phys. Rev. X}\ }\textbf {\bibinfo {volume} {9}},\ \bibinfo {pages} {021012} (\bibinfo {year} {2019})}\BibitemShut {NoStop}%
\bibitem [{\citenamefont {Su}\ \emph {et~al.}(2023)\citenamefont {Su}, \citenamefont {Douglas}, \citenamefont {Szurek}, \citenamefont {Groth}, \citenamefont {Ozturk}, \citenamefont {Krahn}, \citenamefont {Hébert}, \citenamefont {Phelps}, \citenamefont {Ebadi}, \citenamefont {Dickerson}, \citenamefont {Ferlaino}, \citenamefont {Marković},\ and\ \citenamefont {Greiner}}]{su2023}%
  \BibitemOpen
  \bibfield  {author} {\bibinfo {author} {\bibfnamefont {L.}~\bibnamefont {Su}}, \bibinfo {author} {\bibfnamefont {A.}~\bibnamefont {Douglas}}, \bibinfo {author} {\bibfnamefont {M.}~\bibnamefont {Szurek}}, \bibinfo {author} {\bibfnamefont {R.}~\bibnamefont {Groth}}, \bibinfo {author} {\bibfnamefont {S.~F.}\ \bibnamefont {Ozturk}}, \bibinfo {author} {\bibfnamefont {A.}~\bibnamefont {Krahn}}, \bibinfo {author} {\bibfnamefont {A.~H.}\ \bibnamefont {Hébert}}, \bibinfo {author} {\bibfnamefont {G.~A.}\ \bibnamefont {Phelps}}, \bibinfo {author} {\bibfnamefont {S.}~\bibnamefont {Ebadi}}, \bibinfo {author} {\bibfnamefont {S.}~\bibnamefont {Dickerson}}, \bibinfo {author} {\bibfnamefont {F.}~\bibnamefont {Ferlaino}}, \bibinfo {author} {\bibfnamefont {O.}~\bibnamefont {Marković}},\ and\ \bibinfo {author} {\bibfnamefont {M.}~\bibnamefont {Greiner}},\ }\href@noop {} {} (\bibinfo {year} {2023}),\ \Eprint {https://arxiv.org/abs/2306.00888} {arXiv:2306.00888 [cond-mat.quant-gas]} \BibitemShut {NoStop}%
\bibitem [{\citenamefont {Guardado-Sanchez}\ \emph {et~al.}(2021)\citenamefont {Guardado-Sanchez}, \citenamefont {Spar}, \citenamefont {Schauss}, \citenamefont {Belyansky}, \citenamefont {Young}, \citenamefont {Bienias}, \citenamefont {Gorshkov}, \citenamefont {Iadecola},\ and\ \citenamefont {Bakr}}]{Guardado2021}%
  \BibitemOpen
  \bibfield  {author} {\bibinfo {author} {\bibfnamefont {E.}~\bibnamefont {Guardado-Sanchez}}, \bibinfo {author} {\bibfnamefont {B.~M.}\ \bibnamefont {Spar}}, \bibinfo {author} {\bibfnamefont {P.}~\bibnamefont {Schauss}}, \bibinfo {author} {\bibfnamefont {R.}~\bibnamefont {Belyansky}}, \bibinfo {author} {\bibfnamefont {J.~T.}\ \bibnamefont {Young}}, \bibinfo {author} {\bibfnamefont {P.}~\bibnamefont {Bienias}}, \bibinfo {author} {\bibfnamefont {A.~V.}\ \bibnamefont {Gorshkov}}, \bibinfo {author} {\bibfnamefont {T.}~\bibnamefont {Iadecola}},\ and\ \bibinfo {author} {\bibfnamefont {W.~S.}\ \bibnamefont {Bakr}},\ }\href {https://doi.org/10.1103/PhysRevX.11.021036} {\bibfield  {journal} {\bibinfo  {journal} {Phys. Rev. X}\ }\textbf {\bibinfo {volume} {11}},\ \bibinfo {pages} {021036} (\bibinfo {year} {2021})}\BibitemShut {NoStop}%
\bibitem [{\citenamefont {Zahn}\ \emph {et~al.}(2022)\citenamefont {Zahn}, \citenamefont {Singh}, \citenamefont {Kosch}, \citenamefont {Asteria}, \citenamefont {Freystatzky}, \citenamefont {Sengstock}, \citenamefont {Mathey},\ and\ \citenamefont {Weitenberg}}]{Zahn2022}%
  \BibitemOpen
  \bibfield  {author} {\bibinfo {author} {\bibfnamefont {H.~P.}\ \bibnamefont {Zahn}}, \bibinfo {author} {\bibfnamefont {V.~P.}\ \bibnamefont {Singh}}, \bibinfo {author} {\bibfnamefont {M.~N.}\ \bibnamefont {Kosch}}, \bibinfo {author} {\bibfnamefont {L.}~\bibnamefont {Asteria}}, \bibinfo {author} {\bibfnamefont {L.}~\bibnamefont {Freystatzky}}, \bibinfo {author} {\bibfnamefont {K.}~\bibnamefont {Sengstock}}, \bibinfo {author} {\bibfnamefont {L.}~\bibnamefont {Mathey}},\ and\ \bibinfo {author} {\bibfnamefont {C.}~\bibnamefont {Weitenberg}},\ }\href {https://doi.org/10.1103/PhysRevX.12.021014} {\bibfield  {journal} {\bibinfo  {journal} {Phys. Rev. X}\ }\textbf {\bibinfo {volume} {12}},\ \bibinfo {pages} {021014} (\bibinfo {year} {2022})}\BibitemShut {NoStop}%
\bibitem [{\citenamefont {Scholl}\ \emph {et~al.}(2021)\citenamefont {Scholl}, \citenamefont {Schuler}, \citenamefont {Williams}, \citenamefont {Eberharter}, \citenamefont {Barredo}, \citenamefont {Schymik}, \citenamefont {Lienhard}, \citenamefont {Henry}, \citenamefont {Lang}, \citenamefont {Lahaye}, \citenamefont {Läuchli},\ and\ \citenamefont {Browaeys}}]{Scholl2021}%
  \BibitemOpen
  \bibfield  {author} {\bibinfo {author} {\bibfnamefont {P.}~\bibnamefont {Scholl}}, \bibinfo {author} {\bibfnamefont {M.}~\bibnamefont {Schuler}}, \bibinfo {author} {\bibfnamefont {H.~J.}\ \bibnamefont {Williams}}, \bibinfo {author} {\bibfnamefont {A.~A.}\ \bibnamefont {Eberharter}}, \bibinfo {author} {\bibfnamefont {D.}~\bibnamefont {Barredo}}, \bibinfo {author} {\bibfnamefont {K.-N.}\ \bibnamefont {Schymik}}, \bibinfo {author} {\bibfnamefont {V.}~\bibnamefont {Lienhard}}, \bibinfo {author} {\bibfnamefont {L.-P.}\ \bibnamefont {Henry}}, \bibinfo {author} {\bibfnamefont {T.~C.}\ \bibnamefont {Lang}}, \bibinfo {author} {\bibfnamefont {T.}~\bibnamefont {Lahaye}}, \bibinfo {author} {\bibfnamefont {A.~M.}\ \bibnamefont {Läuchli}},\ and\ \bibinfo {author} {\bibfnamefont {A.}~\bibnamefont {Browaeys}},\ }\href {https://doi.org/10.1038/s41586-021-03585-1} {\bibfield  {journal} {\bibinfo  {journal} {Nature}\ }\textbf {\bibinfo {volume} {595}} (\bibinfo {year} {2021})}\BibitemShut {NoStop}%
\bibitem [{\citenamefont {Semeghini}\ and\ \citenamefont {\textit{et al.}}(2021)}]{semeghini2021}%
  \BibitemOpen
  \bibfield  {author} {\bibinfo {author} {\bibfnamefont {G.}~\bibnamefont {Semeghini}}\ and\ \bibinfo {author} {\bibnamefont {\textit{et al.}}},\ }\href {https://doi.org/10.1126/science.abi8794} {\bibfield  {journal} {\bibinfo  {journal} {Science}\ }\textbf {\bibinfo {volume} {374}},\ \bibinfo {pages} {1242} (\bibinfo {year} {2021})}\BibitemShut {NoStop}%
\bibitem [{\citenamefont {Christakis}\ \emph {et~al.}(2023)\citenamefont {Christakis}, \citenamefont {Rosenberg}, \citenamefont {Raj}, \citenamefont {Chi}, \citenamefont {Morningstar}, \citenamefont {Huse}, \citenamefont {Yan},\ and\ \citenamefont {Bakr}}]{Christakis2023}%
  \BibitemOpen
  \bibfield  {author} {\bibinfo {author} {\bibfnamefont {L.}~\bibnamefont {Christakis}}, \bibinfo {author} {\bibfnamefont {J.~S.}\ \bibnamefont {Rosenberg}}, \bibinfo {author} {\bibfnamefont {R.}~\bibnamefont {Raj}}, \bibinfo {author} {\bibfnamefont {S.}~\bibnamefont {Chi}}, \bibinfo {author} {\bibfnamefont {A.}~\bibnamefont {Morningstar}}, \bibinfo {author} {\bibfnamefont {D.~A.}\ \bibnamefont {Huse}}, \bibinfo {author} {\bibfnamefont {Z.~Z.}\ \bibnamefont {Yan}},\ and\ \bibinfo {author} {\bibfnamefont {W.~S.}\ \bibnamefont {Bakr}},\ }\href {https://doi.org/10.1038/s41586-022-05558-4} {\bibfield  {journal} {\bibinfo  {journal} {Nature}\ }\textbf {\bibinfo {volume} {614}},\ \bibinfo {pages} {64} (\bibinfo {year} {2023})}\BibitemShut {NoStop}%
\bibitem [{\citenamefont {Karski}\ \emph {et~al.}(2009)\citenamefont {Karski}, \citenamefont {F\"rster}, \citenamefont {Choi}, \citenamefont {Steffen}, \citenamefont {Alt}, \citenamefont {Meschede},\ and\ \citenamefont {Widera}}]{QuantumWalk}%
  \BibitemOpen
  \bibfield  {author} {\bibinfo {author} {\bibfnamefont {M.}~\bibnamefont {Karski}}, \bibinfo {author} {\bibfnamefont {L.}~\bibnamefont {F\"rster}}, \bibinfo {author} {\bibfnamefont {J.-M.}\ \bibnamefont {Choi}}, \bibinfo {author} {\bibfnamefont {A.}~\bibnamefont {Steffen}}, \bibinfo {author} {\bibfnamefont {W.}~\bibnamefont {Alt}}, \bibinfo {author} {\bibfnamefont {D.}~\bibnamefont {Meschede}},\ and\ \bibinfo {author} {\bibfnamefont {A.}~\bibnamefont {Widera}},\ }\href {https://doi.org/10.1126/science.1174436} {\bibfield  {journal} {\bibinfo  {journal} {Science}\ }\textbf {\bibinfo {volume} {325}},\ \bibinfo {pages} {174} (\bibinfo {year} {2009})}\BibitemShut {NoStop}%
\bibitem [{\citenamefont {Groh}\ \emph {et~al.}(2016)\citenamefont {Groh}, \citenamefont {Brakhane}, \citenamefont {Alt}, \citenamefont {Meschede}, \citenamefont {Asb\'oth},\ and\ \citenamefont {Alberti}}]{PhysRevA.94.013620}%
  \BibitemOpen
  \bibfield  {author} {\bibinfo {author} {\bibfnamefont {T.}~\bibnamefont {Groh}}, \bibinfo {author} {\bibfnamefont {S.}~\bibnamefont {Brakhane}}, \bibinfo {author} {\bibfnamefont {W.}~\bibnamefont {Alt}}, \bibinfo {author} {\bibfnamefont {D.}~\bibnamefont {Meschede}}, \bibinfo {author} {\bibfnamefont {J.~K.}\ \bibnamefont {Asb\'oth}},\ and\ \bibinfo {author} {\bibfnamefont {A.}~\bibnamefont {Alberti}},\ }\href {https://doi.org/10.1103/PhysRevA.94.013620} {\bibfield  {journal} {\bibinfo  {journal} {Phys. Rev. A}\ }\textbf {\bibinfo {volume} {94}},\ \bibinfo {pages} {013620} (\bibinfo {year} {2016})}\BibitemShut {NoStop}%
\bibitem [{\citenamefont {Mandel}\ \emph {et~al.}(2003)\citenamefont {Mandel}, \citenamefont {Greiner}, \citenamefont {Widera}, \citenamefont {Rom}, \citenamefont {H\"ansch},\ and\ \citenamefont {Bloch}}]{mandel_2003}%
  \BibitemOpen
  \bibfield  {author} {\bibinfo {author} {\bibfnamefont {O.}~\bibnamefont {Mandel}}, \bibinfo {author} {\bibfnamefont {M.}~\bibnamefont {Greiner}}, \bibinfo {author} {\bibfnamefont {A.}~\bibnamefont {Widera}}, \bibinfo {author} {\bibfnamefont {T.}~\bibnamefont {Rom}}, \bibinfo {author} {\bibfnamefont {T.~W.}\ \bibnamefont {H\"ansch}},\ and\ \bibinfo {author} {\bibfnamefont {I.}~\bibnamefont {Bloch}},\ }\href {https://doi.org/10.1038/nature02008} {\bibfield  {journal} {\bibinfo  {journal} {Nature}\ }\textbf {\bibinfo {volume} {425}},\ \bibinfo {pages} {937} (\bibinfo {year} {2003})}\BibitemShut {NoStop}%
\bibitem [{\citenamefont {Förster}\ \emph {et~al.}(2009)\citenamefont {Förster}, \citenamefont {Karski}, \citenamefont {Choi}, \citenamefont {Steffen}, \citenamefont {Alt}, \citenamefont {Meschede}, \citenamefont {Widera}, \citenamefont {Montano}, \citenamefont {Lee}, \citenamefont {Rakreungdet},\ and\ \citenamefont {Jessen}}]{forster_microwave_2009}%
  \BibitemOpen
  \bibfield  {author} {\bibinfo {author} {\bibfnamefont {L.}~\bibnamefont {Förster}}, \bibinfo {author} {\bibfnamefont {M.}~\bibnamefont {Karski}}, \bibinfo {author} {\bibfnamefont {J.-M.}\ \bibnamefont {Choi}}, \bibinfo {author} {\bibfnamefont {A.}~\bibnamefont {Steffen}}, \bibinfo {author} {\bibfnamefont {W.}~\bibnamefont {Alt}}, \bibinfo {author} {\bibfnamefont {D.}~\bibnamefont {Meschede}}, \bibinfo {author} {\bibfnamefont {A.}~\bibnamefont {Widera}}, \bibinfo {author} {\bibfnamefont {E.}~\bibnamefont {Montano}}, \bibinfo {author} {\bibfnamefont {J.~H.}\ \bibnamefont {Lee}}, \bibinfo {author} {\bibfnamefont {W.}~\bibnamefont {Rakreungdet}},\ and\ \bibinfo {author} {\bibfnamefont {P.~S.}\ \bibnamefont {Jessen}},\ }\href {https://doi.org/10.1103/PhysRevLett.103.233001} {\bibfield  {journal} {\bibinfo  {journal} {Phys. Rev. Lett.}\ }\textbf {\bibinfo {volume} {103}},\ \bibinfo {pages} {233001} (\bibinfo {year} {2009})}\BibitemShut {NoStop}%
\bibitem [{\citenamefont {Belmechri}\ \emph {et~al.}(2013)\citenamefont {Belmechri}, \citenamefont {Förster}, \citenamefont {Alt}, \citenamefont {Widera}, \citenamefont {Meschede},\ and\ \citenamefont {Alberti}}]{belmechri_microwave_2013}%
  \BibitemOpen
  \bibfield  {author} {\bibinfo {author} {\bibfnamefont {N.}~\bibnamefont {Belmechri}}, \bibinfo {author} {\bibfnamefont {L.}~\bibnamefont {Förster}}, \bibinfo {author} {\bibfnamefont {W.}~\bibnamefont {Alt}}, \bibinfo {author} {\bibfnamefont {A.}~\bibnamefont {Widera}}, \bibinfo {author} {\bibfnamefont {D.}~\bibnamefont {Meschede}},\ and\ \bibinfo {author} {\bibfnamefont {A.}~\bibnamefont {Alberti}},\ }\href {https://doi.org/10.1088/0953-4075/46/10/104006} {\bibfield  {journal} {\bibinfo  {journal} {J. Phys. B: At. Mol. Opt. Phys.}\ }\textbf {\bibinfo {volume} {46}},\ \bibinfo {pages} {104006} (\bibinfo {year} {2013})}\BibitemShut {NoStop}%
\bibitem [{\citenamefont {Riegger}\ \emph {et~al.}(2018)\citenamefont {Riegger}, \citenamefont {Darkwah~Oppong}, \citenamefont {H\"ofer}, \citenamefont {Fernandes}, \citenamefont {Bloch},\ and\ \citenamefont {F\"olling}}]{riegger_2018}%
  \BibitemOpen
  \bibfield  {author} {\bibinfo {author} {\bibfnamefont {L.}~\bibnamefont {Riegger}}, \bibinfo {author} {\bibfnamefont {N.}~\bibnamefont {Darkwah~Oppong}}, \bibinfo {author} {\bibfnamefont {M.}~\bibnamefont {H\"ofer}}, \bibinfo {author} {\bibfnamefont {D.~R.}\ \bibnamefont {Fernandes}}, \bibinfo {author} {\bibfnamefont {I.}~\bibnamefont {Bloch}},\ and\ \bibinfo {author} {\bibfnamefont {S.}~\bibnamefont {F\"olling}},\ }\href {https://doi.org/10.1103/PhysRevLett.120.143601} {\bibfield  {journal} {\bibinfo  {journal} {Phys. Rev. Lett.}\ }\textbf {\bibinfo {volume} {120}},\ \bibinfo {pages} {143601} (\bibinfo {year} {2018})}\BibitemShut {NoStop}%
\bibitem [{\citenamefont {Darkwah~Oppong}\ \emph {et~al.}(2022)\citenamefont {Darkwah~Oppong}, \citenamefont {Pasqualetti}, \citenamefont {Bettermann}, \citenamefont {Zechmann}, \citenamefont {Knap}, \citenamefont {Bloch},\ and\ \citenamefont {F\"olling}}]{oppong_2022}%
  \BibitemOpen
  \bibfield  {author} {\bibinfo {author} {\bibfnamefont {N.}~\bibnamefont {Darkwah~Oppong}}, \bibinfo {author} {\bibfnamefont {G.}~\bibnamefont {Pasqualetti}}, \bibinfo {author} {\bibfnamefont {O.}~\bibnamefont {Bettermann}}, \bibinfo {author} {\bibfnamefont {P.}~\bibnamefont {Zechmann}}, \bibinfo {author} {\bibfnamefont {M.}~\bibnamefont {Knap}}, \bibinfo {author} {\bibfnamefont {I.}~\bibnamefont {Bloch}},\ and\ \bibinfo {author} {\bibfnamefont {S.}~\bibnamefont {F\"olling}},\ }\href {https://doi.org/10.1103/PhysRevLett.120.143601} {\bibfield  {journal} {\bibinfo  {journal} {Phys. Rev. X}\ }\textbf {\bibinfo {volume} {12}},\ \bibinfo {pages} {031026} (\bibinfo {year} {2022})}\BibitemShut {NoStop}%
\bibitem [{\citenamefont {Heinz}\ \emph {et~al.}(2020)\citenamefont {Heinz}, \citenamefont {Park}, \citenamefont {{\v S}anti{\'c}}, \citenamefont {Trautmann}, \citenamefont {Porsev}, \citenamefont {Safronova}, \citenamefont {Bloch},\ and\ \citenamefont {Blatt}}]{heinz_2020}%
  \BibitemOpen
  \bibfield  {author} {\bibinfo {author} {\bibfnamefont {A.}~\bibnamefont {Heinz}}, \bibinfo {author} {\bibfnamefont {A.~J.}\ \bibnamefont {Park}}, \bibinfo {author} {\bibfnamefont {N.}~\bibnamefont {{\v S}anti{\'c}}}, \bibinfo {author} {\bibfnamefont {J.}~\bibnamefont {Trautmann}}, \bibinfo {author} {\bibfnamefont {S.~G.}\ \bibnamefont {Porsev}}, \bibinfo {author} {\bibfnamefont {M.~S.}\ \bibnamefont {Safronova}}, \bibinfo {author} {\bibfnamefont {I.}~\bibnamefont {Bloch}},\ and\ \bibinfo {author} {\bibfnamefont {S.}~\bibnamefont {Blatt}},\ }\href {https://doi.org/10.1103/PhysRevLett.124.203201} {\bibfield  {journal} {\bibinfo  {journal} {Phys. Rev. Lett.}\ }\textbf {\bibinfo {volume} {124}},\ \bibinfo {pages} {203201} (\bibinfo {year} {2020})}\BibitemShut {NoStop}%
\bibitem [{\citenamefont {Höhn}\ \emph {et~al.}(2023)\citenamefont {Höhn}, \citenamefont {Staub}, \citenamefont {Brochier}, \citenamefont {Oppong},\ and\ \citenamefont {Aidelsburger}}]{hohn_state-dependent_2023}%
  \BibitemOpen
  \bibfield  {author} {\bibinfo {author} {\bibfnamefont {T.~O.}\ \bibnamefont {Höhn}}, \bibinfo {author} {\bibfnamefont {E.}~\bibnamefont {Staub}}, \bibinfo {author} {\bibfnamefont {G.}~\bibnamefont {Brochier}}, \bibinfo {author} {\bibfnamefont {N.~D.}\ \bibnamefont {Oppong}},\ and\ \bibinfo {author} {\bibfnamefont {M.}~\bibnamefont {Aidelsburger}},\ }\href {https://doi.org/10.48550/arXiv.2305.20084} {} (\bibinfo {year} {2023}),\ \Eprint {https://arxiv.org/abs/2305.20084} {arXiv:2305.20084 [cond-mat.quant-gas]} \BibitemShut {NoStop}%
\bibitem [{\citenamefont {Frye}\ \emph {et~al.}(2019{\natexlab{a}})\citenamefont {Frye}, \citenamefont {Yang},\ and\ \citenamefont {Hutson}}]{CesiumResonances2019}%
  \BibitemOpen
  \bibfield  {author} {\bibinfo {author} {\bibfnamefont {M.~D.}\ \bibnamefont {Frye}}, \bibinfo {author} {\bibfnamefont {B.~C.}\ \bibnamefont {Yang}},\ and\ \bibinfo {author} {\bibfnamefont {J.~M.}\ \bibnamefont {Hutson}},\ }\href {https://doi.org/10.1103/PhysRevA.100.022702} {\bibfield  {journal} {\bibinfo  {journal} {Phys. Rev. A}\ }\textbf {\bibinfo {volume} {100}},\ \bibinfo {pages} {022702} (\bibinfo {year} {2019}{\natexlab{a}})}\BibitemShut {NoStop}%
\bibitem [{\citenamefont {Drechsler}\ \emph {et~al.}(2007)\citenamefont {Drechsler}, \citenamefont {Volkova}, \citenamefont {Vasiliev}, \citenamefont {Tristan}, \citenamefont {Richter}, \citenamefont {Schmitt}, \citenamefont {Rosner}, \citenamefont {M\'alek}, \citenamefont {Klingeler}, \citenamefont {Zvyagin},\ and\ \citenamefont {B\"uchner}}]{Drechsler2007}%
  \BibitemOpen
  \bibfield  {author} {\bibinfo {author} {\bibfnamefont {S.-L.}\ \bibnamefont {Drechsler}}, \bibinfo {author} {\bibfnamefont {O.}~\bibnamefont {Volkova}}, \bibinfo {author} {\bibfnamefont {A.~N.}\ \bibnamefont {Vasiliev}}, \bibinfo {author} {\bibfnamefont {N.}~\bibnamefont {Tristan}}, \bibinfo {author} {\bibfnamefont {J.}~\bibnamefont {Richter}}, \bibinfo {author} {\bibfnamefont {M.}~\bibnamefont {Schmitt}}, \bibinfo {author} {\bibfnamefont {H.}~\bibnamefont {Rosner}}, \bibinfo {author} {\bibfnamefont {J.}~\bibnamefont {M\'alek}}, \bibinfo {author} {\bibfnamefont {R.}~\bibnamefont {Klingeler}}, \bibinfo {author} {\bibfnamefont {A.~A.}\ \bibnamefont {Zvyagin}},\ and\ \bibinfo {author} {\bibfnamefont {B.}~\bibnamefont {B\"uchner}},\ }\href {https://doi.org/10.1103/PhysRevLett.98.077202} {\bibfield  {journal} {\bibinfo  {journal} {Phys. Rev. Lett.}\ }\textbf {\bibinfo {volume} {98}},\ \bibinfo {pages} {077202} (\bibinfo {year} {2007})}\BibitemShut {NoStop}%
\bibitem [{\citenamefont {Sato}\ \emph {et~al.}(2011)\citenamefont {Sato}, \citenamefont {Furukawa}, \citenamefont {Onoda},\ and\ \citenamefont {Furusaki}}]{SATO2011}%
  \BibitemOpen
  \bibfield  {author} {\bibinfo {author} {\bibfnamefont {M.}~\bibnamefont {Sato}}, \bibinfo {author} {\bibfnamefont {S.}~\bibnamefont {Furukawa}}, \bibinfo {author} {\bibfnamefont {S.}~\bibnamefont {Onoda}},\ and\ \bibinfo {author} {\bibfnamefont {A.}~\bibnamefont {Furusaki}},\ }\href {https://doi.org/10.1142/S0217984911026607} {\bibfield  {journal} {\bibinfo  {journal} {Mod. Phys. Lett. B}\ }\textbf {\bibinfo {volume} {25}},\ \bibinfo {pages} {901} (\bibinfo {year} {2011})}\BibitemShut {NoStop}%
\bibitem [{\citenamefont {Furukawa}\ \emph {et~al.}(2012)\citenamefont {Furukawa}, \citenamefont {Sato}, \citenamefont {Onoda},\ and\ \citenamefont {Furusaki}}]{Furukawa2012}%
  \BibitemOpen
  \bibfield  {author} {\bibinfo {author} {\bibfnamefont {S.}~\bibnamefont {Furukawa}}, \bibinfo {author} {\bibfnamefont {M.}~\bibnamefont {Sato}}, \bibinfo {author} {\bibfnamefont {S.}~\bibnamefont {Onoda}},\ and\ \bibinfo {author} {\bibfnamefont {A.}~\bibnamefont {Furusaki}},\ }\href {https://doi.org/10.1103/PhysRevB.86.094417} {\bibfield  {journal} {\bibinfo  {journal} {Phys. Rev. B}\ }\textbf {\bibinfo {volume} {86}},\ \bibinfo {pages} {094417} (\bibinfo {year} {2012})}\BibitemShut {NoStop}%
\bibitem [{\citenamefont {Wolter}\ \emph {et~al.}(2012)\citenamefont {Wolter}, \citenamefont {Lipps}, \citenamefont {Sch\"apers}, \citenamefont {Drechsler}, \citenamefont {Nishimoto}, \citenamefont {Vogel}, \citenamefont {Kataev}, \citenamefont {B\"uchner}, \citenamefont {Rosner}, \citenamefont {Schmitt}, \citenamefont {Uhlarz}, \citenamefont {Skourski}, \citenamefont {Wosnitza}, \citenamefont {S\"ullow},\ and\ \citenamefont {Rule}}]{Wolter2012}%
  \BibitemOpen
  \bibfield  {author} {\bibinfo {author} {\bibfnamefont {A.~U.~B.}\ \bibnamefont {Wolter}}, \bibinfo {author} {\bibfnamefont {F.}~\bibnamefont {Lipps}}, \bibinfo {author} {\bibfnamefont {M.}~\bibnamefont {Sch\"apers}}, \bibinfo {author} {\bibfnamefont {S.-L.}\ \bibnamefont {Drechsler}}, \bibinfo {author} {\bibfnamefont {S.}~\bibnamefont {Nishimoto}}, \bibinfo {author} {\bibfnamefont {R.}~\bibnamefont {Vogel}}, \bibinfo {author} {\bibfnamefont {V.}~\bibnamefont {Kataev}}, \bibinfo {author} {\bibfnamefont {B.}~\bibnamefont {B\"uchner}}, \bibinfo {author} {\bibfnamefont {H.}~\bibnamefont {Rosner}}, \bibinfo {author} {\bibfnamefont {M.}~\bibnamefont {Schmitt}}, \bibinfo {author} {\bibfnamefont {M.}~\bibnamefont {Uhlarz}}, \bibinfo {author} {\bibfnamefont {Y.}~\bibnamefont {Skourski}}, \bibinfo {author} {\bibfnamefont {J.}~\bibnamefont {Wosnitza}}, \bibinfo {author} {\bibfnamefont {S.}~\bibnamefont {S\"ullow}},\ and\ \bibinfo {author} {\bibfnamefont {K.~C.}\ \bibnamefont {Rule}},\ }\href
  {https://doi.org/10.1103/PhysRevB.85.014407} {\bibfield  {journal} {\bibinfo  {journal} {Phys. Rev. B}\ }\textbf {\bibinfo {volume} {85}},\ \bibinfo {pages} {014407} (\bibinfo {year} {2012})}\BibitemShut {NoStop}%
\bibitem [{\citenamefont {Orlova}\ \emph {et~al.}(2017)\citenamefont {Orlova}, \citenamefont {Green}, \citenamefont {Law}, \citenamefont {Gorbunov}, \citenamefont {Chanda}, \citenamefont {Kr\"amer}, \citenamefont {Horvati\ifmmode~\acute{c}\else \'{c}\fi{}}, \citenamefont {Kremer}, \citenamefont {Wosnitza},\ and\ \citenamefont {Rikken}}]{Orlova2017}%
  \BibitemOpen
  \bibfield  {author} {\bibinfo {author} {\bibfnamefont {A.}~\bibnamefont {Orlova}}, \bibinfo {author} {\bibfnamefont {E.~L.}\ \bibnamefont {Green}}, \bibinfo {author} {\bibfnamefont {J.~M.}\ \bibnamefont {Law}}, \bibinfo {author} {\bibfnamefont {D.~I.}\ \bibnamefont {Gorbunov}}, \bibinfo {author} {\bibfnamefont {G.}~\bibnamefont {Chanda}}, \bibinfo {author} {\bibfnamefont {S.}~\bibnamefont {Kr\"amer}}, \bibinfo {author} {\bibfnamefont {M.}~\bibnamefont {Horvati\ifmmode~\acute{c}\else \'{c}\fi{}}}, \bibinfo {author} {\bibfnamefont {R.~K.}\ \bibnamefont {Kremer}}, \bibinfo {author} {\bibfnamefont {J.}~\bibnamefont {Wosnitza}},\ and\ \bibinfo {author} {\bibfnamefont {G.~L. J.~A.}\ \bibnamefont {Rikken}},\ }\href {https://doi.org/10.1103/PhysRevLett.118.247201} {\bibfield  {journal} {\bibinfo  {journal} {Phys. Rev. Lett.}\ }\textbf {\bibinfo {volume} {118}},\ \bibinfo {pages} {247201} (\bibinfo {year} {2017})}\BibitemShut {NoStop}%
\bibitem [{\citenamefont {Grams}\ \emph {et~al.}(2022)\citenamefont {Grams}, \citenamefont {Brüning}, \citenamefont {Kopatz}, \citenamefont {Lorenz}, \citenamefont {Becker}, \citenamefont {Bohatý},\ and\ \citenamefont {Hemberger}}]{Grams2022}%
  \BibitemOpen
  \bibfield  {author} {\bibinfo {author} {\bibfnamefont {C.~P.}\ \bibnamefont {Grams}}, \bibinfo {author} {\bibfnamefont {D.}~\bibnamefont {Brüning}}, \bibinfo {author} {\bibfnamefont {S.}~\bibnamefont {Kopatz}}, \bibinfo {author} {\bibfnamefont {T.}~\bibnamefont {Lorenz}}, \bibinfo {author} {\bibfnamefont {P.}~\bibnamefont {Becker}}, \bibinfo {author} {\bibfnamefont {L.}~\bibnamefont {Bohatý}},\ and\ \bibinfo {author} {\bibfnamefont {J.}~\bibnamefont {Hemberger}},\ }\href {https://doi.org/10.1038/s42005-022-00811-8} {\bibfield  {journal} {\bibinfo  {journal} {Commun. Phys.}\ }\textbf {\bibinfo {volume} {5}} (\bibinfo {year} {2022})}\BibitemShut {NoStop}%
\bibitem [{\citenamefont {Haegeman}\ \emph {et~al.}(2013)\citenamefont {Haegeman}, \citenamefont {Osborne},\ and\ \citenamefont {Verstraete}}]{Haegeman2013}%
  \BibitemOpen
  \bibfield  {author} {\bibinfo {author} {\bibfnamefont {J.}~\bibnamefont {Haegeman}}, \bibinfo {author} {\bibfnamefont {T.~J.}\ \bibnamefont {Osborne}},\ and\ \bibinfo {author} {\bibfnamefont {F.}~\bibnamefont {Verstraete}},\ }\href {https://doi.org/10.1103/PhysRevB.88.075133} {\bibfield  {journal} {\bibinfo  {journal} {Phys. Rev. B}\ }\textbf {\bibinfo {volume} {88}},\ \bibinfo {pages} {075133} (\bibinfo {year} {2013})}\BibitemShut {NoStop}%
\bibitem [{\citenamefont {Zauner-Stauber}\ \emph {et~al.}(2018)\citenamefont {Zauner-Stauber}, \citenamefont {Vanderstraeten}, \citenamefont {Fishman}, \citenamefont {Verstraete},\ and\ \citenamefont {Haegeman}}]{Zauner2018}%
  \BibitemOpen
  \bibfield  {author} {\bibinfo {author} {\bibfnamefont {V.}~\bibnamefont {Zauner-Stauber}}, \bibinfo {author} {\bibfnamefont {L.}~\bibnamefont {Vanderstraeten}}, \bibinfo {author} {\bibfnamefont {M.~T.}\ \bibnamefont {Fishman}}, \bibinfo {author} {\bibfnamefont {F.}~\bibnamefont {Verstraete}},\ and\ \bibinfo {author} {\bibfnamefont {J.}~\bibnamefont {Haegeman}},\ }\href {https://doi.org/10.1103/PhysRevB.97.045145} {\bibfield  {journal} {\bibinfo  {journal} {Phys. Rev. B}\ }\textbf {\bibinfo {volume} {97}},\ \bibinfo {pages} {045145} (\bibinfo {year} {2018})}\BibitemShut {NoStop}%
\bibitem [{\citenamefont {Landau}\ \emph {et~al.}(1999)\citenamefont {Landau}, \citenamefont {Lifshitz},\ and\ \citenamefont {Pitaevskii}}]{Landau_ssb}%
  \BibitemOpen
  \bibfield  {author} {\bibinfo {author} {\bibfnamefont {L.~D.}\ \bibnamefont {Landau}}, \bibinfo {author} {\bibfnamefont {E.~M.}\ \bibnamefont {Lifshitz}},\ and\ \bibinfo {author} {\bibfnamefont {M.}~\bibnamefont {Pitaevskii}},\ }\href@noop {} {\emph {\bibinfo {title} {Statistical Physics}}}\ (\bibinfo  {publisher} {Butterworth-Heinemann, New York},\ \bibinfo {year} {1999})\BibitemShut {NoStop}%
\bibitem [{\citenamefont {Wilson}\ and\ \citenamefont {Kogut}(1974)}]{wilson_ssb}%
  \BibitemOpen
  \bibfield  {author} {\bibinfo {author} {\bibfnamefont {K.~G.}\ \bibnamefont {Wilson}}\ and\ \bibinfo {author} {\bibfnamefont {J.}~\bibnamefont {Kogut}},\ }\href {https://doi.org/https://doi.org/10.1016/0370-1573(74)90023-4} {\bibfield  {journal} {\bibinfo  {journal} {Phys. Rep.}\ }\textbf {\bibinfo {volume} {12}},\ \bibinfo {pages} {75} (\bibinfo {year} {1974})}\BibitemShut {NoStop}%
\bibitem [{\citenamefont {Senthil}\ \emph {et~al.}(2004{\natexlab{a}})\citenamefont {Senthil}, \citenamefont {Vishwanath}, \citenamefont {Balents}, \citenamefont {Sachdev},\ and\ \citenamefont {Fisher}}]{Senthil2004}%
  \BibitemOpen
  \bibfield  {author} {\bibinfo {author} {\bibfnamefont {T.}~\bibnamefont {Senthil}}, \bibinfo {author} {\bibfnamefont {A.}~\bibnamefont {Vishwanath}}, \bibinfo {author} {\bibfnamefont {L.}~\bibnamefont {Balents}}, \bibinfo {author} {\bibfnamefont {S.}~\bibnamefont {Sachdev}},\ and\ \bibinfo {author} {\bibfnamefont {M.~P.~A.}\ \bibnamefont {Fisher}},\ }\href {https://doi.org/10.1126/science.1091806} {\bibfield  {journal} {\bibinfo  {journal} {Science}\ }\textbf {\bibinfo {volume} {303}},\ \bibinfo {pages} {1490} (\bibinfo {year} {2004}{\natexlab{a}})}\BibitemShut {NoStop}%
\bibitem [{\citenamefont {Senthil}\ \emph {et~al.}(2004{\natexlab{b}})\citenamefont {Senthil}, \citenamefont {Balents}, \citenamefont {Sachdev}, \citenamefont {Vishwanath},\ and\ \citenamefont {Fisher}}]{Senthil2004v2}%
  \BibitemOpen
  \bibfield  {author} {\bibinfo {author} {\bibfnamefont {T.}~\bibnamefont {Senthil}}, \bibinfo {author} {\bibfnamefont {L.}~\bibnamefont {Balents}}, \bibinfo {author} {\bibfnamefont {S.}~\bibnamefont {Sachdev}}, \bibinfo {author} {\bibfnamefont {A.}~\bibnamefont {Vishwanath}},\ and\ \bibinfo {author} {\bibfnamefont {M.~P.~A.}\ \bibnamefont {Fisher}},\ }\href {https://doi.org/10.1103/PhysRevB.70.144407} {\bibfield  {journal} {\bibinfo  {journal} {Phys. Rev. B}\ }\textbf {\bibinfo {volume} {70}},\ \bibinfo {pages} {144407} (\bibinfo {year} {2004}{\natexlab{b}})}\BibitemShut {NoStop}%
\bibitem [{\citenamefont {Senthil}(2023)}]{senthil2023deconfined}%
  \BibitemOpen
  \bibfield  {author} {\bibinfo {author} {\bibfnamefont {T.}~\bibnamefont {Senthil}},\ }\href@noop {} {} (\bibinfo {year} {2023}),\ \Eprint {https://arxiv.org/abs/2306.12638} {arXiv:2306.12638 [cond-mat.str-el]} \BibitemShut {NoStop}%
\bibitem [{\citenamefont {Sandvik}(2007)}]{Sandvik2007}%
  \BibitemOpen
  \bibfield  {author} {\bibinfo {author} {\bibfnamefont {A.~W.}\ \bibnamefont {Sandvik}},\ }\href {https://doi.org/10.1103/PhysRevLett.98.227202} {\bibfield  {journal} {\bibinfo  {journal} {Phys. Rev. Lett.}\ }\textbf {\bibinfo {volume} {98}},\ \bibinfo {pages} {227202} (\bibinfo {year} {2007})}\BibitemShut {NoStop}%
\bibitem [{\citenamefont {Jiang}\ \emph {et~al.}(2008)\citenamefont {Jiang}, \citenamefont {Nyfeler}, \citenamefont {Chandrasekharan},\ and\ \citenamefont {Wiese}}]{Jiang_2008}%
  \BibitemOpen
  \bibfield  {author} {\bibinfo {author} {\bibfnamefont {F.-J.}\ \bibnamefont {Jiang}}, \bibinfo {author} {\bibfnamefont {M.}~\bibnamefont {Nyfeler}}, \bibinfo {author} {\bibfnamefont {S.}~\bibnamefont {Chandrasekharan}},\ and\ \bibinfo {author} {\bibfnamefont {U.-J.}\ \bibnamefont {Wiese}},\ }\href {https://doi.org/10.1088/1742-5468/2008/02/P02009} {\bibfield  {journal} {\bibinfo  {journal} {J. Stat. Mech.: Theory Exp}\ }\textbf {\bibinfo {volume} {2008}},\ \bibinfo {pages} {P02009} (\bibinfo {year} {2008})}\BibitemShut {NoStop}%
\bibitem [{\citenamefont {Motrunich}\ and\ \citenamefont {Vishwanath}(2008)}]{motrunich2008}%
  \BibitemOpen
  \bibfield  {author} {\bibinfo {author} {\bibfnamefont {O.~I.}\ \bibnamefont {Motrunich}}\ and\ \bibinfo {author} {\bibfnamefont {A.}~\bibnamefont {Vishwanath}},\ }\href@noop {} {} (\bibinfo {year} {2008}),\ \Eprint {https://arxiv.org/abs/0805.1494} {arXiv:0805.1494 [cond-mat.stat-mech]} \BibitemShut {NoStop}%
\bibitem [{\citenamefont {Lou}\ \emph {et~al.}(2009)\citenamefont {Lou}, \citenamefont {Sandvik},\ and\ \citenamefont {Kawashima}}]{Lou2009}%
  \BibitemOpen
  \bibfield  {author} {\bibinfo {author} {\bibfnamefont {J.}~\bibnamefont {Lou}}, \bibinfo {author} {\bibfnamefont {A.~W.}\ \bibnamefont {Sandvik}},\ and\ \bibinfo {author} {\bibfnamefont {N.}~\bibnamefont {Kawashima}},\ }\href {https://doi.org/10.1103/PhysRevB.80.180414} {\bibfield  {journal} {\bibinfo  {journal} {Phys. Rev. B}\ }\textbf {\bibinfo {volume} {80}},\ \bibinfo {pages} {180414} (\bibinfo {year} {2009})}\BibitemShut {NoStop}%
\bibitem [{\citenamefont {Banerjee}\ \emph {et~al.}(2010)\citenamefont {Banerjee}, \citenamefont {Damle},\ and\ \citenamefont {Alet}}]{Banerjee2010}%
  \BibitemOpen
  \bibfield  {author} {\bibinfo {author} {\bibfnamefont {A.}~\bibnamefont {Banerjee}}, \bibinfo {author} {\bibfnamefont {K.}~\bibnamefont {Damle}},\ and\ \bibinfo {author} {\bibfnamefont {F.}~\bibnamefont {Alet}},\ }\href {https://doi.org/10.1103/PhysRevB.82.155139} {\bibfield  {journal} {\bibinfo  {journal} {Phys. Rev. B}\ }\textbf {\bibinfo {volume} {82}},\ \bibinfo {pages} {155139} (\bibinfo {year} {2010})}\BibitemShut {NoStop}%
\bibitem [{\citenamefont {Sandvik}(2010)}]{Sandvik2010}%
  \BibitemOpen
  \bibfield  {author} {\bibinfo {author} {\bibfnamefont {A.~W.}\ \bibnamefont {Sandvik}},\ }\href {https://doi.org/10.1103/PhysRevLett.104.177201} {\bibfield  {journal} {\bibinfo  {journal} {Phys. Rev. Lett.}\ }\textbf {\bibinfo {volume} {104}},\ \bibinfo {pages} {177201} (\bibinfo {year} {2010})}\BibitemShut {NoStop}%
\bibitem [{\citenamefont {Harada}\ \emph {et~al.}(2013)\citenamefont {Harada}, \citenamefont {Suzuki}, \citenamefont {Okubo}, \citenamefont {Matsuo}, \citenamefont {Lou}, \citenamefont {Watanabe}, \citenamefont {Todo},\ and\ \citenamefont {Kawashima}}]{Harada2013}%
  \BibitemOpen
  \bibfield  {author} {\bibinfo {author} {\bibfnamefont {K.}~\bibnamefont {Harada}}, \bibinfo {author} {\bibfnamefont {T.}~\bibnamefont {Suzuki}}, \bibinfo {author} {\bibfnamefont {T.}~\bibnamefont {Okubo}}, \bibinfo {author} {\bibfnamefont {H.}~\bibnamefont {Matsuo}}, \bibinfo {author} {\bibfnamefont {J.}~\bibnamefont {Lou}}, \bibinfo {author} {\bibfnamefont {H.}~\bibnamefont {Watanabe}}, \bibinfo {author} {\bibfnamefont {S.}~\bibnamefont {Todo}},\ and\ \bibinfo {author} {\bibfnamefont {N.}~\bibnamefont {Kawashima}},\ }\href {https://doi.org/10.1103/PhysRevB.88.220408} {\bibfield  {journal} {\bibinfo  {journal} {Phys. Rev. B}\ }\textbf {\bibinfo {volume} {88}},\ \bibinfo {pages} {220408} (\bibinfo {year} {2013})}\BibitemShut {NoStop}%
\bibitem [{\citenamefont {Chen}\ \emph {et~al.}(2013)\citenamefont {Chen}, \citenamefont {Huang}, \citenamefont {Deng}, \citenamefont {Kuklov}, \citenamefont {Prokof'ev},\ and\ \citenamefont {Svistunov}}]{Chen2013}%
  \BibitemOpen
  \bibfield  {author} {\bibinfo {author} {\bibfnamefont {K.}~\bibnamefont {Chen}}, \bibinfo {author} {\bibfnamefont {Y.}~\bibnamefont {Huang}}, \bibinfo {author} {\bibfnamefont {Y.}~\bibnamefont {Deng}}, \bibinfo {author} {\bibfnamefont {A.~B.}\ \bibnamefont {Kuklov}}, \bibinfo {author} {\bibfnamefont {N.~V.}\ \bibnamefont {Prokof'ev}},\ and\ \bibinfo {author} {\bibfnamefont {B.~V.}\ \bibnamefont {Svistunov}},\ }\href {https://doi.org/10.1103/PhysRevLett.110.185701} {\bibfield  {journal} {\bibinfo  {journal} {Phys. Rev. Lett.}\ }\textbf {\bibinfo {volume} {110}},\ \bibinfo {pages} {185701} (\bibinfo {year} {2013})}\BibitemShut {NoStop}%
\bibitem [{\citenamefont {Nahum}\ \emph {et~al.}(2015)\citenamefont {Nahum}, \citenamefont {Chalker}, \citenamefont {Serna}, \citenamefont {Ortu\~no},\ and\ \citenamefont {Somoza}}]{Nahum2015}%
  \BibitemOpen
  \bibfield  {author} {\bibinfo {author} {\bibfnamefont {A.}~\bibnamefont {Nahum}}, \bibinfo {author} {\bibfnamefont {J.~T.}\ \bibnamefont {Chalker}}, \bibinfo {author} {\bibfnamefont {P.}~\bibnamefont {Serna}}, \bibinfo {author} {\bibfnamefont {M.}~\bibnamefont {Ortu\~no}},\ and\ \bibinfo {author} {\bibfnamefont {A.~M.}\ \bibnamefont {Somoza}},\ }\href {https://doi.org/10.1103/PhysRevX.5.041048} {\bibfield  {journal} {\bibinfo  {journal} {Phys. Rev. X}\ }\textbf {\bibinfo {volume} {5}},\ \bibinfo {pages} {041048} (\bibinfo {year} {2015})}\BibitemShut {NoStop}%
\bibitem [{\citenamefont {Shao}\ \emph {et~al.}(2016)\citenamefont {Shao}, \citenamefont {Guo},\ and\ \citenamefont {Sandvik}}]{Shao2016}%
  \BibitemOpen
  \bibfield  {author} {\bibinfo {author} {\bibfnamefont {H.}~\bibnamefont {Shao}}, \bibinfo {author} {\bibfnamefont {W.}~\bibnamefont {Guo}},\ and\ \bibinfo {author} {\bibfnamefont {A.~W.}\ \bibnamefont {Sandvik}},\ }\href {https://doi.org/10.1126/science.aad5007} {\bibfield  {journal} {\bibinfo  {journal} {Science}\ }\textbf {\bibinfo {volume} {352}},\ \bibinfo {pages} {213} (\bibinfo {year} {2016})}\BibitemShut {NoStop}%
\bibitem [{\citenamefont {Lee}\ \emph {et~al.}(2019)\citenamefont {Lee}, \citenamefont {You}, \citenamefont {Sachdev},\ and\ \citenamefont {Vishwanath}}]{Lee2019}%
  \BibitemOpen
  \bibfield  {author} {\bibinfo {author} {\bibfnamefont {J.~Y.}\ \bibnamefont {Lee}}, \bibinfo {author} {\bibfnamefont {Y.-Z.}\ \bibnamefont {You}}, \bibinfo {author} {\bibfnamefont {S.}~\bibnamefont {Sachdev}},\ and\ \bibinfo {author} {\bibfnamefont {A.}~\bibnamefont {Vishwanath}},\ }\href {https://doi.org/10.1103/PhysRevX.9.041037} {\bibfield  {journal} {\bibinfo  {journal} {Phys. Rev. X}\ }\textbf {\bibinfo {volume} {9}},\ \bibinfo {pages} {041037} (\bibinfo {year} {2019})}\BibitemShut {NoStop}%
\bibitem [{\citenamefont {Song}\ \emph {et~al.}(2023)\citenamefont {Song}, \citenamefont {Zhao}, \citenamefont {Janssen}, \citenamefont {Scherer},\ and\ \citenamefont {Meng}}]{song2023deconfined}%
  \BibitemOpen
  \bibfield  {author} {\bibinfo {author} {\bibfnamefont {M.}~\bibnamefont {Song}}, \bibinfo {author} {\bibfnamefont {J.}~\bibnamefont {Zhao}}, \bibinfo {author} {\bibfnamefont {L.}~\bibnamefont {Janssen}}, \bibinfo {author} {\bibfnamefont {M.~M.}\ \bibnamefont {Scherer}},\ and\ \bibinfo {author} {\bibfnamefont {Z.~Y.}\ \bibnamefont {Meng}},\ }\href@noop {} {} (\bibinfo {year} {2023}),\ \Eprint {https://arxiv.org/abs/2307.02547} {arXiv:2307.02547 [cond-mat.str-el]} \BibitemShut {NoStop}%
\bibitem [{\citenamefont {Li}\ \emph {et~al.}(2019)\citenamefont {Li}, \citenamefont {Jian},\ and\ \citenamefont {Yao}}]{Zi2019}%
  \BibitemOpen
  \bibfield  {author} {\bibinfo {author} {\bibfnamefont {Z.-X.}\ \bibnamefont {Li}}, \bibinfo {author} {\bibfnamefont {S.-K.}\ \bibnamefont {Jian}},\ and\ \bibinfo {author} {\bibfnamefont {H.}~\bibnamefont {Yao}},\ }\href@noop {} {} (\bibinfo {year} {2019}),\ \Eprint {https://arxiv.org/abs/1904.10975} {arXiv:1904.10975 [cond-mat.str-el]} \BibitemShut {NoStop}%
\bibitem [{\citenamefont {Assaad}\ and\ \citenamefont {Grover}(2016)}]{Assaad2016}%
  \BibitemOpen
  \bibfield  {author} {\bibinfo {author} {\bibfnamefont {F.~F.}\ \bibnamefont {Assaad}}\ and\ \bibinfo {author} {\bibfnamefont {T.}~\bibnamefont {Grover}},\ }\href {https://doi.org/10.1103/PhysRevX.6.041049} {\bibfield  {journal} {\bibinfo  {journal} {Phys. Rev. X}\ }\textbf {\bibinfo {volume} {6}},\ \bibinfo {pages} {041049} (\bibinfo {year} {2016})}\BibitemShut {NoStop}%
\bibitem [{\citenamefont {Liu}\ \emph {et~al.}(2022)\citenamefont {Liu}, \citenamefont {Jiang}, \citenamefont {Chen}, \citenamefont {Rong}, \citenamefont {Cheng}, \citenamefont {Sun}, \citenamefont {Meng},\ and\ \citenamefont {Assaad}}]{liu2022}%
  \BibitemOpen
  \bibfield  {author} {\bibinfo {author} {\bibfnamefont {Z.~H.}\ \bibnamefont {Liu}}, \bibinfo {author} {\bibfnamefont {W.}~\bibnamefont {Jiang}}, \bibinfo {author} {\bibfnamefont {B.-B.}\ \bibnamefont {Chen}}, \bibinfo {author} {\bibfnamefont {J.}~\bibnamefont {Rong}}, \bibinfo {author} {\bibfnamefont {M.}~\bibnamefont {Cheng}}, \bibinfo {author} {\bibfnamefont {K.}~\bibnamefont {Sun}}, \bibinfo {author} {\bibfnamefont {Z.~Y.}\ \bibnamefont {Meng}},\ and\ \bibinfo {author} {\bibfnamefont {F.~F.}\ \bibnamefont {Assaad}},\ }\href@noop {} {} (\bibinfo {year} {2022}),\ \Eprint {https://arxiv.org/abs/2212.11821} {arXiv:2212.11821 [cond-mat.str-el]} \BibitemShut {NoStop}%
\bibitem [{\citenamefont {Liao}\ \emph {et~al.}(2023)\citenamefont {Liao}, \citenamefont {Pan}, \citenamefont {Jiang}, \citenamefont {Qi},\ and\ \citenamefont {Meng}}]{Yuan2023}%
  \BibitemOpen
  \bibfield  {author} {\bibinfo {author} {\bibfnamefont {Y.~D.}\ \bibnamefont {Liao}}, \bibinfo {author} {\bibfnamefont {G.}~\bibnamefont {Pan}}, \bibinfo {author} {\bibfnamefont {W.}~\bibnamefont {Jiang}}, \bibinfo {author} {\bibfnamefont {Y.}~\bibnamefont {Qi}},\ and\ \bibinfo {author} {\bibfnamefont {Z.~Y.}\ \bibnamefont {Meng}},\ }\href@noop {} {} (\bibinfo {year} {2023}),\ \Eprint {https://arxiv.org/abs/2302.11742} {arXiv:2302.11742 [cond-mat.str-el]} \BibitemShut {NoStop}%
\bibitem [{\citenamefont {Charrier}\ \emph {et~al.}(2008)\citenamefont {Charrier}, \citenamefont {Alet},\ and\ \citenamefont {Pujol}}]{Charrier2008}%
  \BibitemOpen
  \bibfield  {author} {\bibinfo {author} {\bibfnamefont {D.}~\bibnamefont {Charrier}}, \bibinfo {author} {\bibfnamefont {F.}~\bibnamefont {Alet}},\ and\ \bibinfo {author} {\bibfnamefont {P.}~\bibnamefont {Pujol}},\ }\href {https://doi.org/10.1103/PhysRevLett.101.167205} {\bibfield  {journal} {\bibinfo  {journal} {Phys. Rev. Lett.}\ }\textbf {\bibinfo {volume} {101}},\ \bibinfo {pages} {167205} (\bibinfo {year} {2008})}\BibitemShut {NoStop}%
\bibitem [{\citenamefont {Sreejith}\ and\ \citenamefont {Powell}(2015)}]{Sreejith2015}%
  \BibitemOpen
  \bibfield  {author} {\bibinfo {author} {\bibfnamefont {G.~J.}\ \bibnamefont {Sreejith}}\ and\ \bibinfo {author} {\bibfnamefont {S.}~\bibnamefont {Powell}},\ }\href {https://doi.org/10.1103/PhysRevB.92.184413} {\bibfield  {journal} {\bibinfo  {journal} {Phys. Rev. B}\ }\textbf {\bibinfo {volume} {92}},\ \bibinfo {pages} {184413} (\bibinfo {year} {2015})}\BibitemShut {NoStop}%
\bibitem [{\citenamefont {Jiang}\ and\ \citenamefont {Motrunich}(2019)}]{Jiang2019}%
  \BibitemOpen
  \bibfield  {author} {\bibinfo {author} {\bibfnamefont {S.}~\bibnamefont {Jiang}}\ and\ \bibinfo {author} {\bibfnamefont {O.}~\bibnamefont {Motrunich}},\ }\href {https://doi.org/10.1103/PhysRevB.99.075103} {\bibfield  {journal} {\bibinfo  {journal} {Phys. Rev. B}\ }\textbf {\bibinfo {volume} {99}},\ \bibinfo {pages} {075103} (\bibinfo {year} {2019})}\BibitemShut {NoStop}%
\bibitem [{\citenamefont {Roberts}\ \emph {et~al.}(2019)\citenamefont {Roberts}, \citenamefont {Jiang},\ and\ \citenamefont {Motrunich}}]{Roberts2019}%
  \BibitemOpen
  \bibfield  {author} {\bibinfo {author} {\bibfnamefont {B.}~\bibnamefont {Roberts}}, \bibinfo {author} {\bibfnamefont {S.}~\bibnamefont {Jiang}},\ and\ \bibinfo {author} {\bibfnamefont {O.~I.}\ \bibnamefont {Motrunich}},\ }\href {https://doi.org/10.1103/PhysRevB.99.165143} {\bibfield  {journal} {\bibinfo  {journal} {Phys. Rev. B}\ }\textbf {\bibinfo {volume} {99}},\ \bibinfo {pages} {165143} (\bibinfo {year} {2019})}\BibitemShut {NoStop}%
\bibitem [{\citenamefont {Huang}\ \emph {et~al.}(2019)\citenamefont {Huang}, \citenamefont {Lu}, \citenamefont {You}, \citenamefont {Meng},\ and\ \citenamefont {Xiang}}]{Huang2019}%
  \BibitemOpen
  \bibfield  {author} {\bibinfo {author} {\bibfnamefont {R.-Z.}\ \bibnamefont {Huang}}, \bibinfo {author} {\bibfnamefont {D.-C.}\ \bibnamefont {Lu}}, \bibinfo {author} {\bibfnamefont {Y.-Z.}\ \bibnamefont {You}}, \bibinfo {author} {\bibfnamefont {Z.~Y.}\ \bibnamefont {Meng}},\ and\ \bibinfo {author} {\bibfnamefont {T.}~\bibnamefont {Xiang}},\ }\href {https://doi.org/10.1103/PhysRevB.100.125137} {\bibfield  {journal} {\bibinfo  {journal} {Phys. Rev. B}\ }\textbf {\bibinfo {volume} {100}},\ \bibinfo {pages} {125137} (\bibinfo {year} {2019})}\BibitemShut {NoStop}%
\bibitem [{\citenamefont {Mudry}\ \emph {et~al.}(2019)\citenamefont {Mudry}, \citenamefont {Furusaki}, \citenamefont {Morimoto},\ and\ \citenamefont {Hikihara}}]{Mudry2019}%
  \BibitemOpen
  \bibfield  {author} {\bibinfo {author} {\bibfnamefont {C.}~\bibnamefont {Mudry}}, \bibinfo {author} {\bibfnamefont {A.}~\bibnamefont {Furusaki}}, \bibinfo {author} {\bibfnamefont {T.}~\bibnamefont {Morimoto}},\ and\ \bibinfo {author} {\bibfnamefont {T.}~\bibnamefont {Hikihara}},\ }\href {https://doi.org/10.1103/PhysRevB.99.205153} {\bibfield  {journal} {\bibinfo  {journal} {Phys. Rev. B}\ }\textbf {\bibinfo {volume} {99}},\ \bibinfo {pages} {205153} (\bibinfo {year} {2019})}\BibitemShut {NoStop}%
\bibitem [{\citenamefont {Roberts}\ \emph {et~al.}(2021)\citenamefont {Roberts}, \citenamefont {Jiang},\ and\ \citenamefont {Motrunich}}]{Roberts2021}%
  \BibitemOpen
  \bibfield  {author} {\bibinfo {author} {\bibfnamefont {B.}~\bibnamefont {Roberts}}, \bibinfo {author} {\bibfnamefont {S.}~\bibnamefont {Jiang}},\ and\ \bibinfo {author} {\bibfnamefont {O.~I.}\ \bibnamefont {Motrunich}},\ }\href {https://doi.org/10.1103/PhysRevB.103.155143} {\bibfield  {journal} {\bibinfo  {journal} {Phys. Rev. B}\ }\textbf {\bibinfo {volume} {103}},\ \bibinfo {pages} {155143} (\bibinfo {year} {2021})}\BibitemShut {NoStop}%
\bibitem [{\citenamefont {Lee}\ \emph {et~al.}(2022)\citenamefont {Lee}, \citenamefont {Ramette}, \citenamefont {Metlitski}, \citenamefont {Vuletic}, \citenamefont {Ho},\ and\ \citenamefont {Choi}}]{lee2022}%
  \BibitemOpen
  \bibfield  {author} {\bibinfo {author} {\bibfnamefont {J.~Y.}\ \bibnamefont {Lee}}, \bibinfo {author} {\bibfnamefont {J.}~\bibnamefont {Ramette}}, \bibinfo {author} {\bibfnamefont {M.~A.}\ \bibnamefont {Metlitski}}, \bibinfo {author} {\bibfnamefont {V.}~\bibnamefont {Vuletic}}, \bibinfo {author} {\bibfnamefont {W.~W.}\ \bibnamefont {Ho}},\ and\ \bibinfo {author} {\bibfnamefont {S.}~\bibnamefont {Choi}},\ }\href@noop {} {} (\bibinfo {year} {2022}),\ \Eprint {https://arxiv.org/abs/2207.08829} {arXiv:2207.08829 [cond-mat.str-el]} \BibitemShut {NoStop}%
\bibitem [{\citenamefont {Prembabu}\ \emph {et~al.}(2022)\citenamefont {Prembabu}, \citenamefont {Thorngren},\ and\ \citenamefont {Verresen}}]{prembabu2022}%
  \BibitemOpen
  \bibfield  {author} {\bibinfo {author} {\bibfnamefont {S.}~\bibnamefont {Prembabu}}, \bibinfo {author} {\bibfnamefont {R.}~\bibnamefont {Thorngren}},\ and\ \bibinfo {author} {\bibfnamefont {R.}~\bibnamefont {Verresen}},\ }\href@noop {} {} (\bibinfo {year} {2022}),\ \Eprint {https://arxiv.org/abs/2208.12258} {arXiv:2208.12258 [cond-mat.str-el]} \BibitemShut {NoStop}%
\bibitem [{\citenamefont {Zayed}\ \emph {et~al.}(2017)\citenamefont {Zayed}, \citenamefont {Rüegg}, \citenamefont {Larrea~J.}, \citenamefont {Läuchli}, \citenamefont {Panagopoulos}, \citenamefont {Saxena}, \citenamefont {Ellerby}, \citenamefont {McMorrow}, \citenamefont {Strässle}, \citenamefont {Klotz}, \citenamefont {Hamel}, \citenamefont {Sadykov}, \citenamefont {Pomjakushin}, \citenamefont {Boehm}, \citenamefont {Jiménez–Ruiz}, \citenamefont {Schneidewind}, \citenamefont {Pomjakushina}, \citenamefont {Stingaciu}, \citenamefont {Conder},\ and\ \citenamefont {Rønnow}}]{Zayed2017}%
  \BibitemOpen
  \bibfield  {author} {\bibinfo {author} {\bibfnamefont {M.~E.}\ \bibnamefont {Zayed}}, \bibinfo {author} {\bibfnamefont {C.}~\bibnamefont {Rüegg}}, \bibinfo {author} {\bibfnamefont {J.}~\bibnamefont {Larrea~J.}}, \bibinfo {author} {\bibfnamefont {A.~M.}\ \bibnamefont {Läuchli}}, \bibinfo {author} {\bibfnamefont {C.}~\bibnamefont {Panagopoulos}}, \bibinfo {author} {\bibfnamefont {S.~S.}\ \bibnamefont {Saxena}}, \bibinfo {author} {\bibfnamefont {M.}~\bibnamefont {Ellerby}}, \bibinfo {author} {\bibfnamefont {D.~F.}\ \bibnamefont {McMorrow}}, \bibinfo {author} {\bibfnamefont {T.}~\bibnamefont {Strässle}}, \bibinfo {author} {\bibfnamefont {S.}~\bibnamefont {Klotz}}, \bibinfo {author} {\bibfnamefont {G.}~\bibnamefont {Hamel}}, \bibinfo {author} {\bibfnamefont {R.~A.}\ \bibnamefont {Sadykov}}, \bibinfo {author} {\bibfnamefont {V.}~\bibnamefont {Pomjakushin}}, \bibinfo {author} {\bibfnamefont {M.}~\bibnamefont {Boehm}}, \bibinfo {author} {\bibfnamefont {M.}~\bibnamefont {Jiménez–Ruiz}}, \bibinfo {author}
  {\bibfnamefont {A.}~\bibnamefont {Schneidewind}}, \bibinfo {author} {\bibfnamefont {E.}~\bibnamefont {Pomjakushina}}, \bibinfo {author} {\bibfnamefont {M.}~\bibnamefont {Stingaciu}}, \bibinfo {author} {\bibfnamefont {K.}~\bibnamefont {Conder}},\ and\ \bibinfo {author} {\bibfnamefont {H.~M.}\ \bibnamefont {Rønnow}},\ }\href {https://doi.org/10.1038/nphys4190} {\bibfield  {journal} {\bibinfo  {journal} {Nat. Commun.}\ }\textbf {\bibinfo {volume} {13}},\ \bibinfo {pages} {962} (\bibinfo {year} {2017})}\BibitemShut {NoStop}%
\bibitem [{\citenamefont {Guo}\ \emph {et~al.}(2020)\citenamefont {Guo}, \citenamefont {Sun}, \citenamefont {Zhao}, \citenamefont {Wang}, \citenamefont {Hong}, \citenamefont {Sidorov}, \citenamefont {Ma}, \citenamefont {Wu}, \citenamefont {Li}, \citenamefont {Meng}, \citenamefont {Sandvik},\ and\ \citenamefont {Sun}}]{Guo2020}%
  \BibitemOpen
  \bibfield  {author} {\bibinfo {author} {\bibfnamefont {J.}~\bibnamefont {Guo}}, \bibinfo {author} {\bibfnamefont {G.}~\bibnamefont {Sun}}, \bibinfo {author} {\bibfnamefont {B.}~\bibnamefont {Zhao}}, \bibinfo {author} {\bibfnamefont {L.}~\bibnamefont {Wang}}, \bibinfo {author} {\bibfnamefont {W.}~\bibnamefont {Hong}}, \bibinfo {author} {\bibfnamefont {V.~A.}\ \bibnamefont {Sidorov}}, \bibinfo {author} {\bibfnamefont {N.}~\bibnamefont {Ma}}, \bibinfo {author} {\bibfnamefont {Q.}~\bibnamefont {Wu}}, \bibinfo {author} {\bibfnamefont {S.}~\bibnamefont {Li}}, \bibinfo {author} {\bibfnamefont {Z.~Y.}\ \bibnamefont {Meng}}, \bibinfo {author} {\bibfnamefont {A.~W.}\ \bibnamefont {Sandvik}},\ and\ \bibinfo {author} {\bibfnamefont {L.}~\bibnamefont {Sun}},\ }\href {https://doi.org/10.1103/PhysRevLett.124.206602} {\bibfield  {journal} {\bibinfo  {journal} {Phys. Rev. Lett.}\ }\textbf {\bibinfo {volume} {124}},\ \bibinfo {pages} {206602} (\bibinfo {year} {2020})}\BibitemShut {NoStop}%
\bibitem [{\citenamefont {Cui}\ \emph {et~al.}(2022)\citenamefont {Cui}, \citenamefont {Liu}, \citenamefont {Lin}, \citenamefont {Wu}, \citenamefont {Hong}, \citenamefont {Liu}, \citenamefont {Li}, \citenamefont {Hu}, \citenamefont {Xi}, \citenamefont {Li}, \citenamefont {Yu}, \citenamefont {Sandvik},\ and\ \citenamefont {Yu}}]{cui2022}%
  \BibitemOpen
  \bibfield  {author} {\bibinfo {author} {\bibfnamefont {Y.}~\bibnamefont {Cui}}, \bibinfo {author} {\bibfnamefont {L.}~\bibnamefont {Liu}}, \bibinfo {author} {\bibfnamefont {H.}~\bibnamefont {Lin}}, \bibinfo {author} {\bibfnamefont {K.-H.}\ \bibnamefont {Wu}}, \bibinfo {author} {\bibfnamefont {W.}~\bibnamefont {Hong}}, \bibinfo {author} {\bibfnamefont {X.}~\bibnamefont {Liu}}, \bibinfo {author} {\bibfnamefont {C.}~\bibnamefont {Li}}, \bibinfo {author} {\bibfnamefont {Z.}~\bibnamefont {Hu}}, \bibinfo {author} {\bibfnamefont {N.}~\bibnamefont {Xi}}, \bibinfo {author} {\bibfnamefont {S.}~\bibnamefont {Li}}, \bibinfo {author} {\bibfnamefont {R.}~\bibnamefont {Yu}}, \bibinfo {author} {\bibfnamefont {A.~W.}\ \bibnamefont {Sandvik}},\ and\ \bibinfo {author} {\bibfnamefont {W.}~\bibnamefont {Yu}},\ }\href@noop {} {} (\bibinfo {year} {2022}),\ \Eprint {https://arxiv.org/abs/2204.08133} {arXiv:2204.08133 [cond-mat.str-el]} \BibitemShut {NoStop}%
\bibitem [{\citenamefont {Hong}\ \emph {et~al.}(2022)\citenamefont {Hong}, \citenamefont {Ying}, \citenamefont {Huang}, \citenamefont {Dissanayake}, \citenamefont {Qiu}, \citenamefont {Turnbull}, \citenamefont {Podlesnyak}, \citenamefont {Wu}, \citenamefont {Cao}, \citenamefont {Liu}, \citenamefont {Umehara}, \citenamefont {Gouchi}, \citenamefont {Uwatoko}, \citenamefont {Matsuda}, \citenamefont {Tennant}, \citenamefont {Chern}, \citenamefont {Schmidt},\ and\ \citenamefont {Wessel}}]{Tao2022}%
  \BibitemOpen
  \bibfield  {author} {\bibinfo {author} {\bibfnamefont {T.}~\bibnamefont {Hong}}, \bibinfo {author} {\bibfnamefont {T.}~\bibnamefont {Ying}}, \bibinfo {author} {\bibfnamefont {Q.}~\bibnamefont {Huang}}, \bibinfo {author} {\bibfnamefont {S.~E.}\ \bibnamefont {Dissanayake}}, \bibinfo {author} {\bibfnamefont {Y.}~\bibnamefont {Qiu}}, \bibinfo {author} {\bibfnamefont {M.~M.}\ \bibnamefont {Turnbull}}, \bibinfo {author} {\bibfnamefont {A.~A.}\ \bibnamefont {Podlesnyak}}, \bibinfo {author} {\bibfnamefont {Y.}~\bibnamefont {Wu}}, \bibinfo {author} {\bibfnamefont {H.}~\bibnamefont {Cao}}, \bibinfo {author} {\bibfnamefont {Y.}~\bibnamefont {Liu}}, \bibinfo {author} {\bibfnamefont {I.}~\bibnamefont {Umehara}}, \bibinfo {author} {\bibfnamefont {J.}~\bibnamefont {Gouchi}}, \bibinfo {author} {\bibfnamefont {Y.}~\bibnamefont {Uwatoko}}, \bibinfo {author} {\bibfnamefont {M.}~\bibnamefont {Matsuda}}, \bibinfo {author} {\bibfnamefont {D.~A.}\ \bibnamefont {Tennant}}, \bibinfo {author} {\bibfnamefont {G.-W.}\ \bibnamefont
  {Chern}}, \bibinfo {author} {\bibfnamefont {K.~P.}\ \bibnamefont {Schmidt}},\ and\ \bibinfo {author} {\bibfnamefont {S.}~\bibnamefont {Wessel}},\ }\href {https://doi.org/10.1038/s41467-022-30769-8} {\bibfield  {journal} {\bibinfo  {journal} {Nat. Commun.}\ }\textbf {\bibinfo {volume} {13}},\ \bibinfo {pages} {3073} (\bibinfo {year} {2022})}\BibitemShut {NoStop}%
\bibitem [{\citenamefont {Jaksch}\ and\ \citenamefont {Zoller}(2003)}]{Jaksch_2003}%
  \BibitemOpen
  \bibfield  {author} {\bibinfo {author} {\bibfnamefont {D.}~\bibnamefont {Jaksch}}\ and\ \bibinfo {author} {\bibfnamefont {P.}~\bibnamefont {Zoller}},\ }\href {https://doi.org/10.1088/1367-2630/5/1/356} {\bibfield  {journal} {\bibinfo  {journal} {New J. Phys.}\ }\textbf {\bibinfo {volume} {5}},\ \bibinfo {pages} {56} (\bibinfo {year} {2003})}\BibitemShut {NoStop}%
\bibitem [{\citenamefont {Gerbier}\ and\ \citenamefont {Dalibard}(2010)}]{Gerbier_2010}%
  \BibitemOpen
  \bibfield  {author} {\bibinfo {author} {\bibfnamefont {F.}~\bibnamefont {Gerbier}}\ and\ \bibinfo {author} {\bibfnamefont {J.}~\bibnamefont {Dalibard}},\ }\href {https://doi.org/10.1088/1367-2630/12/3/033007} {\bibfield  {journal} {\bibinfo  {journal} {New J. Phys.}\ }\textbf {\bibinfo {volume} {12}},\ \bibinfo {pages} {033007} (\bibinfo {year} {2010})}\BibitemShut {NoStop}%
\bibitem [{SM()}]{SM}%
  \BibitemOpen
  \href@noop {} {}\bibinfo {note} {See Supplementary material, containing Refs.~\cite{SATO2011,Furukawa2012,Drechsler2007,Wolter2012,Orlova2017,Grams2022,GRIMM200095,klostermann22,Chin2004,Frye2019,Impertro2022,anismovas2016,kessler2014,atala14a,Julia2022,Murthy2014,Henning2018,Bohrdt2018,Brown2020,Atala2014} for details on the effective strong-coupling limit spin model of $H_{FEBH}$, the proposed experimental implementation of the frustrated geometry and the detection schemes for the quantities defined in the text.}\BibitemShut {Stop}%
\bibitem [{\citenamefont {Dalla~Torre}\ \emph {et~al.}(2006)\citenamefont {Dalla~Torre}, \citenamefont {Berg},\ and\ \citenamefont {Altman}}]{dallatorre2006}%
  \BibitemOpen
  \bibfield  {author} {\bibinfo {author} {\bibfnamefont {E.~G.}\ \bibnamefont {Dalla~Torre}}, \bibinfo {author} {\bibfnamefont {E.}~\bibnamefont {Berg}},\ and\ \bibinfo {author} {\bibfnamefont {E.}~\bibnamefont {Altman}},\ }\href {https://doi.org/10.1103/PhysRevLett.97.260401} {\bibfield  {journal} {\bibinfo  {journal} {Phys. Rev. Lett.}\ }\textbf {\bibinfo {volume} {97}},\ \bibinfo {pages} {260401} (\bibinfo {year} {2006})}\BibitemShut {NoStop}%
\bibitem [{\citenamefont {Dhar}\ \emph {et~al.}(2012)\citenamefont {Dhar}, \citenamefont {Maji}, \citenamefont {Mishra}, \citenamefont {Pai}, \citenamefont {Mukerjee},\ and\ \citenamefont {Paramekanti}}]{Dhar2012}%
  \BibitemOpen
  \bibfield  {author} {\bibinfo {author} {\bibfnamefont {A.}~\bibnamefont {Dhar}}, \bibinfo {author} {\bibfnamefont {M.}~\bibnamefont {Maji}}, \bibinfo {author} {\bibfnamefont {T.}~\bibnamefont {Mishra}}, \bibinfo {author} {\bibfnamefont {R.~V.}\ \bibnamefont {Pai}}, \bibinfo {author} {\bibfnamefont {S.}~\bibnamefont {Mukerjee}},\ and\ \bibinfo {author} {\bibfnamefont {A.}~\bibnamefont {Paramekanti}},\ }\href {https://doi.org/10.1103/PhysRevA.85.041602} {\bibfield  {journal} {\bibinfo  {journal} {Phys. Rev. A}\ }\textbf {\bibinfo {volume} {85}},\ \bibinfo {pages} {041602} (\bibinfo {year} {2012})}\BibitemShut {NoStop}%
\bibitem [{\citenamefont {Greschner}\ \emph {et~al.}(2013)\citenamefont {Greschner}, \citenamefont {Santos},\ and\ \citenamefont {Vekua}}]{Greschner2013}%
  \BibitemOpen
  \bibfield  {author} {\bibinfo {author} {\bibfnamefont {S.}~\bibnamefont {Greschner}}, \bibinfo {author} {\bibfnamefont {L.}~\bibnamefont {Santos}},\ and\ \bibinfo {author} {\bibfnamefont {T.}~\bibnamefont {Vekua}},\ }\href {https://doi.org/10.1103/PhysRevA.87.033609} {\bibfield  {journal} {\bibinfo  {journal} {Phys. Rev. A}\ }\textbf {\bibinfo {volume} {87}},\ \bibinfo {pages} {033609} (\bibinfo {year} {2013})}\BibitemShut {NoStop}%
\bibitem [{\citenamefont {Zaletel}\ \emph {et~al.}(2014)\citenamefont {Zaletel}, \citenamefont {Parameswaran}, \citenamefont {R\"uegg},\ and\ \citenamefont {Altman}}]{Zaletel2014}%
  \BibitemOpen
  \bibfield  {author} {\bibinfo {author} {\bibfnamefont {M.~P.}\ \bibnamefont {Zaletel}}, \bibinfo {author} {\bibfnamefont {S.~A.}\ \bibnamefont {Parameswaran}}, \bibinfo {author} {\bibfnamefont {A.}~\bibnamefont {R\"uegg}},\ and\ \bibinfo {author} {\bibfnamefont {E.}~\bibnamefont {Altman}},\ }\href {https://doi.org/10.1103/PhysRevB.89.155142} {\bibfield  {journal} {\bibinfo  {journal} {Phys. Rev. B}\ }\textbf {\bibinfo {volume} {89}},\ \bibinfo {pages} {155142} (\bibinfo {year} {2014})}\BibitemShut {NoStop}%
\bibitem [{\citenamefont {Mishra}\ \emph {et~al.}(2015)\citenamefont {Mishra}, \citenamefont {Greschner},\ and\ \citenamefont {Santos}}]{Mishra2015}%
  \BibitemOpen
  \bibfield  {author} {\bibinfo {author} {\bibfnamefont {T.}~\bibnamefont {Mishra}}, \bibinfo {author} {\bibfnamefont {S.}~\bibnamefont {Greschner}},\ and\ \bibinfo {author} {\bibfnamefont {L.}~\bibnamefont {Santos}},\ }\href {https://doi.org/10.1103/PhysRevA.91.043614} {\bibfield  {journal} {\bibinfo  {journal} {Phys. Rev. A}\ }\textbf {\bibinfo {volume} {91}},\ \bibinfo {pages} {043614} (\bibinfo {year} {2015})}\BibitemShut {NoStop}%
\bibitem [{\citenamefont {Romen}\ and\ \citenamefont {L\"auchli}(2018)}]{Romen2018}%
  \BibitemOpen
  \bibfield  {author} {\bibinfo {author} {\bibfnamefont {C.}~\bibnamefont {Romen}}\ and\ \bibinfo {author} {\bibfnamefont {A.~M.}\ \bibnamefont {L\"auchli}},\ }\href {https://doi.org/10.1103/PhysRevB.98.054519} {\bibfield  {journal} {\bibinfo  {journal} {Phys. Rev. B}\ }\textbf {\bibinfo {volume} {98}},\ \bibinfo {pages} {054519} (\bibinfo {year} {2018})}\BibitemShut {NoStop}%
\bibitem [{\citenamefont {Fraxanet}\ \emph {et~al.}(2022)\citenamefont {Fraxanet}, \citenamefont {Gonz\'alez-Cuadra}, \citenamefont {Pfau}, \citenamefont {Lewenstein}, \citenamefont {Langen},\ and\ \citenamefont {Barbiero}}]{Fraxanet2022}%
  \BibitemOpen
  \bibfield  {author} {\bibinfo {author} {\bibfnamefont {J.}~\bibnamefont {Fraxanet}}, \bibinfo {author} {\bibfnamefont {D.}~\bibnamefont {Gonz\'alez-Cuadra}}, \bibinfo {author} {\bibfnamefont {T.}~\bibnamefont {Pfau}}, \bibinfo {author} {\bibfnamefont {M.}~\bibnamefont {Lewenstein}}, \bibinfo {author} {\bibfnamefont {T.}~\bibnamefont {Langen}},\ and\ \bibinfo {author} {\bibfnamefont {L.}~\bibnamefont {Barbiero}},\ }\href {https://doi.org/10.1103/PhysRevLett.128.043402} {\bibfield  {journal} {\bibinfo  {journal} {Phys. Rev. Lett.}\ }\textbf {\bibinfo {volume} {128}},\ \bibinfo {pages} {043402} (\bibinfo {year} {2022})}\BibitemShut {NoStop}%
\bibitem [{\citenamefont {Singha~Roy}\ \emph {et~al.}(2022)\citenamefont {Singha~Roy}, \citenamefont {Carl},\ and\ \citenamefont {Hauke}}]{Roy2022}%
  \BibitemOpen
  \bibfield  {author} {\bibinfo {author} {\bibfnamefont {S.}~\bibnamefont {Singha~Roy}}, \bibinfo {author} {\bibfnamefont {L.}~\bibnamefont {Carl}},\ and\ \bibinfo {author} {\bibfnamefont {P.}~\bibnamefont {Hauke}},\ }\href {https://doi.org/10.1103/PhysRevB.106.195158} {\bibfield  {journal} {\bibinfo  {journal} {Phys. Rev. B}\ }\textbf {\bibinfo {volume} {106}},\ \bibinfo {pages} {195158} (\bibinfo {year} {2022})}\BibitemShut {NoStop}%
\bibitem [{\citenamefont {Halati}\ and\ \citenamefont {Giamarchi}(2023)}]{Halati2023}%
  \BibitemOpen
  \bibfield  {author} {\bibinfo {author} {\bibfnamefont {C.-M.}\ \bibnamefont {Halati}}\ and\ \bibinfo {author} {\bibfnamefont {T.}~\bibnamefont {Giamarchi}},\ }\href {https://doi.org/10.1103/PhysRevResearch.5.013126} {\bibfield  {journal} {\bibinfo  {journal} {Phys. Rev. Res.}\ }\textbf {\bibinfo {volume} {5}},\ \bibinfo {pages} {013126} (\bibinfo {year} {2023})}\BibitemShut {NoStop}%
\bibitem [{\citenamefont {Giamarchi}(2004)}]{Giamarchi2004}%
  \BibitemOpen
  \bibfield  {author} {\bibinfo {author} {\bibfnamefont {T.}~\bibnamefont {Giamarchi}},\ }\href {https://www.ebook.de/de/product/3263637/thierry_giamarchi_quantum_physics_in_one_dimension.html} {\emph {\bibinfo {title} {Quantum Physics in One Dimension}}}\ (\bibinfo  {publisher} {Oxford University Press},\ \bibinfo {year} {2004})\BibitemShut {NoStop}%
\bibitem [{\citenamefont {Peierls}(1996)}]{peierls_96}%
  \BibitemOpen
  \bibfield  {author} {\bibinfo {author} {\bibfnamefont {R.~E.}\ \bibnamefont {Peierls}},\ }\href@noop {} {\emph {\bibinfo {title} {Quantum theory of solids}}}\ (\bibinfo  {publisher} {Clarendon Press},\ \bibinfo {year} {1996})\BibitemShut {NoStop}%
\bibitem [{\citenamefont {Juli\`a-Farr\'e}\ \emph {et~al.}(2022)\citenamefont {Juli\`a-Farr\'e}, \citenamefont {Gonz\'alez-Cuadra}, \citenamefont {Patscheider}, \citenamefont {Mark}, \citenamefont {Ferlaino}, \citenamefont {Lewenstein}, \citenamefont {Barbiero},\ and\ \citenamefont {Dauphin}}]{Julia2022}%
  \BibitemOpen
  \bibfield  {author} {\bibinfo {author} {\bibfnamefont {S.}~\bibnamefont {Juli\`a-Farr\'e}}, \bibinfo {author} {\bibfnamefont {D.}~\bibnamefont {Gonz\'alez-Cuadra}}, \bibinfo {author} {\bibfnamefont {A.}~\bibnamefont {Patscheider}}, \bibinfo {author} {\bibfnamefont {M.~J.}\ \bibnamefont {Mark}}, \bibinfo {author} {\bibfnamefont {F.}~\bibnamefont {Ferlaino}}, \bibinfo {author} {\bibfnamefont {M.}~\bibnamefont {Lewenstein}}, \bibinfo {author} {\bibfnamefont {L.}~\bibnamefont {Barbiero}},\ and\ \bibinfo {author} {\bibfnamefont {A.}~\bibnamefont {Dauphin}},\ }\href {https://doi.org/10.1103/PhysRevResearch.4.L032005} {\bibfield  {journal} {\bibinfo  {journal} {Phys. Rev. Res.}\ }\textbf {\bibinfo {volume} {4}},\ \bibinfo {pages} {L032005} (\bibinfo {year} {2022})}\BibitemShut {NoStop}%
\bibitem [{\citenamefont {Fraxanet}\ \emph {et~al.}(2023)\citenamefont {Fraxanet}, \citenamefont {Dauphin}, \citenamefont {Lewenstein}, \citenamefont {Barbiero},\ and\ \citenamefont {González-Cuadra}}]{Fraxanet2023}%
  \BibitemOpen
  \bibfield  {author} {\bibinfo {author} {\bibfnamefont {J.}~\bibnamefont {Fraxanet}}, \bibinfo {author} {\bibfnamefont {A.}~\bibnamefont {Dauphin}}, \bibinfo {author} {\bibfnamefont {M.}~\bibnamefont {Lewenstein}}, \bibinfo {author} {\bibfnamefont {L.}~\bibnamefont {Barbiero}},\ and\ \bibinfo {author} {\bibfnamefont {D.}~\bibnamefont {González-Cuadra}},\ }\href@noop {} {} (\bibinfo {year} {2023}),\ \Eprint {https://arxiv.org/abs/2305.03409} {arXiv:2305.03409 [cond-mat.quant-gas]} \BibitemShut {NoStop}%
\bibitem [{\citenamefont {Gross}\ and\ \citenamefont {Bakr}(2021)}]{Gross2021}%
  \BibitemOpen
  \bibfield  {author} {\bibinfo {author} {\bibfnamefont {C.}~\bibnamefont {Gross}}\ and\ \bibinfo {author} {\bibfnamefont {W.~S.}\ \bibnamefont {Bakr}},\ }\href {https://doi.org/10.1038/s41567-021-01370-5} {\bibfield  {journal} {\bibinfo  {journal} {Nat. Phys.}\ }\textbf {\bibinfo {volume} {17}},\ \bibinfo {pages} {1316} (\bibinfo {year} {2021})}\BibitemShut {NoStop}%
\bibitem [{\citenamefont {Pollmann}\ \emph {et~al.}(2009)\citenamefont {Pollmann}, \citenamefont {Mukerjee}, \citenamefont {Turner},\ and\ \citenamefont {Moore}}]{Pollmann2009}%
  \BibitemOpen
  \bibfield  {author} {\bibinfo {author} {\bibfnamefont {F.}~\bibnamefont {Pollmann}}, \bibinfo {author} {\bibfnamefont {S.}~\bibnamefont {Mukerjee}}, \bibinfo {author} {\bibfnamefont {A.~M.}\ \bibnamefont {Turner}},\ and\ \bibinfo {author} {\bibfnamefont {J.~E.}\ \bibnamefont {Moore}},\ }\href {https://doi.org/10.1103/PhysRevLett.102.255701} {\bibfield  {journal} {\bibinfo  {journal} {Phys. Rev. Lett.}\ }\textbf {\bibinfo {volume} {102}},\ \bibinfo {pages} {255701} (\bibinfo {year} {2009})}\BibitemShut {NoStop}%
\bibitem [{\citenamefont {Grimm}\ \emph {et~al.}(2000)\citenamefont {Grimm}, \citenamefont {Weidemüller},\ and\ \citenamefont {Ovchinnikov}}]{GRIMM200095}%
  \BibitemOpen
  \bibfield  {author} {\bibinfo {author} {\bibfnamefont {R.}~\bibnamefont {Grimm}}, \bibinfo {author} {\bibfnamefont {M.}~\bibnamefont {Weidemüller}},\ and\ \bibinfo {author} {\bibfnamefont {Y.~B.}\ \bibnamefont {Ovchinnikov}}\ }(\bibinfo  {publisher} {Academic Press},\ \bibinfo {year} {2000})\ pp.\ \bibinfo {pages} {95--170}\BibitemShut {NoStop}%
\bibitem [{\citenamefont {Klostermann}(2022)}]{klostermann22}%
  \BibitemOpen
  \bibfield  {author} {\bibinfo {author} {\bibfnamefont {T.~M.}\ \bibnamefont {Klostermann}},\ }\emph {\bibinfo {title} {Construction of a caesium quantum gas microscope.}},\ \href {http://nbn-resolving.de/urn:nbn:de:bvb:19-295213} {Ph.D. thesis},\ \bibinfo  {school} {Ludwig-Maximilians-Universit{\"a}t M{\"u}nchen} (\bibinfo {year} {2022})\BibitemShut {NoStop}%
\bibitem [{\citenamefont {Chin}\ \emph {et~al.}(2004)\citenamefont {Chin}, \citenamefont {Vuleti\ifmmode~\acute{c}\else \'{c}\fi{}}, \citenamefont {Kerman}, \citenamefont {Chu}, \citenamefont {Tiesinga}, \citenamefont {Leo},\ and\ \citenamefont {Williams}}]{Chin2004}%
  \BibitemOpen
  \bibfield  {author} {\bibinfo {author} {\bibfnamefont {C.}~\bibnamefont {Chin}}, \bibinfo {author} {\bibfnamefont {V.}~\bibnamefont {Vuleti\ifmmode~\acute{c}\else \'{c}\fi{}}}, \bibinfo {author} {\bibfnamefont {A.~J.}\ \bibnamefont {Kerman}}, \bibinfo {author} {\bibfnamefont {S.}~\bibnamefont {Chu}}, \bibinfo {author} {\bibfnamefont {E.}~\bibnamefont {Tiesinga}}, \bibinfo {author} {\bibfnamefont {P.~J.}\ \bibnamefont {Leo}},\ and\ \bibinfo {author} {\bibfnamefont {C.~J.}\ \bibnamefont {Williams}},\ }\href {https://doi.org/10.1103/PhysRevA.70.032701} {\bibfield  {journal} {\bibinfo  {journal} {Phys. Rev. A}\ }\textbf {\bibinfo {volume} {70}},\ \bibinfo {pages} {032701} (\bibinfo {year} {2004})}\BibitemShut {NoStop}%
\bibitem [{\citenamefont {Frye}\ \emph {et~al.}(2019{\natexlab{b}})\citenamefont {Frye}, \citenamefont {Yang},\ and\ \citenamefont {Hutson}}]{Frye2019}%
  \BibitemOpen
  \bibfield  {author} {\bibinfo {author} {\bibfnamefont {M.~D.}\ \bibnamefont {Frye}}, \bibinfo {author} {\bibfnamefont {B.~C.}\ \bibnamefont {Yang}},\ and\ \bibinfo {author} {\bibfnamefont {J.~M.}\ \bibnamefont {Hutson}},\ }\href {https://doi.org/10.1103/PhysRevA.100.022702} {\bibfield  {journal} {\bibinfo  {journal} {Phys. Rev. A}\ }\textbf {\bibinfo {volume} {100}},\ \bibinfo {pages} {022702} (\bibinfo {year} {2019}{\natexlab{b}})}\BibitemShut {NoStop}%
\bibitem [{\citenamefont {Impertro}\ \emph {et~al.}(2023)\citenamefont {Impertro}, \citenamefont {Wienand}, \citenamefont {H{\"a}fele}, \citenamefont {von Raven}, \citenamefont {Hubele}, \citenamefont {Klostermann}, \citenamefont {Cabrera}, \citenamefont {Bloch},\ and\ \citenamefont {Aidelsburger}}]{Impertro2022}%
  \BibitemOpen
  \bibfield  {author} {\bibinfo {author} {\bibfnamefont {A.}~\bibnamefont {Impertro}}, \bibinfo {author} {\bibfnamefont {J.~F.}\ \bibnamefont {Wienand}}, \bibinfo {author} {\bibfnamefont {S.}~\bibnamefont {H{\"a}fele}}, \bibinfo {author} {\bibfnamefont {H.}~\bibnamefont {von Raven}}, \bibinfo {author} {\bibfnamefont {S.}~\bibnamefont {Hubele}}, \bibinfo {author} {\bibfnamefont {T.}~\bibnamefont {Klostermann}}, \bibinfo {author} {\bibfnamefont {C.~R.}\ \bibnamefont {Cabrera}}, \bibinfo {author} {\bibfnamefont {I.}~\bibnamefont {Bloch}},\ and\ \bibinfo {author} {\bibfnamefont {M.}~\bibnamefont {Aidelsburger}},\ }\href {https://doi.org/10.1038/s42005-023-01287-w} {\bibfield  {journal} {\bibinfo  {journal} {Communications Physics}\ }\textbf {\bibinfo {volume} {6}},\ \bibinfo {pages} {166} (\bibinfo {year} {2023})}\BibitemShut {NoStop}%
\bibitem [{\citenamefont {Anisimovas}\ \emph {et~al.}(2016{\natexlab{b}})\citenamefont {Anisimovas}, \citenamefont {Ra\ifmmode \check{c}\else \v{c}\fi{}i\ifmmode~\bar{u}\else \={u}\fi{}nas}, \citenamefont {Str\"ater}, \citenamefont {Eckardt}, \citenamefont {Spielman},\ and\ \citenamefont {Juzeli\ifmmode~\bar{u}\else \={u}\fi{}nas}}]{anismovas2016}%
  \BibitemOpen
  \bibfield  {author} {\bibinfo {author} {\bibfnamefont {E.}~\bibnamefont {Anisimovas}}, \bibinfo {author} {\bibfnamefont {M.}~\bibnamefont {Ra\ifmmode \check{c}\else \v{c}\fi{}i\ifmmode~\bar{u}\else \={u}\fi{}nas}}, \bibinfo {author} {\bibfnamefont {C.}~\bibnamefont {Str\"ater}}, \bibinfo {author} {\bibfnamefont {A.}~\bibnamefont {Eckardt}}, \bibinfo {author} {\bibfnamefont {I.~B.}\ \bibnamefont {Spielman}},\ and\ \bibinfo {author} {\bibfnamefont {G.}~\bibnamefont {Juzeli\ifmmode~\bar{u}\else \={u}\fi{}nas}},\ }\href {https://doi.org/10.1103/PhysRevA.94.063632} {\bibfield  {journal} {\bibinfo  {journal} {Phys. Rev. A}\ }\textbf {\bibinfo {volume} {94}},\ \bibinfo {pages} {063632} (\bibinfo {year} {2016}{\natexlab{b}})}\BibitemShut {NoStop}%
\bibitem [{\citenamefont {Ke\ss{}ler}\ and\ \citenamefont {Marquardt}(2014)}]{kessler2014}%
  \BibitemOpen
  \bibfield  {author} {\bibinfo {author} {\bibfnamefont {S.}~\bibnamefont {Ke\ss{}ler}}\ and\ \bibinfo {author} {\bibfnamefont {F.}~\bibnamefont {Marquardt}},\ }\href {https://doi.org/10.1103/PhysRevA.89.061601} {\bibfield  {journal} {\bibinfo  {journal} {Phys. Rev. A}\ }\textbf {\bibinfo {volume} {89}},\ \bibinfo {pages} {061601} (\bibinfo {year} {2014})}\BibitemShut {NoStop}%
\bibitem [{\citenamefont {Atala}\ \emph {et~al.}(2014{\natexlab{a}})\citenamefont {Atala}, \citenamefont {Aidelsburger}, \citenamefont {Lohse}, \citenamefont {Barreiro}, \citenamefont {Paredes},\ and\ \citenamefont {Bloch}}]{atala14a}%
  \BibitemOpen
  \bibfield  {author} {\bibinfo {author} {\bibfnamefont {M.}~\bibnamefont {Atala}}, \bibinfo {author} {\bibfnamefont {M.}~\bibnamefont {Aidelsburger}}, \bibinfo {author} {\bibfnamefont {M.}~\bibnamefont {Lohse}}, \bibinfo {author} {\bibfnamefont {J.~T.}\ \bibnamefont {Barreiro}}, \bibinfo {author} {\bibfnamefont {B.}~\bibnamefont {Paredes}},\ and\ \bibinfo {author} {\bibfnamefont {I.}~\bibnamefont {Bloch}},\ }\href {https://doi.org/10.1038/nphys2998} {\bibfield  {journal} {\bibinfo  {journal} {Nat. Phys.}\ }\textbf {\bibinfo {volume} {10}},\ \bibinfo {pages} {588} (\bibinfo {year} {2014}{\natexlab{a}})}\BibitemShut {NoStop}%
\bibitem [{\citenamefont {Murthy}\ \emph {et~al.}(2014)\citenamefont {Murthy}, \citenamefont {Kedar}, \citenamefont {Lompe}, \citenamefont {Neidig}, \citenamefont {Ries}, \citenamefont {Wenz}, \citenamefont {Z\"urn},\ and\ \citenamefont {Jochim}}]{Murthy2014}%
  \BibitemOpen
  \bibfield  {author} {\bibinfo {author} {\bibfnamefont {P.~A.}\ \bibnamefont {Murthy}}, \bibinfo {author} {\bibfnamefont {D.}~\bibnamefont {Kedar}}, \bibinfo {author} {\bibfnamefont {T.}~\bibnamefont {Lompe}}, \bibinfo {author} {\bibfnamefont {M.}~\bibnamefont {Neidig}}, \bibinfo {author} {\bibfnamefont {M.~G.}\ \bibnamefont {Ries}}, \bibinfo {author} {\bibfnamefont {A.~N.}\ \bibnamefont {Wenz}}, \bibinfo {author} {\bibfnamefont {G.}~\bibnamefont {Z\"urn}},\ and\ \bibinfo {author} {\bibfnamefont {S.}~\bibnamefont {Jochim}},\ }\href {https://doi.org/10.1103/PhysRevA.90.043611} {\bibfield  {journal} {\bibinfo  {journal} {Phys. Rev. A}\ }\textbf {\bibinfo {volume} {90}},\ \bibinfo {pages} {043611} (\bibinfo {year} {2014})}\BibitemShut {NoStop}%
\bibitem [{\citenamefont {Hueck}\ \emph {et~al.}(2018)\citenamefont {Hueck}, \citenamefont {Luick}, \citenamefont {Sobirey}, \citenamefont {Siegl}, \citenamefont {Lompe},\ and\ \citenamefont {Moritz}}]{Henning2018}%
  \BibitemOpen
  \bibfield  {author} {\bibinfo {author} {\bibfnamefont {K.}~\bibnamefont {Hueck}}, \bibinfo {author} {\bibfnamefont {N.}~\bibnamefont {Luick}}, \bibinfo {author} {\bibfnamefont {L.}~\bibnamefont {Sobirey}}, \bibinfo {author} {\bibfnamefont {J.}~\bibnamefont {Siegl}}, \bibinfo {author} {\bibfnamefont {T.}~\bibnamefont {Lompe}},\ and\ \bibinfo {author} {\bibfnamefont {H.}~\bibnamefont {Moritz}},\ }\href {https://doi.org/10.1103/PhysRevLett.120.060402} {\bibfield  {journal} {\bibinfo  {journal} {Phys. Rev. Lett.}\ }\textbf {\bibinfo {volume} {120}},\ \bibinfo {pages} {060402} (\bibinfo {year} {2018})}\BibitemShut {NoStop}%
\bibitem [{\citenamefont {Bohrdt}\ \emph {et~al.}(2018)\citenamefont {Bohrdt}, \citenamefont {Greif}, \citenamefont {Demler}, \citenamefont {Knap},\ and\ \citenamefont {Grusdt}}]{Bohrdt2018}%
  \BibitemOpen
  \bibfield  {author} {\bibinfo {author} {\bibfnamefont {A.}~\bibnamefont {Bohrdt}}, \bibinfo {author} {\bibfnamefont {D.}~\bibnamefont {Greif}}, \bibinfo {author} {\bibfnamefont {E.}~\bibnamefont {Demler}}, \bibinfo {author} {\bibfnamefont {M.}~\bibnamefont {Knap}},\ and\ \bibinfo {author} {\bibfnamefont {F.}~\bibnamefont {Grusdt}},\ }\href {https://doi.org/10.1103/PhysRevB.97.125117} {\bibfield  {journal} {\bibinfo  {journal} {Phys. Rev. B}\ }\textbf {\bibinfo {volume} {97}},\ \bibinfo {pages} {125117} (\bibinfo {year} {2018})}\BibitemShut {NoStop}%
\bibitem [{\citenamefont {Brown}\ \emph {et~al.}(2020)\citenamefont {Brown}, \citenamefont {Guardado-Sanchez}, \citenamefont {Spar}, \citenamefont {Huang}, \citenamefont {Devereaux},\ and\ \citenamefont {Bakr}}]{Brown2020}%
  \BibitemOpen
  \bibfield  {author} {\bibinfo {author} {\bibfnamefont {P.~T.}\ \bibnamefont {Brown}}, \bibinfo {author} {\bibfnamefont {E.}~\bibnamefont {Guardado-Sanchez}}, \bibinfo {author} {\bibfnamefont {B.~M.}\ \bibnamefont {Spar}}, \bibinfo {author} {\bibfnamefont {E.~W.}\ \bibnamefont {Huang}}, \bibinfo {author} {\bibfnamefont {T.~P.}\ \bibnamefont {Devereaux}},\ and\ \bibinfo {author} {\bibfnamefont {W.~S.}\ \bibnamefont {Bakr}},\ }\href {https://doi.org/10.1038/s41567-019-0696-0} {\bibfield  {journal} {\bibinfo  {journal} {Nat. Phys.}\ }\textbf {\bibinfo {volume} {16}},\ \bibinfo {pages} {26} (\bibinfo {year} {2020})}\BibitemShut {NoStop}%
\bibitem [{\citenamefont {Atala}\ \emph {et~al.}(2014{\natexlab{b}})\citenamefont {Atala}, \citenamefont {Aidelsburger}, \citenamefont {Lohse}, \citenamefont {Barreiro}, \citenamefont {Paredes},\ and\ \citenamefont {Bloch}}]{Atala2014}%
  \BibitemOpen
  \bibfield  {author} {\bibinfo {author} {\bibfnamefont {M.}~\bibnamefont {Atala}}, \bibinfo {author} {\bibfnamefont {M.}~\bibnamefont {Aidelsburger}}, \bibinfo {author} {\bibfnamefont {M.}~\bibnamefont {Lohse}}, \bibinfo {author} {\bibfnamefont {J.~T.}\ \bibnamefont {Barreiro}}, \bibinfo {author} {\bibfnamefont {B.}~\bibnamefont {Paredes}},\ and\ \bibinfo {author} {\bibfnamefont {I.}~\bibnamefont {Bloch}},\ }\href {https://doi.org/10.1038/nphys2998} {\bibfield  {journal} {\bibinfo  {journal} {Nature Physics}\ }\textbf {\bibinfo {volume} {10}},\ \bibinfo {pages} {588} (\bibinfo {year} {2014}{\natexlab{b}})}\BibitemShut {NoStop}%
\end{thebibliography}%


\begin{thebibliography}{22}%
\makeatletter
\providecommand \@ifxundefined [1]{%
 \@ifx{#1\undefined}
}%
\providecommand \@ifnum [1]{%
 \ifnum #1\expandafter \@firstoftwo
 \else \expandafter \@secondoftwo
 \fi
}%
\providecommand \@ifx [1]{%
 \ifx #1\expandafter \@firstoftwo
 \else \expandafter \@secondoftwo
 \fi
}%
\providecommand \natexlab [1]{#1}%
\providecommand \enquote  [1]{``#1''}%
\providecommand \bibnamefont  [1]{#1}%
\providecommand \bibfnamefont [1]{#1}%
\providecommand \citenamefont [1]{#1}%
\providecommand \href@noop [0]{\@secondoftwo}%
\providecommand \href [0]{\begingroup \@sanitize@url \@href}%
\providecommand \@href[1]{\@@startlink{#1}\@@href}%
\providecommand \@@href[1]{\endgroup#1\@@endlink}%
\providecommand \@sanitize@url [0]{\catcode `\\12\catcode `\$12\catcode `\&12\catcode `\#12\catcode `\^12\catcode `\_12\catcode `\%12\relax}%
\providecommand \@@startlink[1]{}%
\providecommand \@@endlink[0]{}%
\providecommand \url  [0]{\begingroup\@sanitize@url \@url }%
\providecommand \@url [1]{\endgroup\@href {#1}{\urlprefix }}%
\providecommand \urlprefix  [0]{URL }%
\providecommand \Eprint [0]{\href }%
\providecommand \doibase [0]{https://doi.org/}%
\providecommand \selectlanguage [0]{\@gobble}%
\providecommand \bibinfo  [0]{\@secondoftwo}%
\providecommand \bibfield  [0]{\@secondoftwo}%
\providecommand \translation [1]{[#1]}%
\providecommand \BibitemOpen [0]{}%
\providecommand \bibitemStop [0]{}%
\providecommand \bibitemNoStop [0]{.\EOS\space}%
\providecommand \EOS [0]{\spacefactor3000\relax}%
\providecommand \BibitemShut  [1]{\csname bibitem#1\endcsname}%
\let\auto@bib@innerbib\@empty
\bibitem [{\citenamefont {Sato}\ \emph {et~al.}(2011)\citenamefont {Sato}, \citenamefont {Furukawa}, \citenamefont {Onoda},\ and\ \citenamefont {Furusaki}}]{SATO2011}%
  \BibitemOpen
  \bibfield  {author} {\bibinfo {author} {\bibfnamefont {M.}~\bibnamefont {Sato}}, \bibinfo {author} {\bibfnamefont {S.}~\bibnamefont {Furukawa}}, \bibinfo {author} {\bibfnamefont {S.}~\bibnamefont {Onoda}},\ and\ \bibinfo {author} {\bibfnamefont {A.}~\bibnamefont {Furusaki}},\ }\href {https://doi.org/10.1142/S0217984911026607} {\bibfield  {journal} {\bibinfo  {journal} {Mod. Phys. Lett. B}\ }\textbf {\bibinfo {volume} {25}},\ \bibinfo {pages} {901} (\bibinfo {year} {2011})}\BibitemShut {NoStop}%
\bibitem [{\citenamefont {Furukawa}\ \emph {et~al.}(2012)\citenamefont {Furukawa}, \citenamefont {Sato}, \citenamefont {Onoda},\ and\ \citenamefont {Furusaki}}]{Furukawa2012}%
  \BibitemOpen
  \bibfield  {author} {\bibinfo {author} {\bibfnamefont {S.}~\bibnamefont {Furukawa}}, \bibinfo {author} {\bibfnamefont {M.}~\bibnamefont {Sato}}, \bibinfo {author} {\bibfnamefont {S.}~\bibnamefont {Onoda}},\ and\ \bibinfo {author} {\bibfnamefont {A.}~\bibnamefont {Furusaki}},\ }\href {https://doi.org/10.1103/PhysRevB.86.094417} {\bibfield  {journal} {\bibinfo  {journal} {Phys. Rev. B}\ }\textbf {\bibinfo {volume} {86}},\ \bibinfo {pages} {094417} (\bibinfo {year} {2012})}\BibitemShut {NoStop}%
\bibitem [{\citenamefont {Drechsler}\ \emph {et~al.}(2007)\citenamefont {Drechsler}, \citenamefont {Volkova}, \citenamefont {Vasiliev}, \citenamefont {Tristan}, \citenamefont {Richter}, \citenamefont {Schmitt}, \citenamefont {Rosner}, \citenamefont {M\'alek}, \citenamefont {Klingeler}, \citenamefont {Zvyagin},\ and\ \citenamefont {B\"uchner}}]{Drechsler2007}%
  \BibitemOpen
  \bibfield  {author} {\bibinfo {author} {\bibfnamefont {S.-L.}\ \bibnamefont {Drechsler}}, \bibinfo {author} {\bibfnamefont {O.}~\bibnamefont {Volkova}}, \bibinfo {author} {\bibfnamefont {A.~N.}\ \bibnamefont {Vasiliev}}, \bibinfo {author} {\bibfnamefont {N.}~\bibnamefont {Tristan}}, \bibinfo {author} {\bibfnamefont {J.}~\bibnamefont {Richter}}, \bibinfo {author} {\bibfnamefont {M.}~\bibnamefont {Schmitt}}, \bibinfo {author} {\bibfnamefont {H.}~\bibnamefont {Rosner}}, \bibinfo {author} {\bibfnamefont {J.}~\bibnamefont {M\'alek}}, \bibinfo {author} {\bibfnamefont {R.}~\bibnamefont {Klingeler}}, \bibinfo {author} {\bibfnamefont {A.~A.}\ \bibnamefont {Zvyagin}},\ and\ \bibinfo {author} {\bibfnamefont {B.}~\bibnamefont {B\"uchner}},\ }\href {https://doi.org/10.1103/PhysRevLett.98.077202} {\bibfield  {journal} {\bibinfo  {journal} {Phys. Rev. Lett.}\ }\textbf {\bibinfo {volume} {98}},\ \bibinfo {pages} {077202} (\bibinfo {year} {2007})}\BibitemShut {NoStop}%
\bibitem [{\citenamefont {Wolter}\ \emph {et~al.}(2012)\citenamefont {Wolter}, \citenamefont {Lipps}, \citenamefont {Sch\"apers}, \citenamefont {Drechsler}, \citenamefont {Nishimoto}, \citenamefont {Vogel}, \citenamefont {Kataev}, \citenamefont {B\"uchner}, \citenamefont {Rosner}, \citenamefont {Schmitt}, \citenamefont {Uhlarz}, \citenamefont {Skourski}, \citenamefont {Wosnitza}, \citenamefont {S\"ullow},\ and\ \citenamefont {Rule}}]{Wolter2012}%
  \BibitemOpen
  \bibfield  {author} {\bibinfo {author} {\bibfnamefont {A.~U.~B.}\ \bibnamefont {Wolter}}, \bibinfo {author} {\bibfnamefont {F.}~\bibnamefont {Lipps}}, \bibinfo {author} {\bibfnamefont {M.}~\bibnamefont {Sch\"apers}}, \bibinfo {author} {\bibfnamefont {S.-L.}\ \bibnamefont {Drechsler}}, \bibinfo {author} {\bibfnamefont {S.}~\bibnamefont {Nishimoto}}, \bibinfo {author} {\bibfnamefont {R.}~\bibnamefont {Vogel}}, \bibinfo {author} {\bibfnamefont {V.}~\bibnamefont {Kataev}}, \bibinfo {author} {\bibfnamefont {B.}~\bibnamefont {B\"uchner}}, \bibinfo {author} {\bibfnamefont {H.}~\bibnamefont {Rosner}}, \bibinfo {author} {\bibfnamefont {M.}~\bibnamefont {Schmitt}}, \bibinfo {author} {\bibfnamefont {M.}~\bibnamefont {Uhlarz}}, \bibinfo {author} {\bibfnamefont {Y.}~\bibnamefont {Skourski}}, \bibinfo {author} {\bibfnamefont {J.}~\bibnamefont {Wosnitza}}, \bibinfo {author} {\bibfnamefont {S.}~\bibnamefont {S\"ullow}},\ and\ \bibinfo {author} {\bibfnamefont {K.~C.}\ \bibnamefont {Rule}},\ }\href
  {https://doi.org/10.1103/PhysRevB.85.014407} {\bibfield  {journal} {\bibinfo  {journal} {Phys. Rev. B}\ }\textbf {\bibinfo {volume} {85}},\ \bibinfo {pages} {014407} (\bibinfo {year} {2012})}\BibitemShut {NoStop}%
\bibitem [{\citenamefont {Orlova}\ \emph {et~al.}(2017)\citenamefont {Orlova}, \citenamefont {Green}, \citenamefont {Law}, \citenamefont {Gorbunov}, \citenamefont {Chanda}, \citenamefont {Kr\"amer}, \citenamefont {Horvati\ifmmode~\acute{c}\else \'{c}\fi{}}, \citenamefont {Kremer}, \citenamefont {Wosnitza},\ and\ \citenamefont {Rikken}}]{Orlova2017}%
  \BibitemOpen
  \bibfield  {author} {\bibinfo {author} {\bibfnamefont {A.}~\bibnamefont {Orlova}}, \bibinfo {author} {\bibfnamefont {E.~L.}\ \bibnamefont {Green}}, \bibinfo {author} {\bibfnamefont {J.~M.}\ \bibnamefont {Law}}, \bibinfo {author} {\bibfnamefont {D.~I.}\ \bibnamefont {Gorbunov}}, \bibinfo {author} {\bibfnamefont {G.}~\bibnamefont {Chanda}}, \bibinfo {author} {\bibfnamefont {S.}~\bibnamefont {Kr\"amer}}, \bibinfo {author} {\bibfnamefont {M.}~\bibnamefont {Horvati\ifmmode~\acute{c}\else \'{c}\fi{}}}, \bibinfo {author} {\bibfnamefont {R.~K.}\ \bibnamefont {Kremer}}, \bibinfo {author} {\bibfnamefont {J.}~\bibnamefont {Wosnitza}},\ and\ \bibinfo {author} {\bibfnamefont {G.~L. J.~A.}\ \bibnamefont {Rikken}},\ }\href {https://doi.org/10.1103/PhysRevLett.118.247201} {\bibfield  {journal} {\bibinfo  {journal} {Phys. Rev. Lett.}\ }\textbf {\bibinfo {volume} {118}},\ \bibinfo {pages} {247201} (\bibinfo {year} {2017})}\BibitemShut {NoStop}%
\bibitem [{\citenamefont {Grams}\ \emph {et~al.}(2022)\citenamefont {Grams}, \citenamefont {Brüning}, \citenamefont {Kopatz}, \citenamefont {Lorenz}, \citenamefont {Becker}, \citenamefont {Bohatý},\ and\ \citenamefont {Hemberger}}]{Grams2022}%
  \BibitemOpen
  \bibfield  {author} {\bibinfo {author} {\bibfnamefont {C.~P.}\ \bibnamefont {Grams}}, \bibinfo {author} {\bibfnamefont {D.}~\bibnamefont {Brüning}}, \bibinfo {author} {\bibfnamefont {S.}~\bibnamefont {Kopatz}}, \bibinfo {author} {\bibfnamefont {T.}~\bibnamefont {Lorenz}}, \bibinfo {author} {\bibfnamefont {P.}~\bibnamefont {Becker}}, \bibinfo {author} {\bibfnamefont {L.}~\bibnamefont {Bohatý}},\ and\ \bibinfo {author} {\bibfnamefont {J.}~\bibnamefont {Hemberger}},\ }\href {https://doi.org/10.1038/s42005-022-00811-8} {\bibfield  {journal} {\bibinfo  {journal} {Commun. Phys.}\ }\textbf {\bibinfo {volume} {5}} (\bibinfo {year} {2022})}\BibitemShut {NoStop}%
\bibitem [{\citenamefont {Grimm}\ \emph {et~al.}(2000)\citenamefont {Grimm}, \citenamefont {Weidemüller},\ and\ \citenamefont {Ovchinnikov}}]{GRIMM200095}%
  \BibitemOpen
  \bibfield  {author} {\bibinfo {author} {\bibfnamefont {R.}~\bibnamefont {Grimm}}, \bibinfo {author} {\bibfnamefont {M.}~\bibnamefont {Weidemüller}},\ and\ \bibinfo {author} {\bibfnamefont {Y.~B.}\ \bibnamefont {Ovchinnikov}}\ }(\bibinfo  {publisher} {Academic Press},\ \bibinfo {year} {2000})\ pp.\ \bibinfo {pages} {95--170}\BibitemShut {NoStop}%
\bibitem [{\citenamefont {Klostermann}(2022)}]{klostermann22}%
  \BibitemOpen
  \bibfield  {author} {\bibinfo {author} {\bibfnamefont {T.~M.}\ \bibnamefont {Klostermann}},\ }\emph {\bibinfo {title} {Construction of a caesium quantum gas microscope.}},\ \href {http://nbn-resolving.de/urn:nbn:de:bvb:19-295213} {Ph.D. thesis},\ \bibinfo  {school} {Ludwig-Maximilians-Universit{\"a}t M{\"u}nchen} (\bibinfo {year} {2022})\BibitemShut {NoStop}%
\bibitem [{\citenamefont {Chin}\ \emph {et~al.}(2004)\citenamefont {Chin}, \citenamefont {Vuleti\ifmmode~\acute{c}\else \'{c}\fi{}}, \citenamefont {Kerman}, \citenamefont {Chu}, \citenamefont {Tiesinga}, \citenamefont {Leo},\ and\ \citenamefont {Williams}}]{Chin2004}%
  \BibitemOpen
  \bibfield  {author} {\bibinfo {author} {\bibfnamefont {C.}~\bibnamefont {Chin}}, \bibinfo {author} {\bibfnamefont {V.}~\bibnamefont {Vuleti\ifmmode~\acute{c}\else \'{c}\fi{}}}, \bibinfo {author} {\bibfnamefont {A.~J.}\ \bibnamefont {Kerman}}, \bibinfo {author} {\bibfnamefont {S.}~\bibnamefont {Chu}}, \bibinfo {author} {\bibfnamefont {E.}~\bibnamefont {Tiesinga}}, \bibinfo {author} {\bibfnamefont {P.~J.}\ \bibnamefont {Leo}},\ and\ \bibinfo {author} {\bibfnamefont {C.~J.}\ \bibnamefont {Williams}},\ }\href {https://doi.org/10.1103/PhysRevA.70.032701} {\bibfield  {journal} {\bibinfo  {journal} {Phys. Rev. A}\ }\textbf {\bibinfo {volume} {70}},\ \bibinfo {pages} {032701} (\bibinfo {year} {2004})}\BibitemShut {NoStop}%
\bibitem [{\citenamefont {Frye}\ \emph {et~al.}(2019)\citenamefont {Frye}, \citenamefont {Yang},\ and\ \citenamefont {Hutson}}]{Frye2019}%
  \BibitemOpen
  \bibfield  {author} {\bibinfo {author} {\bibfnamefont {M.~D.}\ \bibnamefont {Frye}}, \bibinfo {author} {\bibfnamefont {B.~C.}\ \bibnamefont {Yang}},\ and\ \bibinfo {author} {\bibfnamefont {J.~M.}\ \bibnamefont {Hutson}},\ }\href {https://doi.org/10.1103/PhysRevA.100.022702} {\bibfield  {journal} {\bibinfo  {journal} {Phys. Rev. A}\ }\textbf {\bibinfo {volume} {100}},\ \bibinfo {pages} {022702} (\bibinfo {year} {2019})}\BibitemShut {NoStop}%
\bibitem [{\citenamefont {Impertro}\ \emph {et~al.}(2023)\citenamefont {Impertro}, \citenamefont {Wienand}, \citenamefont {H{\"a}fele}, \citenamefont {von Raven}, \citenamefont {Hubele}, \citenamefont {Klostermann}, \citenamefont {Cabrera}, \citenamefont {Bloch},\ and\ \citenamefont {Aidelsburger}}]{Impertro2022}%
  \BibitemOpen
  \bibfield  {author} {\bibinfo {author} {\bibfnamefont {A.}~\bibnamefont {Impertro}}, \bibinfo {author} {\bibfnamefont {J.~F.}\ \bibnamefont {Wienand}}, \bibinfo {author} {\bibfnamefont {S.}~\bibnamefont {H{\"a}fele}}, \bibinfo {author} {\bibfnamefont {H.}~\bibnamefont {von Raven}}, \bibinfo {author} {\bibfnamefont {S.}~\bibnamefont {Hubele}}, \bibinfo {author} {\bibfnamefont {T.}~\bibnamefont {Klostermann}}, \bibinfo {author} {\bibfnamefont {C.~R.}\ \bibnamefont {Cabrera}}, \bibinfo {author} {\bibfnamefont {I.}~\bibnamefont {Bloch}},\ and\ \bibinfo {author} {\bibfnamefont {M.}~\bibnamefont {Aidelsburger}},\ }\href {https://doi.org/10.1038/s42005-023-01287-w} {\bibfield  {journal} {\bibinfo  {journal} {Communications Physics}\ }\textbf {\bibinfo {volume} {6}},\ \bibinfo {pages} {166} (\bibinfo {year} {2023})}\BibitemShut {NoStop}%
\bibitem [{\citenamefont {Anisimovas}\ \emph {et~al.}(2016)\citenamefont {Anisimovas}, \citenamefont {Ra\ifmmode \check{c}\else \v{c}\fi{}i\ifmmode~\bar{u}\else \={u}\fi{}nas}, \citenamefont {Str\"ater}, \citenamefont {Eckardt}, \citenamefont {Spielman},\ and\ \citenamefont {Juzeli\ifmmode~\bar{u}\else \={u}\fi{}nas}}]{anismovas2016}%
  \BibitemOpen
  \bibfield  {author} {\bibinfo {author} {\bibfnamefont {E.}~\bibnamefont {Anisimovas}}, \bibinfo {author} {\bibfnamefont {M.}~\bibnamefont {Ra\ifmmode \check{c}\else \v{c}\fi{}i\ifmmode~\bar{u}\else \={u}\fi{}nas}}, \bibinfo {author} {\bibfnamefont {C.}~\bibnamefont {Str\"ater}}, \bibinfo {author} {\bibfnamefont {A.}~\bibnamefont {Eckardt}}, \bibinfo {author} {\bibfnamefont {I.~B.}\ \bibnamefont {Spielman}},\ and\ \bibinfo {author} {\bibfnamefont {G.}~\bibnamefont {Juzeli\ifmmode~\bar{u}\else \={u}\fi{}nas}},\ }\href {https://doi.org/10.1103/PhysRevA.94.063632} {\bibfield  {journal} {\bibinfo  {journal} {Phys. Rev. A}\ }\textbf {\bibinfo {volume} {94}},\ \bibinfo {pages} {063632} (\bibinfo {year} {2016})}\BibitemShut {NoStop}%
\bibitem [{\citenamefont {Ke\ss{}ler}\ and\ \citenamefont {Marquardt}(2014)}]{kessler2014}%
  \BibitemOpen
  \bibfield  {author} {\bibinfo {author} {\bibfnamefont {S.}~\bibnamefont {Ke\ss{}ler}}\ and\ \bibinfo {author} {\bibfnamefont {F.}~\bibnamefont {Marquardt}},\ }\href {https://doi.org/10.1103/PhysRevA.89.061601} {\bibfield  {journal} {\bibinfo  {journal} {Phys. Rev. A}\ }\textbf {\bibinfo {volume} {89}},\ \bibinfo {pages} {061601} (\bibinfo {year} {2014})}\BibitemShut {NoStop}%
\bibitem [{\citenamefont {Atala}\ \emph {et~al.}(2014{\natexlab{a}})\citenamefont {Atala}, \citenamefont {Aidelsburger}, \citenamefont {Lohse}, \citenamefont {Barreiro}, \citenamefont {Paredes},\ and\ \citenamefont {Bloch}}]{atala14a}%
  \BibitemOpen
  \bibfield  {author} {\bibinfo {author} {\bibfnamefont {M.}~\bibnamefont {Atala}}, \bibinfo {author} {\bibfnamefont {M.}~\bibnamefont {Aidelsburger}}, \bibinfo {author} {\bibfnamefont {M.}~\bibnamefont {Lohse}}, \bibinfo {author} {\bibfnamefont {J.~T.}\ \bibnamefont {Barreiro}}, \bibinfo {author} {\bibfnamefont {B.}~\bibnamefont {Paredes}},\ and\ \bibinfo {author} {\bibfnamefont {I.}~\bibnamefont {Bloch}},\ }\href {https://doi.org/10.1038/nphys2998} {\bibfield  {journal} {\bibinfo  {journal} {Nat. Phys.}\ }\textbf {\bibinfo {volume} {10}},\ \bibinfo {pages} {588} (\bibinfo {year} {2014}{\natexlab{a}})}\BibitemShut {NoStop}%
\bibitem [{\citenamefont {Juli\`a-Farr\'e}\ \emph {et~al.}(2022)\citenamefont {Juli\`a-Farr\'e}, \citenamefont {Gonz\'alez-Cuadra}, \citenamefont {Patscheider}, \citenamefont {Mark}, \citenamefont {Ferlaino}, \citenamefont {Lewenstein}, \citenamefont {Barbiero},\ and\ \citenamefont {Dauphin}}]{Julia2022}%
  \BibitemOpen
  \bibfield  {author} {\bibinfo {author} {\bibfnamefont {S.}~\bibnamefont {Juli\`a-Farr\'e}}, \bibinfo {author} {\bibfnamefont {D.}~\bibnamefont {Gonz\'alez-Cuadra}}, \bibinfo {author} {\bibfnamefont {A.}~\bibnamefont {Patscheider}}, \bibinfo {author} {\bibfnamefont {M.~J.}\ \bibnamefont {Mark}}, \bibinfo {author} {\bibfnamefont {F.}~\bibnamefont {Ferlaino}}, \bibinfo {author} {\bibfnamefont {M.}~\bibnamefont {Lewenstein}}, \bibinfo {author} {\bibfnamefont {L.}~\bibnamefont {Barbiero}},\ and\ \bibinfo {author} {\bibfnamefont {A.}~\bibnamefont {Dauphin}},\ }\href {https://doi.org/10.1103/PhysRevResearch.4.L032005} {\bibfield  {journal} {\bibinfo  {journal} {Phys. Rev. Res.}\ }\textbf {\bibinfo {volume} {4}},\ \bibinfo {pages} {L032005} (\bibinfo {year} {2022})}\BibitemShut {NoStop}%
\bibitem [{\citenamefont {Murthy}\ \emph {et~al.}(2014)\citenamefont {Murthy}, \citenamefont {Kedar}, \citenamefont {Lompe}, \citenamefont {Neidig}, \citenamefont {Ries}, \citenamefont {Wenz}, \citenamefont {Z\"urn},\ and\ \citenamefont {Jochim}}]{Murthy2014}%
  \BibitemOpen
  \bibfield  {author} {\bibinfo {author} {\bibfnamefont {P.~A.}\ \bibnamefont {Murthy}}, \bibinfo {author} {\bibfnamefont {D.}~\bibnamefont {Kedar}}, \bibinfo {author} {\bibfnamefont {T.}~\bibnamefont {Lompe}}, \bibinfo {author} {\bibfnamefont {M.}~\bibnamefont {Neidig}}, \bibinfo {author} {\bibfnamefont {M.~G.}\ \bibnamefont {Ries}}, \bibinfo {author} {\bibfnamefont {A.~N.}\ \bibnamefont {Wenz}}, \bibinfo {author} {\bibfnamefont {G.}~\bibnamefont {Z\"urn}},\ and\ \bibinfo {author} {\bibfnamefont {S.}~\bibnamefont {Jochim}},\ }\href {https://doi.org/10.1103/PhysRevA.90.043611} {\bibfield  {journal} {\bibinfo  {journal} {Phys. Rev. A}\ }\textbf {\bibinfo {volume} {90}},\ \bibinfo {pages} {043611} (\bibinfo {year} {2014})}\BibitemShut {NoStop}%
\bibitem [{\citenamefont {Hueck}\ \emph {et~al.}(2018)\citenamefont {Hueck}, \citenamefont {Luick}, \citenamefont {Sobirey}, \citenamefont {Siegl}, \citenamefont {Lompe},\ and\ \citenamefont {Moritz}}]{Henning2018}%
  \BibitemOpen
  \bibfield  {author} {\bibinfo {author} {\bibfnamefont {K.}~\bibnamefont {Hueck}}, \bibinfo {author} {\bibfnamefont {N.}~\bibnamefont {Luick}}, \bibinfo {author} {\bibfnamefont {L.}~\bibnamefont {Sobirey}}, \bibinfo {author} {\bibfnamefont {J.}~\bibnamefont {Siegl}}, \bibinfo {author} {\bibfnamefont {T.}~\bibnamefont {Lompe}},\ and\ \bibinfo {author} {\bibfnamefont {H.}~\bibnamefont {Moritz}},\ }\href {https://doi.org/10.1103/PhysRevLett.120.060402} {\bibfield  {journal} {\bibinfo  {journal} {Phys. Rev. Lett.}\ }\textbf {\bibinfo {volume} {120}},\ \bibinfo {pages} {060402} (\bibinfo {year} {2018})}\BibitemShut {NoStop}%
\bibitem [{\citenamefont {Bohrdt}\ \emph {et~al.}(2018)\citenamefont {Bohrdt}, \citenamefont {Greif}, \citenamefont {Demler}, \citenamefont {Knap},\ and\ \citenamefont {Grusdt}}]{Bohrdt2018}%
  \BibitemOpen
  \bibfield  {author} {\bibinfo {author} {\bibfnamefont {A.}~\bibnamefont {Bohrdt}}, \bibinfo {author} {\bibfnamefont {D.}~\bibnamefont {Greif}}, \bibinfo {author} {\bibfnamefont {E.}~\bibnamefont {Demler}}, \bibinfo {author} {\bibfnamefont {M.}~\bibnamefont {Knap}},\ and\ \bibinfo {author} {\bibfnamefont {F.}~\bibnamefont {Grusdt}},\ }\href {https://doi.org/10.1103/PhysRevB.97.125117} {\bibfield  {journal} {\bibinfo  {journal} {Phys. Rev. B}\ }\textbf {\bibinfo {volume} {97}},\ \bibinfo {pages} {125117} (\bibinfo {year} {2018})}\BibitemShut {NoStop}%
\bibitem [{\citenamefont {Brown}\ \emph {et~al.}(2020)\citenamefont {Brown}, \citenamefont {Guardado-Sanchez}, \citenamefont {Spar}, \citenamefont {Huang}, \citenamefont {Devereaux},\ and\ \citenamefont {Bakr}}]{Brown2020}%
  \BibitemOpen
  \bibfield  {author} {\bibinfo {author} {\bibfnamefont {P.~T.}\ \bibnamefont {Brown}}, \bibinfo {author} {\bibfnamefont {E.}~\bibnamefont {Guardado-Sanchez}}, \bibinfo {author} {\bibfnamefont {B.~M.}\ \bibnamefont {Spar}}, \bibinfo {author} {\bibfnamefont {E.~W.}\ \bibnamefont {Huang}}, \bibinfo {author} {\bibfnamefont {T.~P.}\ \bibnamefont {Devereaux}},\ and\ \bibinfo {author} {\bibfnamefont {W.~S.}\ \bibnamefont {Bakr}},\ }\href {https://doi.org/10.1038/s41567-019-0696-0} {\bibfield  {journal} {\bibinfo  {journal} {Nat. Phys.}\ }\textbf {\bibinfo {volume} {16}},\ \bibinfo {pages} {26} (\bibinfo {year} {2020})}\BibitemShut {NoStop}%
\bibitem [{\citenamefont {Atala}\ \emph {et~al.}(2014{\natexlab{b}})\citenamefont {Atala}, \citenamefont {Aidelsburger}, \citenamefont {Lohse}, \citenamefont {Barreiro}, \citenamefont {Paredes},\ and\ \citenamefont {Bloch}}]{Atala2014}%
  \BibitemOpen
  \bibfield  {author} {\bibinfo {author} {\bibfnamefont {M.}~\bibnamefont {Atala}}, \bibinfo {author} {\bibfnamefont {M.}~\bibnamefont {Aidelsburger}}, \bibinfo {author} {\bibfnamefont {M.}~\bibnamefont {Lohse}}, \bibinfo {author} {\bibfnamefont {J.~T.}\ \bibnamefont {Barreiro}}, \bibinfo {author} {\bibfnamefont {B.}~\bibnamefont {Paredes}},\ and\ \bibinfo {author} {\bibfnamefont {I.}~\bibnamefont {Bloch}},\ }\href {https://doi.org/10.1038/nphys2998} {\bibfield  {journal} {\bibinfo  {journal} {Nature Physics}\ }\textbf {\bibinfo {volume} {10}},\ \bibinfo {pages} {588} (\bibinfo {year} {2014}{\natexlab{b}})}\BibitemShut {NoStop}%
\bibitem [{not()}]{note}%
  \BibitemOpen
  \href@noop {} {}\bibinfo {note} {The VUMPS simulations are perfomed by employing the ITensor library \cite{Fishman2022}. We perform sweeps increasing gradually the bond dimension using a subspace expansion up to $\chi = 400$. The sweeps were performed until a convergence of $10^{-8}$ in the standard convergence error for VUMPS was achieved.}\BibitemShut {Stop}%
\bibitem [{\citenamefont {Fishman}\ \emph {et~al.}(2022)\citenamefont {Fishman}, \citenamefont {White},\ and\ \citenamefont {Stoudenmire}}]{Fishman2022}%
  \BibitemOpen
  \bibfield  {author} {\bibinfo {author} {\bibfnamefont {M.}~\bibnamefont {Fishman}}, \bibinfo {author} {\bibfnamefont {S.~R.}\ \bibnamefont {White}},\ and\ \bibinfo {author} {\bibfnamefont {E.~M.}\ \bibnamefont {Stoudenmire}},\ }\href {https://doi.org/10.21468/SciPostPhysCodeb.4} {\bibfield  {journal} {\bibinfo  {journal} {SciPost Phys. Codebases}\ ,\ \bibinfo {pages} {4}} (\bibinfo {year} {2022})}\BibitemShut {NoStop}%
\end{thebibliography}%

\end{document}


\title{
Supplemental Material: \textit{Frustrated extended Bose-Hubbard model and deconfined quantum critical points with optical lattices at the anti-magic wavelength}}

\author{Niccol\`o Baldelli}
\email{niccolo.baldelli@icfo.eu}
\thanks{These two authors contributed equally}
\affiliation{ICFO - Institut de Ci\`encies Fot\`oniques, The Barcelona Institute of Science and Technology, 08860 Castelldefels (Barcelona), Spain}

\author{Cesar R. Cabrera}
\thanks{These two authors contributed equally}
\affiliation{Institut f\"ur Laserphysik, Universit\"at Hamburg,
Luruper Chaussee 149, 22761 Hamburg, Germany}

\author{Sergi Juli\`a-Farr\'e}
\affiliation{ICFO - Institut de Ci\`encies Fot\`oniques, The Barcelona Institute of Science and Technology, 08860 Castelldefels (Barcelona), Spain}

\author{Monika Aidelsburger}
\affiliation{Max-Planck-Institut f\"ur Quantenoptik, 85748 Garching, Germany}
\affiliation{Ludwig-Maximilians-Universit\"at M\"unchen, Schellingstr. 4, D-80799 Munich, Germany}
\affiliation{Munich Center for Quantum Science and Technology (MCQST), Schellingstr. 4, D-80799 Munich, Germany}

\author{Luca Barbiero}
\email{luca.barbiero@polito.it}
\affiliation{Institute for Condensed Matter Physics and Complex Systems,
DISAT, Politecnico di Torino, I-10129 Torino, Italy}

\date{\today}
\maketitle

\section{Effective Heisenberg model}
\begin{figure}
    \centering
    \includegraphics{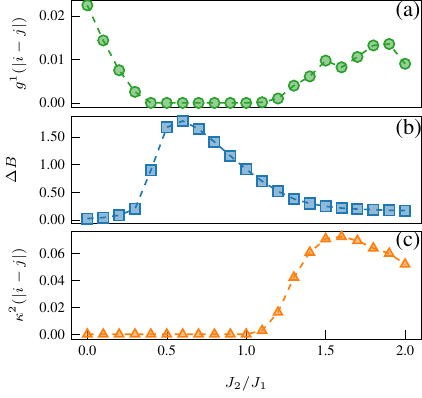}
    \caption{Phase diagram of (Eq. \ref{hamHM}) corresponding to $H_{\text{FEHB}}$ with $U=\infty$ and $\bar{n}=0.5$ where we show in (a) superfluid correlator $g^1(|i-j|)$ for $|i-j|=100$; (b) bond-order-wave order parameter $\Delta B$; (c) chiral superfluid correlator $\kappa^2(|i-j|)$ for $|i-j|=100$; The VUMPS simulation have been performed by using a bond dimension $\chi$=400 }
    \label{figSC}
\end{figure}
In the strong coupling limit $J_1/U, J_2/U\rightarrow 0$ $H_{\text{FEBH}}$ maps into the a triangular spin-$1/2$ $XY$ model
\begin{eqnarray}
\begin{split}
    H=&-2J_2 \sum_{j}(S_i^xS_{j+2}^x+S_j^yS_{j+2}^y)-\\
    &-2J_1\sum_{j}(-1)^j(S_j^xS_{j+1}^x+S_j^yS_{j+1}^y),
\end{split}
\label{hamHM}
\end{eqnarray}
where $S_j^x=\frac{1}{2}(b_j^\dagger+b_j)$ and $S_j^y=\frac{1}{2\imath}(b_j^\dagger-b_j)$ and where we fix $V=0$. Within a different gauge sector \cite{SATO2011,Furukawa2012}, this Heisenberg model accurately describes triangular frustrated quantum magnets \cite{Drechsler2007,Wolter2012,Orlova2017,Grams2022}. In Fig.~\ref{figSC} as a function of the frustration strength $J_2/J_1$ we recover the three phases SF, BOW and CSF detected in $H_{\text{FEBH}}$. This proves our experimental proposal to be highly suitable to investigate the properties of frustrated quantum magnets.

\section{Experimental implementation}

The experimental implementation involves the use of Cesium atoms confined in a one-dimensional (1D) state-dependent optical lattice. Our primary focus is on the combination of two specific atomic states: $\vert b \rangle \equiv \vert F = 3, m_\mathrm{F} = 3 \rangle$ and $\vert g \rangle \equiv \vert F = 3, m_\mathrm{F} = 2 \rangle$. The optical potential $U_{\mathrm{dip}}(\mathbf {r} ) \propto \alpha \ I (\mathbf{r}) $ experienced by the atoms is directly proportional to the laser intensity $I (\mathbf{r})$ (which follows a spatial periodic pattern for an optical lattice), and the polarizability $\alpha$. The polarizability is defined for a particular internal state ($ F, m_\mathrm{F}$), polarization and wavelength of the laser field \cite{GRIMM200095}. The term ‘‘anti-magic’’ wavelength is introduced to describe the situation where the two internal states ($\vert b \rangle$ and $\vert g \rangle$) experience an optical potential with equal amplitude and opposite sign. For this system, this happens in the vicinity of $\lambda = 871$ nm for $\sigma^+$ polarized light \cite{klostermann22}.

To achieve a large and nearly symmetric inter- and intra-species interaction, we operate at a magnetic field strength of approximately 56 G. At this field strength, the background scattering length of the three Feshbach resonances (inter- and intra-species) is $ a\sim 1000 \ a_0$ \cite{Chin2004, Frye2019}. Additionally, a narrow resonance located at approximately 56.9 G provides additional flexibility in tuning the inter-species interaction strength $a_{bg}$, allowing the control of the next-nearest (NN) interaction.

Following the lattice configuration in the Cs quantum gas microscope presented in \cite{Impertro2022}, we compute the on-site and NN interaction in the presence of a state-dependent potential. Here, two directions strongly confine the system restricting the dynamics to one dimensions, along which we apply the state-dependent lattice. The on-site $U$ and NN interaction $V$ are computed by the overlap of the Wannier functions $w_0(x)$ on each lattice site \cite{anismovas2016}:

\begin{equation}
U= \frac{4 \pi \hbar^2 a}{m} \int\left|w_0(x)\right|^4 d x,
\end{equation}
\begin{equation}
V=\frac{4 \pi \hbar^2 a_{bg}}{m}\int\left|w_0(x+\lambda/ 4)\right|^2\left|w_0(x)\right|^2 d x.
\end{equation}

\begin{figure}
    \centering
    \includegraphics{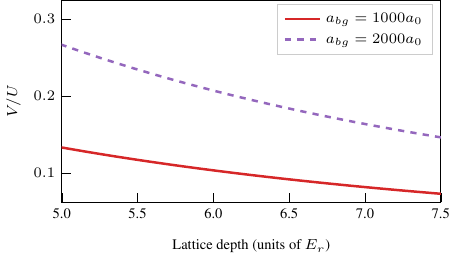}
    \caption{Ratio of nearest-neighbour $V$ and on-site $U$ interactions for two different inter-species scattering length $a_{bg}$. This is shown as a function of the lattice depth at the ‘‘anti-magic’’ wavelength in units of the recoil energy $E_R$. The intra-species scattering lengths $a_{bb}, a_{gg}$ were set to $1000 a_0$. }
    \label{figs1}
\end{figure}

The ratio of NN and on-site interactions is shown in Fig. \ref{figs1}.

\section{Detection scheme and probes}

\begin{figure}
    \centering
    \includegraphics{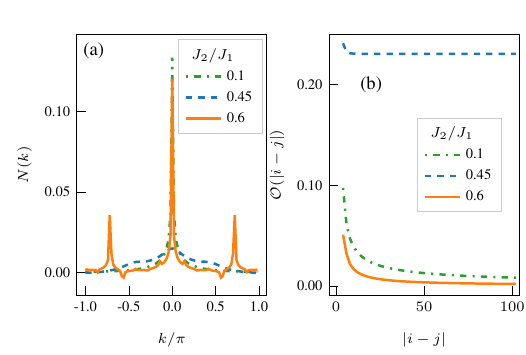}
    \caption{Momentum distribution $N(k)$ (a) and ${\cal{O}}(|i-j|)$ (b) for $U/J_1=6$, $\bar{n}=0.5$, and $J_2/J_1=0.1,0.4,0.8$ corresponding to the SF, BOW and CSF respectively, see \cite{note} for the simulations details}
    \label{fig6}
\end{figure}

The experimental detection of the DW phase can be performed in a very accurate way. Indeed the measurement of the DW local order parameter $\delta N=\frac{1}{L}\sum_j(-1)^j(n_j-\bar{n})$ requires resolving the local density $n_j$. Quantum gas microscopy allows for an impressive precision in measuring the bosonic occupation and therefore to accurately unveil the presence of the DW regime. On the other hand, the detection of the SF, BOW and CSF phases would require measurements of nearest-neighbor correlations and local currents, see Eqs. (3), (4) and (5) of the main text. While this is indeed possible \cite{kessler2014, atala14a}, we provide an alternative way to characterize the mentioned regimes. Following the results in~\cite{Julia2022}, we reveal the BOW via the string correlator 
\begin{equation}
{\cal{O}}(|i-j|)=\langle\delta n_{2i}e^{\imath\pi\sum_{k=2i+1}^{2j-2}\delta n_k}\delta n_{2j-1}\rangle,
\label{string}
\end{equation}
where $\delta n_j=(n_j-\bar{n})$. The results reported in Fig.~\ref{fig6} prove this string correlator to be highly suitable to detect the BOW. In particular, this strategy offers the fundamental advantage that Eq.~(\ref{string}) depends uniquely on the local occupation $n_j$ which, as mentioned, can be accurately measured with a quantum gas microscope. Moreover, we find that the asymptotic value of ${\cal{O}}(|i-j|)$ perfectly reflects the behavior of $\Delta B$ and thus making this strategy realistic also to detect the DQCPs.\\In order to distinguish between the two gapless phases, we can make use of the structure of the dispersion relation. In particular, while in the SF regime the dispersion relation has one minimum at the momentum $k=0$, in the CSF regime two minima at incommensurate $k$ occur. The structure of the dispersion relation can be accurately probed by the momentum distribution 
\begin{equation}
N(k)=\frac{1}{L^2}\sum_{i,j} e^{\imath|i-j|k}g^1(|i-j|).   
\label{eq:mom_dist}
\end{equation}
Our numerical analysis in Fig. \ref{fig6} reveals specifically this behavior as we find one $k=0$ peak in the SF and two peaks at different $k$s, when in the CSF. To access the momentum distribution $N(k)$, we propose using a matter wave focusing technique. Here, the momentum-space is mapped into real space after a quarter period ($T$) evolution in a harmonic trap \cite{Murthy2014, Henning2018}. In the presence of an optical lattice, the experimental protocol would consist of mapping the band population and the quasi-momentum states into real-space momentum components, followed by the expansion in the harmonic potential for a time $T/4$. Finally, the atomic distribution is frozen in a deep optical lattice for single-site imaging. Similar protocols have been proposed and implemented in \cite{Bohrdt2018,Brown2020}. Although not yet proved in optical lattice at the anti-magic wavelength, a possible alternative protocol would consist in extracting the local current $\kappa_i$ following the scheme derived in \cite{Atala2014}. In this way the CSF phase could be easily distinguished from the normal SF phase.

Based on the previous results, combining quantum gas microscopy with matter wave focusing techniques makes it possible to accurately extract the previous quantities and detect the phases reported in the main text.

\bibliographystyle{apsrev4-2}
\bibliography{biblio2.bib}